\documentclass[12pt,fleqn]{article}

\usepackage{amsthm,amsfonts,amssymb}
\usepackage[authoryear]{natbib}
\usepackage{amsmath}               
\usepackage{url}
\usepackage{authblk}
\usepackage{enumitem}
\usepackage{epstopdf}
\usepackage{graphicx}
\usepackage[font=small,labelfont=bf,margin=\parindent,tableposition=top]{caption}
\usepackage[official]{eurosym}
\usepackage[ansinew]{inputenc}
\usepackage{multirow}
\usepackage{caption}
\usepackage{subfigure}
\usepackage{placeins}
\usepackage{footmisc}
\usepackage[nouppercase]{scrpage2}
\usepackage{rotating}
\usepackage{float}
\usepackage{xcolor}
\usepackage{array}
\usepackage{bbm}
\usepackage{booktabs}
\usepackage{setspace}
\usepackage{geometry}
\usepackage[hyperfootnotes=false]{hyperref}
\numberwithin{equation}{section}
\usepackage{lmodern}
\usepackage{xifthen}
\usepackage{comment}
\usepackage{makecell}
\usepackage{dsfont}

\usepackage[title]{appendix}
\restylefloat{figure}

\hypersetup{pdfborder = 0 0 0,%
    pdftitle =  Understanding the explosive trend in EU ETS prices -- fundamentals or speculation?,%
    pdfauthor = {Marina Friedrich},%
    setpagesize = false,%
    pdfpagelayout = SinglePage,%
    bookmarksopen = true,%
    bookmarksnumbered = true,%
    pdfstartview = Fit,%
    linkcolor = black,%
    anchorcolor = black,%
    filecolor = black,%
    citecolor = black,%
    menucolor = black,%
    urlcolor = blue,%
    colorlinks = true,%
}
\newtheorem{algorithm}{Algorithm}


\title{\vspace{-20mm}\sc Understanding the explosive trend in EU ETS prices -- fundamentals or speculation?\thanks{A previous version of this paper entitled \textit{Allowance prices in the EU ETS -- fundamental price drivers and the recent upward trend} has been presented at the conferences Econometric Models of Climate Change III in Rome and EAERE in Manchester. The authors thank the participants for helpful discussions. The authors additionally thank Samuel Okullo, Oliver Tietjen, Sean Telg and Mikkel Bennedsen for valuable comments. The research leading to these results has received funding from BMBF under the research project START and the European Union's Horizon 2020 research and innovation programme under grant agreement No 730403 (INNOPATHS).}}
\author[a]{Marina Friedrich\thanks{corresponding author at: Potsdam Institute for Climate Impact Research (PIK), Member of the Leibniz Association, P.O. Box 601203, 14412 Potsdam, Germany. E-mail: \href{mailto:friedrich@pik-potsdam.de}{friedrich@pik-potsdam.de}.}}
\author[b]{S\'{e}bastien Fries}
\author[a]{Michael Pahle}
\author[a,c,d]{Ottmar Edenhofer}

\affil[a]{Potsdam Institute for Climate Impact Research -- Member of the Leibniz Association}
\affil[b]{Vrije Universiteit Amsterdam}
\affil[c]{Mercator Research Institute on Global Commons and Climate Change}
\affil[d]{Technical University Berlin}
\begin{document}
\date{}

\maketitle
\vspace{-10mm}
\begin{abstract}
In 2018, allowance prices in the EU Emission Trading Scheme (EU ETS) experienced a run-up from persistently low levels in previous years. Regulators attribute this to a comprehensive reform in the same year, and are confident the new price level reflects an anticipated tighter supply of allowances. We ask if this is indeed the case, or if it is an overreaction of the market driven by speculation. We combine several econometric methods -- time-varying coefficient regression, formal bubble detection as well as time stamping and crash odds prediction -- to juxtapose the regulators' claim versus the concurrent explanation. We find evidence of a long period of explosive behaviour in allowance prices, starting in March 2018 when the reform was adopted. Our results suggest that the reform triggered market participants into speculation, and question regulators' confidence in its long-term outcome. This has implications for both the further development of the EU ETS, and the long lasting debate about taxes versus emission trading schemes.
\end{abstract}


\thispagestyle{empty}

\clearpage
\setcounter{page}{1}
\doublespacing
\section{Introduction}
\label{sec:intro}

\begin{figure}[t]
\centering
\includegraphics[width=0.8\textwidth,trim=0cm 1cm 0 1cm, clip]{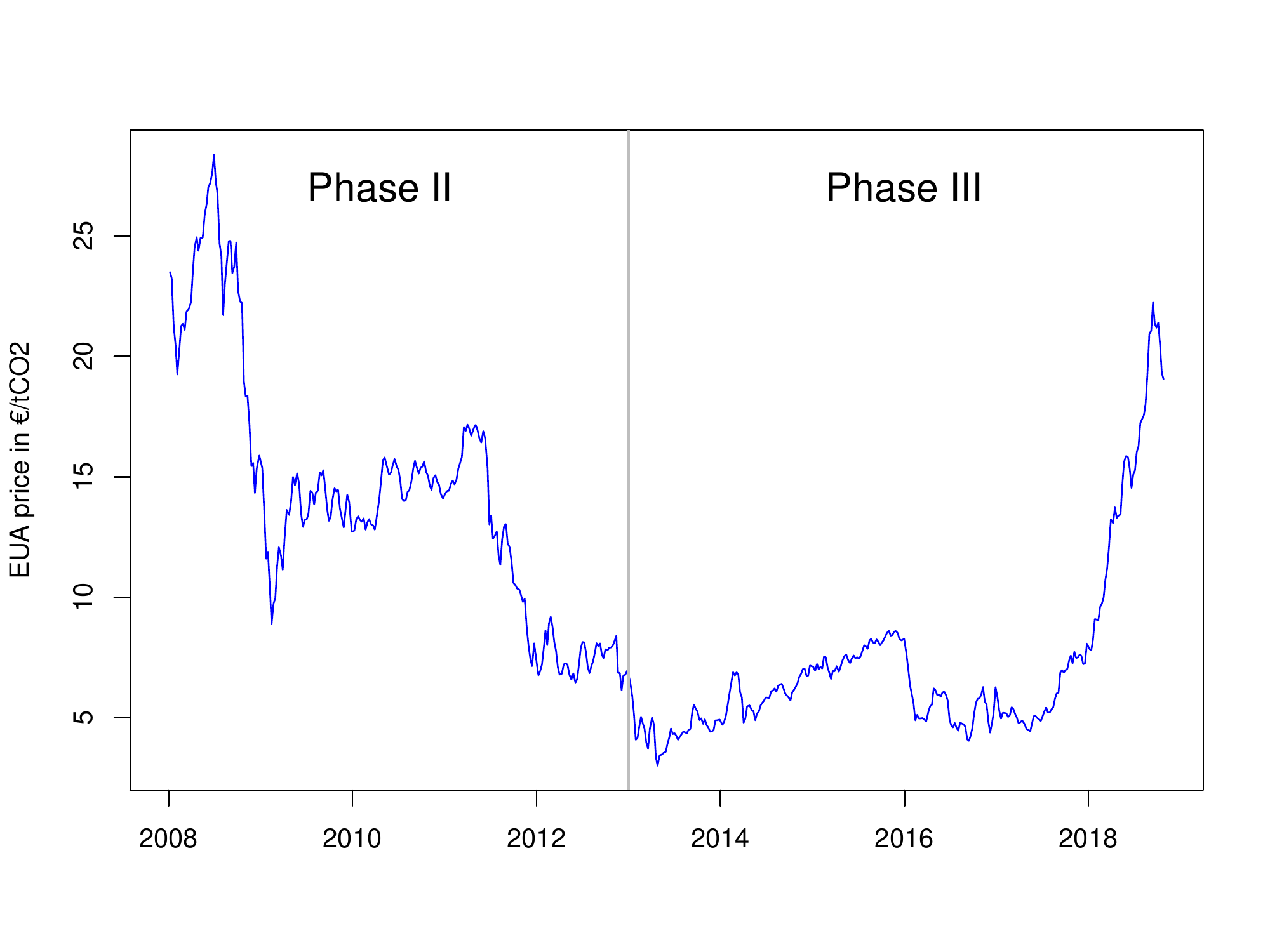}
\caption{EUA price development -- nearest December Futures Contract\\\textit{Source}: EEX Leipzig; see Section 3.1 for details.}
\label{fig:EUAprice}
\end{figure}

The EU Emissions Trading Scheme (EU ETS) is the flagship policy for regulating carbon dioxide (CO2) emissions in Europe. Being implemented in 2005, allowance prices in the EU ETS went through periods of highs and lows. Following the onset of the third trading period in 2013, prices reached an all-time low of around 3 Euro per tonne of CO2 (\euro{}/tCO2). This led EU policy makers to initiate a series of reforms primarily intended, in their own words, to "reduce the surplus of emission allowances in the carbon market".\footnote{\url{https://ec.europa.eu/clima/policies/ets/revision_en}} The reform process was concluded with an amendment of the EU ETS directive in March 2018. Its  main new feature is the cancellation of allowances from 2023 on through the Market Stability Reserve (MSR), which will tighten the cap to a yet unknown extent. In the wake of the reform, prices rallied from a level of around 10 \euro{}/tCO2 in March 2018 to 25 \euro{}/tCO2 by the end of 2018. This is shown in Figure \ref{fig:EUAprice} which plots the prices for European Emission Allowances (EUA) from 2008 to 2018.

This paper examines potential causes of the recent price run-up: we assess if market fundamentals have gained importance in allowance price determination; additionally, we investigate if the price increase represents the inflationary phase of a bubble. A bubble could have been caused by an overreaction of the market to the reform. According to recent surveys, market participants view the anticipated cancellation of allowances through the MSR as the most important driver of the run-up \citep{ZEW2019,ThomsonReuters2019}. Yet in both surveys, also speculative buying ranks high. Furthermore, rumor in media and market intelligence has it that much of the price run-up was due to trading activities from speculative investors like banks and hedge funds in anticipation of the (price) effect of the reform.\footnote{See for example \url{https://www.ft.com/content/6e60b6ec-b10b-11e8-99ca-68cf89602132} and \url{https://carbon-pulse.com/93184/}} Indeed, overall trading volumes in 2018 increased by 42\% compared to 2017 \citep{Marcu2019}. While this does not rule out market fundamentals as potential price driver, it deserves further investigation.

Taking an empirical approach, this paper obtains three main results. First, our analysis suggests that typical abatement-related fundamentals can be excluded as driver of the price increase. This conclusion is based on two observations: (i) demand-side fundamentals have not gained importance as price determinants; (ii) none of the price drivers show explosive behaviour coinciding with the increase in allowance prices. The fact that allowance prices underwent an explosive period constitutes our second main result. Our empirical approach timestamps the onset of this period to March 2018, the month in which the reform was adopted. Third, the estimated crash odds of the explosive period after October 2018 are quite high, suggesting that the rate of increase was unsustainable and that a potential bubble is likely to collapse within a year. Looking at the out-of-sample price developments from October 2018 onward does not corroborate this finding. Nevertheless, the model underlying our approach fits the explosive period very well. This suggests, together with the previous test results, that the reform triggered market participants into speculation. The subsequent consolidation of prices is in line with two potential explanations. Either, there has been a stabilisation of beliefs about the future impact of the reform, or a price collapse to pre-bubble levels is overdue. 

To our knowledge, we are the first to provide empirical evidence in this direction. Our paper expands the empirical literature in three key ways. First, we conduct an analysis to clarify the role of the fundamentals. While theory predicts that, in particular, coal and natural gas prices are important fundamental drivers, previous research could not provide clear empirical evidence. As a possible explanation for this we find evidence that the relationship between the allowance price and its fundamental drivers is unstable over time. We employ a regression approach with time-varying coefficients in which the form of the time-variation does not have to be specified in advance. Our particular interest is to see whether the price run-up is reflected in a parallel increase of the coefficients. To that end we analyse prices from the beginning of Phase II on to see if the coefficients undergo a structural change in the period after the reform compared to the years before it. Second, pursuing the question of a speculative bubble, we formally test for periods of explosive behaviour in allowance prices as proposed by \cite{PSY}. Such behaviour could be a primary indicator for the inflationary phase of a speculative bubble. The flexible date stamping approach allows us to conclude that the adoption of the reform coincides with the onset of the explosive period. Third, we estimate the crash odds of the explosive episode using the method proposed by \cite{fries2018conditional} which we can compare to the current (out-of-sample) evolution of prices.

These findings are of particular relevance in light of EU regulators' stance towards the outcome of the reform. Notably, the EU's leading climate policy maker is confident that after the reform allowance prices will "only go in the right [upward] direction"\footnote{Satement by Frans Timermans, Executive Vice-President of the EC; see  \url{https://www.euractiv.com/section/climate-environment/news/timmermans-talks-the-talk-in-ep-hearing-the-main-points/}}.  If anything, our results put this confidence into question by providing empirical evidence that prices may well collapse again. Taking this risk seriously has important implications for the further development of the EU ETS. It in particular emphasises the need to introduce mechanisms to stabilise expectations about long-term prices - in a way that can also remedy the more profound long-term policy commitment problem haunting the EU ETS.

The paper is structured as follows. Section \ref{sec:lit} reviews the related literature. In Section \ref{sec:data}, we introduce our dataset. Section \ref{sec:fundamentals} analyses the relationship between allowance prices and its fundamental price drivers. Subsequently, Section \ref{sec:trend} investigates potential explosive behaviour in allowance prices. First, we test for explosive behaviour. Second, based on the findings which deliver evidence of such behaviour in our series, we model the explosive period and estimate probabilities of collapse at different time horizons. Section \ref{sec:conclusion} concludes.

\section{Related literature}
\label{sec:lit}
This section is devoted to a discussion of the literature related to this paper. It is divided in two parts. First, we review the empirical literature on fundamental price drivers. Second, we move to the literature on speculative bubbles. We briefly treat bubble formation and mention related papers using bubble testing as well as bubble modeling. 

\subsection{Fundamental price drivers}
According to economic theory, market fundamentals such as coal and gas prices as well as economic activity should have a major effect on allowance prices. A study of the related literature shows, however, that empirical evidence is mixed. Previous studies indicate that standard approaches, such as linear regression models, need to be adapted by splitting the sample, including breaks or dummy variables in order to improve their findings. This is discussed in a review of the empirical literature of price formation in the EU ETS by \citet{Friedrich2019}. 

Time constant models have frequently been used in this context; for instance, in \citet{Hintermann2010}, \citet{Koch2014a} and \citet{Aatola2013}. The question whether it might be more appropriate to account for potential time-variation when modeling the relationship between allowance prices and fundamentals has been raised in \citet{Lutz2013} who consider potential non-linearities. Using a regime-switching model, they distinguish two different pricing regimes - one applies during periods of high volatility and the other during periods of low volatility. By construction, the impact of explanatory variables on the allowance price can differ among the two regimes. In both regimes, they find the same set of relevant price drivers. Coal and gas prices, oil prices and the stock index are statistically significant determinants of the EUA price. In Regime 2, which is characterised by low and constant volatility, all significant price drivers show the anticipated sign. Regime 1, however, shows a positive impact of the coal price. This goes against economic considerations that predict, as in the second regime, a negative effect of the coal price on allowance prices. In a recent contribution, \citet{Rodriguez2019} apply a time-varying coefficient model to assess the impact of stock price indices on the allowance price and find a time-varying effect.\footnote{This paper differs from our work as it does not include other price drivers apart from the stock price indices. Moreover, a different model and estimation method are used and no confidence intervals are provided.} Both papers provide further evidence that the relationship between the allowance price and its fundamentals might not be constant over time but can be subject to (structural) changes. 

The effect of the coal price on allowance prices causes further disagreement in findings. Similar to the results in the second regime of \citet{Lutz2013}, \citet{Rickels2014a} find a positive effect of the coal price on the allowance price in their single variable analysis. The paper by \citet{Rickels2014a} differentiates itself from previous studies, because the authors investigate the effect of the choice of data series by performing various regressions with only one explanatory variable. They consider multiple data series for the different factors, from different sources and with different sampling frequency (daily and weekly). In their final regression specification, they do not include the coal price as a separate explanatory factor but as part of the switching price. Further evidence is found in \citet{Aatola2013} for the period 2005-2010 who find a negative coefficient of coal, while \citet{Hintermann2010} finds it to be insignificant in Phase I data. In addition, \citet{Koch2014a} look at the entire second Phase II and the first year of Phase III and find an insignificant coefficient of coal. However, the explicitly calculated fuel switching price is found to have a significant effect. Regarding the effect of the gas price on allowance prices, there is no ambiguity. All studies find a positive and significant coefficient of the gas price independent which approach is used. In particular, in \citet{Hintermann2010} it is the only explanatory variable with a significant effect throughout all considered specifications.

Our paper adds to this strand of the empirical literature by estimating a flexible regression model with time-varying coefficients to a more recent dataset. In our approach, the researcher does not have to determine the form of the time-variation, it is estimated from the data under mild assumptions on the coefficient curves. 

\subsection{Testing for and modeling explosive episodes}
Regarding the second part of our paper, the theoretical literature discusses a number of bubble generating mechanisms. \citet{PSY} provide a short overview of these mechanisms including rational bubbles, intrinsic bubbles, herd behaviour, and time-varying discount factor fundamentals. For most of them, information in the form of news and stories play an important role. In particular, herd behaviour can lead to short-term mispricing when there is uncertainty about the effect of a shock to the asset value and the quality of traders' information about it \citep{Avery1998}. Experiments confirm that if some traders are better informed than others, bubbles can occur \citep{Oechssler2011}. Moreover, \citet{Shiller2017} proposes that contagious stories can lead to bubbles. They can occur when "contagion is altered by the public attention to price increases: rapid price increases boost the contagion rate of popular stories justifying that increase, heightening demand and more price increases" (p.983). In a similar vein, recent theoretical work finds that good news about fundamentals can trigger large price bubbles when investors "waver" between extrapolating from past prices and measuring the difference between current price and rational valuation of the final cash flow when forming their demand \citep{Barberis2018}.

There is a number of related papers in the empirical literature which use the same test for explosive behaviour as in this paper. Mostly, they consider different markets, like e.g. \citet{Corbet2018} who study bubbles in Bitcoin price series as well as \citet{Shi2017} who investigates bubbles in the US housing market. \citet{Sharma2018} study the explosive behaviour of 8 energy sector series. They find evidence of such behaviour in many of their considered series using weekly data until December 2015. One related paper by \citet{Creti2017} investigates EU ETS prices. The authors find evidence of short explosive episodes in their sample from 2005 to 2014. However, the detected periods do not last longer than a few days -- only 2 out of 11 last longer than 5 days, among them one negative bubble lasting 12 days and one positive bubble lasting 9 days. Our paper extends this finding by applying the test to an updated dataset which includes the rapid price increase in allowance prices which is the most interesting period to study in the context of this test. \\
\indent As we obtain statistical evidence for an explosive behaviour towards the end of our sample, we propose a modelling approach of the recent upward trend using noncausal processes.
Noncausal autoregressive processes were recently proposed by \cite{gourieroux2017local} to model speculative bubbles in financial markets because of their ability to mimic locally explosive patterns in time series data.
A rapidly emerging econometric literature has been applying these models to diverse bubble phenomena appearing for instance in time series of inflation rates \citep{hecq2019identification},
stock indexes \citep{gourieroux2017local, fries2018conditional}, 
cryptocurrency exchange rates \citep{hencic2015noncausal,cavaliere2020bootstrapping},
volatilities \citep{hecq2016identification}, 
and commodity prices \citep{fries2019mixed,voisin2019forecasting,hecq2019predicting}.
We leverage the analytical expression of the crash odds of explosive episodes obtained by \cite{fries2018conditional} in the framework of noncausal processes  to propose a quantification of the (un)sustainability of the recent explosive trend in EU ETS prices.

\vspace{-4mm}
\section{The data}
In this section, we first present our data set. In a second part, we carefully investigate the stationarity properties of each time series. This is an important starting point before applying our proposed methods as some of them require the data to be stationary.
\label{sec:data}
\subsection{EUA prices and explanatory variables}
We consider weekly data for the period from January 2008 to October 2018 resulting in $T=538$ observations. This sample period covers the entire Phase II and a large part of Phase III. As allowance price series, we use the December futures contract traded on the European Energy Exchange (EEX). The contract is rolled over to the next contract at the end of October each year. The resulting continuous price series is displayed in Figure \ref{fig:EUAprice}. Our sample captures the entire period of rapid increase in allowance prices which started in mid 2017. It ends with the first point of considerable decline at the end of October 2018. This allows us to investigate the entire trending period while leaving enough room for out-of-sample prediction of crash odds in the final step of our empirical approach. 

We choose December futures prices like most related papers since they are frequently traded \citep[e.g.][]{Koch2014a,Lutz2013,Aatola2013}. As our main set of explanatory variables, we include natural gas and coal prices as month-ahead futures from the same platform as well as the stock index STOXX Europe 50 as an indicator of current and expected economic activity.\footnote{Other commonly used indicators such as the Industrial Production Index or the Economic Sentiment Indicator are measured at a lower frequency such that we do not consider them here. Another promising indicator could be nighttime light from satellite observations.} As an alternative, we use data on a comparable index, which is sometimes used in this context, the STOXX Europe 600 index.\footnote{The stock index data are retrieved  on 21.01.2019 from \url{https://quotes.wsj.com/index/XX/SXXP/historical-prices} (STOXX Eur 600) and \url{http://quotes.wsj.com/index/XX/SX5E/historical-prices} (STOXX Eur 50).} Further, we consider the oil price. In the related literature, there is no clear agreement on whether its effect is due to being a proxy for economic activity or if it comes from the (limited) fuel switching from oil to gas \citep{Hintermann2016}. In addition, we use daily mean temperature data for seven European cities from the European Climate Assessment \& Dataset (ECA\&D) presented in \citet{KleinTank}. We transform the data into a series of weekly temperature averages and we take out the seasonal component with the help of Fourier terms. Details on the location of the stations and the Fourier term regression can be found in Appendix \ref{sec:temp_data}. 
\begin{figure}[!h]
	\centering
	\subfigure[Coal]
    {
    \includegraphics[width=0.47\linewidth, clip, trim = {0 1cm 0 2cm}]{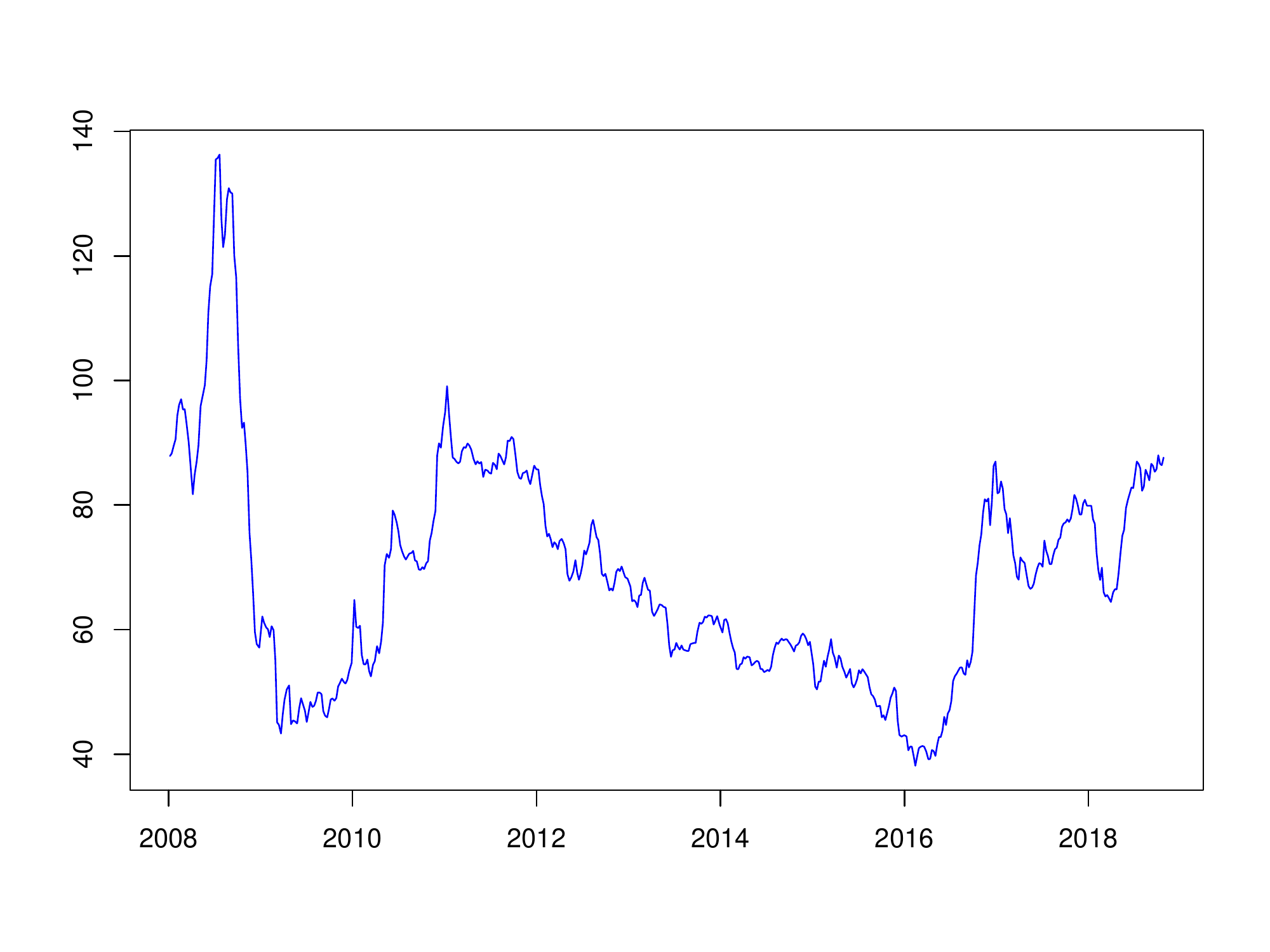}	
    \label{fig:coal_prices}
     }
     \subfigure[Gas]
     {
      \includegraphics[width=0.47\linewidth, clip, trim = {0 1cm 0 2cm}]{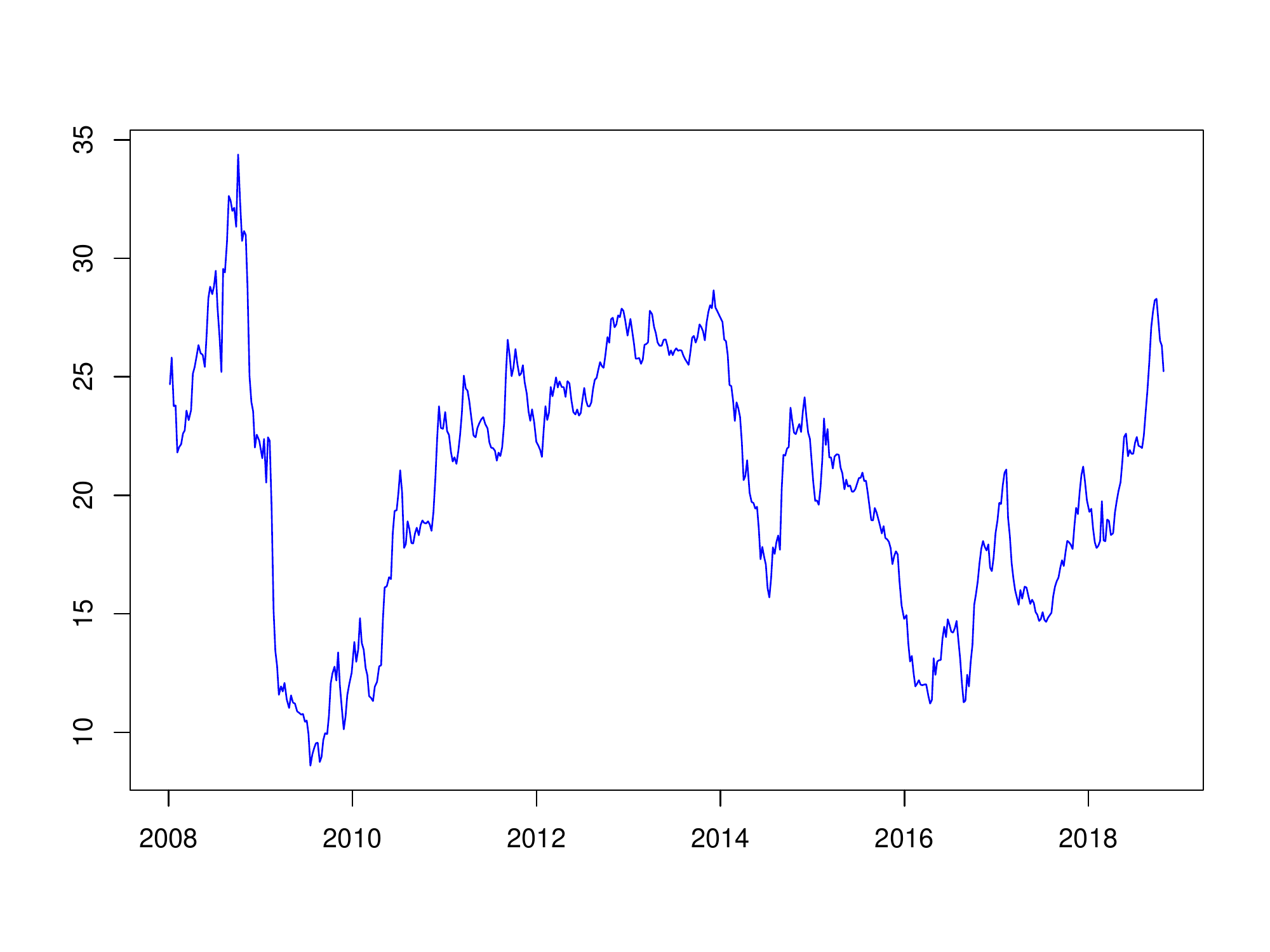}
      \label{fig:gas_prices}
     }\\
		\subfigure[Oil]
    {
    \includegraphics[width=0.47\linewidth, clip, trim = {0 1cm 0 1.5cm}]{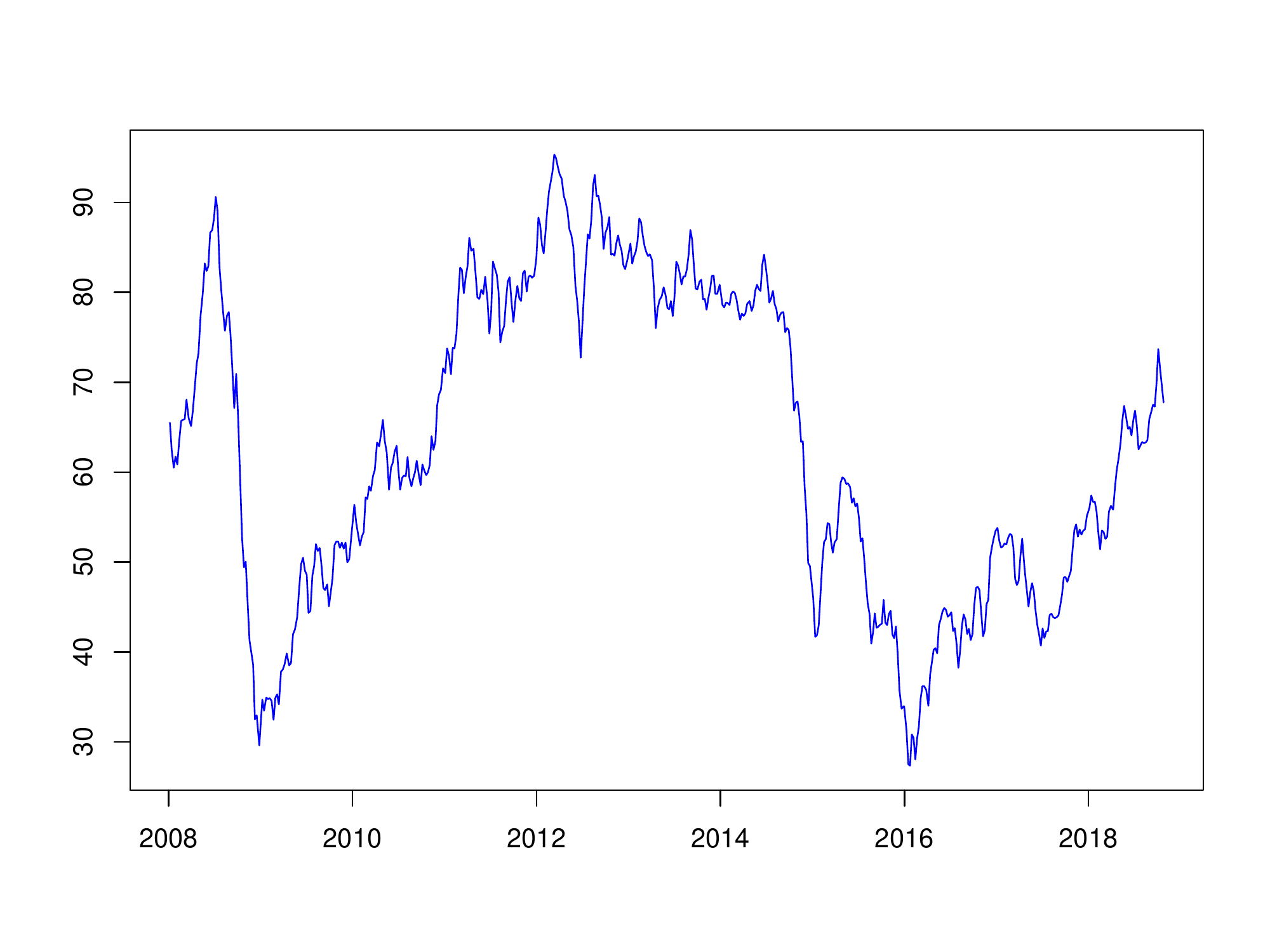}	
    \label{fig:oil_prices}
     }
     \subfigure[Stoxx 50]
     {
      \includegraphics[width=0.47\linewidth, clip, trim = {0 1cm 0 1.5cm}]{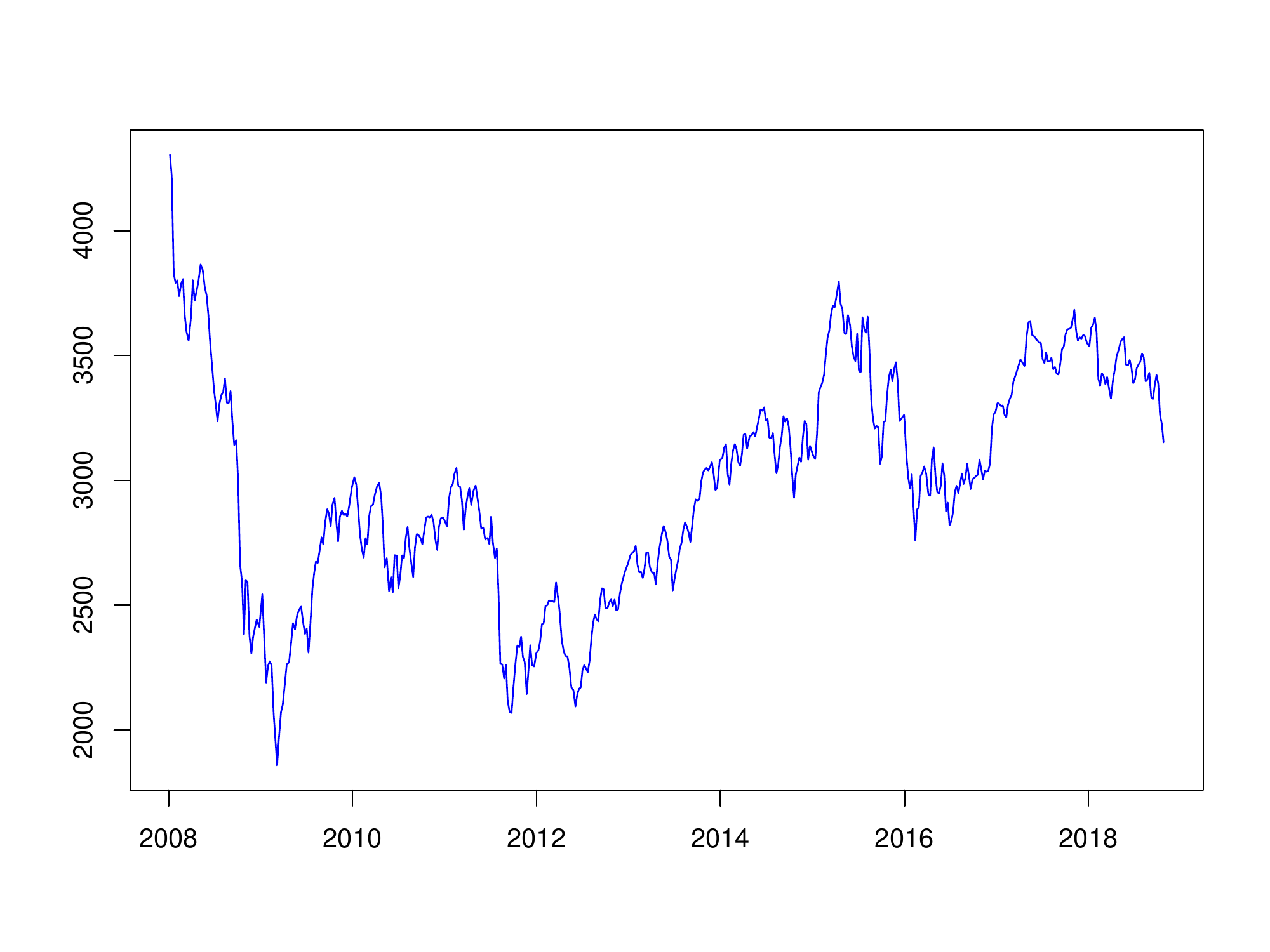}
      \label{fig:stoxx50_data}
     }
     \subfigure[Stoxx 600]
     {
      \includegraphics[width=0.47\linewidth, clip, trim = {0 1cm 0 1.5cm}]{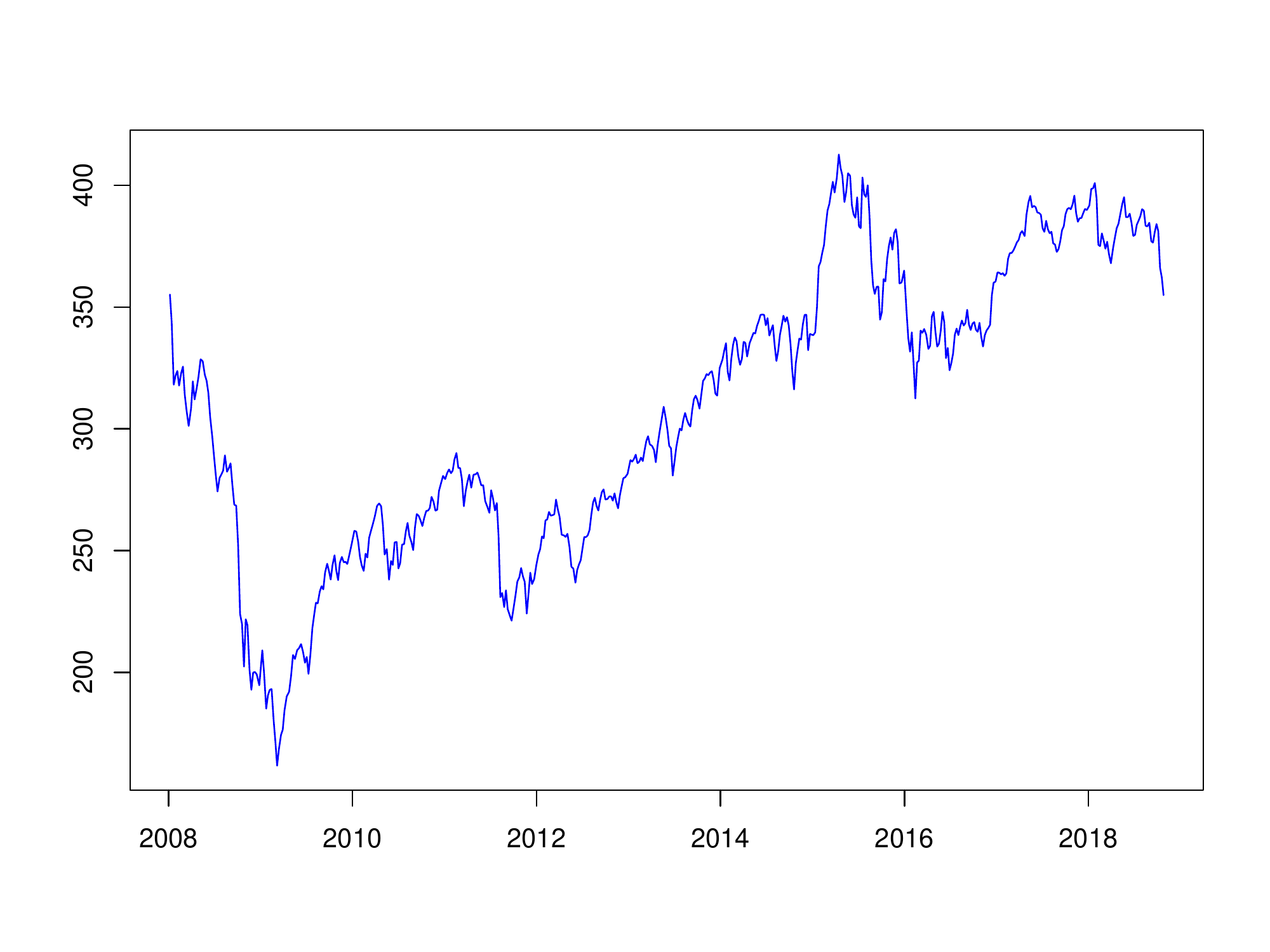}
      \label{fig:stoxx600_data}
     }
     \subfigure[Temperature averages]
     {
      \includegraphics[width=0.47\linewidth, clip, trim = {0 1cm 0 1.5cm}]{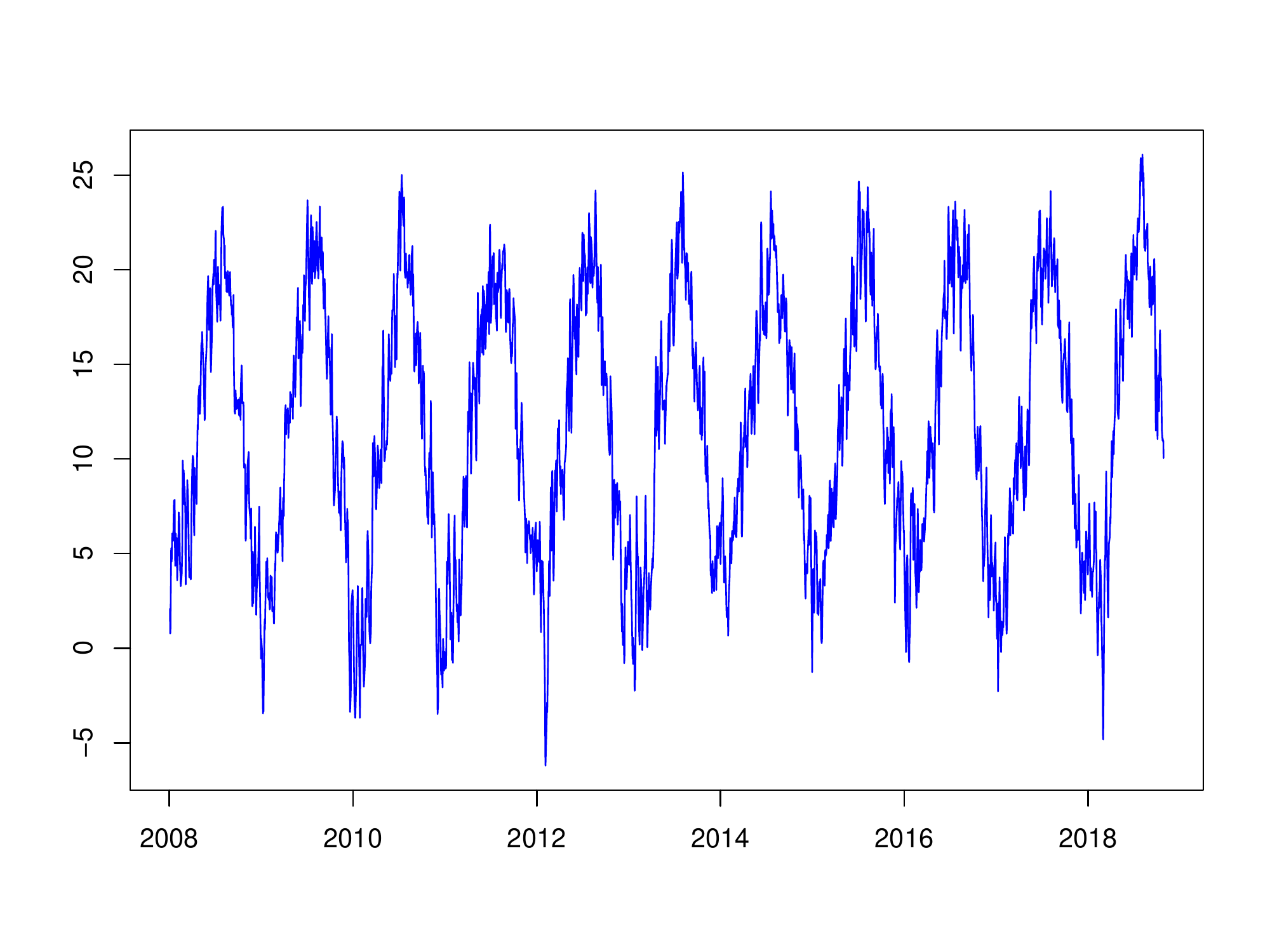}
      \label{fig:temp_data}
     }
		\caption{Plot of the time series in our data set\\\textit{Source:} For a detailed description of the sources we refer to Section 3.1.\\\textit{Notes:} In Panels (a) to (c), prices are given in EUR, while the coal and the oil price was transformed from USD into EUR before plotting. Panels (d) and (e) are given in index values and (f) in degrees Celsius.}
		\label{fig:timeseries}
\end{figure}

The gas price is the settlement price of month-ahead Dutch TTF futures, denoted in EUR/MWh. TTF stand for Title Transfer Facility and is a virtual trading point for natural gas in the Netherlands. The TTF futures contract is a frequently used series for gas prices in the related literature. Similarly, the coal price we consider is the settlement price of month-ahead futures based on the API2 index of the ARA region (Amsterdam-Rotterdam-Antwerp). The contract size is 1,000 metric tonnes of thermal coal.\footnote{To convert the coal price data into EUR/MWh, one simply has to divide the series by the conversion factor of 8.14. Since the conversion factor is constant and we consider first differences, this would not change our results.} Both are obtained from the EEX. For oil we rely on the historical futures prices (continuous contract) of Brent crude oil based on data from the Intercontinental Exchange (ICE), retrieved from Quandl.\footnote{Retrieved from \url{https://www.quandl.com/data/CHRIS/ICE_B1} on 21.01.2019} The contract size is 1,000 barrels. The coal and the oil prices need to be converted into EUR, as they are denoted in USD. This is done using USD/EUR exchange rate data from Tullett Prebon.\footnote{Retrieved from \url{https://quotes.wsj.com/fx/EURUSD/historical-prices} on 21.01.2019}  

Various other variables -- e.g. data on renewable energy production, issued Certified Emission Reductions and fuel switching prices -- appear in the literature. Apart from the switching price, these variables are often found to be insignificant or to have a negligible effect in terms of the magnitude of the coefficients. Further limitations arise, since data on wind, solar or hydro production is only available for specific countries or regions. In our preliminary regression analysis we consider data on energy supply from hydro power in Norway from the Norwegian Water Resources and Energy Directorate\footnote{Retrieved from \url{http://vannmagasinfylling.nve.no/Default.aspx?ViewType=AllYearsTable&Omr=EL} on 21.01.2019} as well as data on electricity generation from wind for Germany obtained from the database of the European Network of Transmission System Operators for Electricity (ENTSO-E). None of the time series showed a significant effect in our regression analysis.

We also calculate a fuel switching price from our coal and gas price data. However, since it is a linear combination of the two series, the effect is already captured with the inclusion of the individual price series. We show this in Appendix \ref{sec:add_var} where we also present the results from the inclusion of the additional variables hydro and wind power generation. In the main text, we focus on the set of classical abatement-related price drivers: coal and gas prices as well as economic activity and temperature. The time series are plotted in Figure \ref{fig:timeseries}. We observe in Panels (b) and (c) that both the gas and the oil price display a similar upward trend as the allowance price at the end of our sample period. As both variables influence the allowance price, their development could be a potential driver of the trend in allowance prices. However, our further analysis shows that this is not the case. 

\subsection{Stationarity properties}
\label{sec:stationarity}
One of the assumptions of the regression analysis we perform in Section \ref{sec:fundamentals} is that the data are stationary. Previous research finds that the allowance price series contains a unit root and therefore, the log return series, calculated as $r_{yt}=ln(y_t/y_{t-1})$, is used whenever stationarity is needed.\footnote{In the remainder of the paper we often leave out the term "log" and use the terms "log returns" and "returns" interchangeably.} 

To investigate the stationarity properties of our data, we use a battery of unit root tests. Most of them consider a unit root under the null hypothesis versus stationarity around a linear trend under the alternative. Standard unit root tests, e.g. the Augmented-Dickey-Fuller test \citep{ADF}, which have been applied in the previous literature, consider a fixed mean or a linear trend under the alternative. It has been shown that this can quite frequently lead to spurious rejections of the unit root null hypothesis. Working with data in first differences if the data are trend stationary can substantially change the results. Therefore, it is crucial to carefully select the alternative hypothesis of unit root tests. 

Next to the Augmented-Dickey-Fuller (ADF) test with a linear trend, we apply the tests by \citet{PP}, \citet{LNV} and \citet{KPSS}. While the Phillips-Perron (PP) test has very similar properties, the test by \citet{LNV} (LNV) considers a smoothly varying time trend under the alternative hypothesis. The trend can undergo one transition and the time point as well as speed of the transition is determined endogenously by the test. This modification allows for much more flexibility under the alternative which is suitable for our complex data series.The KPSS test by \citet{KPSS} is the only test that considers (trend) stationarity under the null hypothesis and a unit root process under the alternative.

\begin{table}[h!]
\centering
\caption{Unit root tests}
\begin{tabular}{lcccccccc}
\midrule\midrule
 & \multicolumn{2}{c}{ADF test} & \multicolumn{2}{c}{PP test} & \multicolumn{2}{c}{LNV test} & \multicolumn{2}{c}{KPSS test} \\\midrule
 & $y_t$ & $r_{y_t}$ & $y_t$ & $r_{y_t}$ & $y_t$ & $r_{y_t}$ & $y_t$ & $r_{y_t}$\\\cmidrule(lr){2-3}  \cmidrule(lr){4-5} \cmidrule(lr){6-7}  \cmidrule(lr){8-9} 
 EUA & -0.56 & -14.13 & -0.68 & -17.64 & -1.01  & -14.34  & 1.31 & 0.07\\
 Coal & -2.24 & -12.88 & -2.09  & -16.49 & -2.34  & -13.05 & 0.55 & 0.06 \\
 Gas & -2.00 & -13.59 & -2.11  & -18.68  & -2.03 & 13.72  & 0.57 & 0.07 \\
 Oil & -1.78 & -13.88 &  -1.76 & -18.59  & -2.033 & -13.96 & 1.13 & 0.08 \\
 Stoxx 50 & -3.42 & -15.97 & -4.34  & -19.11  & -3.71 & -16.21  & 0.74 & 0.10 \\
 Stoxx 600 & -3.45 & -15.90 & -4.05  & -19.87 & -3.50 &  -16.30 & 0.45 & 0.07\\
 Temp & -13.05 & -- &  -15.03 & -- & -13.15 &  -- & 0.07 & --\\\cmidrule(lr){1-9}
 & \multicolumn{8}{c}{Critical values: (90\%, 95\%, 99\%)} \\\cmidrule(lr){1-9} 
& \multicolumn{4}{c}{(-3.13, -3.42, -3.98)} &  \multicolumn{2}{c}{(-4.55, -4.83, -5.42)} & \multicolumn{2}{c}{(0.12, 0.15, 0.22)}\\
\midrule\midrule
\end{tabular}
\label{tab:unitroot}
\caption*{\small\textit{Source:} Test statistics from own calculations, critical values from the original papers.\\\textit{Notes:} Results from four different unit root tests. The top part gives the test statistics for the data in levels ($y_t$) and in log returns ($r_{yt}$). The bottom part shows the 90\%, 95\% and 99\% critical values. The critical values for the ADF and PP tests coincide. The tests were performed with the Bayesian Information Criterion for lag length selection and a maximum number of 8 considered lags. The chosen lag length equals 1 in all cases.}
\end{table}

The results are presented in Table \ref{tab:unitroot}. The ADF test results are presented in the first two columns. They show that in most cases, the unit root null hypothesis cannot be rejected at a 1\% significance level when we look at the data in levels. Exceptions are the temperature series and the two stock indices (at a 10\% level for STOXX 50 and at a 5\% level for STOXX 600). The return series are all stationary. The PP test which is given in the next two columns are very similar. The unit root null hypothesis is now also rejected for the two stock indices at the 1\% significance level. Results of the more flexible LNV are given in the next two columns. Again, all series but the temperature data contain a unit root. The KPSS test, which is presented in the last two columns, comes to the same conclusion. 

Given these results, we conclude that all fuel prices and stock indices contain a unit root and the temperature data are stationary.

\vspace{-4mm}
\section{Fundamental price drivers -- revisited}
\label{sec:fundamentals}
If the price run up reflects an increasing scarcity of allowances, fundamentals should have gained importance as price drivers. Accordingly, coefficients would have increased gradually, which we examine using a linear regression model with time-varying coefficients. Such models are flexible, and allow uncovering changes in the relationship without restrictive assumptions regarding the form of the change. We analyse the full time series and not simply the period of the price run up between 2017 and 2018. In fact, we find significant changes in coefficients in earlier years that have not been identified in the previous literature. 

The longer time span of our dataset compared to previous studies provides us with an opportunity for a better understanding of the relationship between allowance prices and abatement-related price drivers. In addition, to be able to compare our results to previous findings, we also run a linear regression model with constant coefficients. 

Both approaches, constant and time-varying coefficients, can be captured using the following linear regression framework. For $t=1,\dots,T$, let 
\begin{equation}
r_{EUA,t}=\beta_{0,t}+\beta_{1,t}x_{1,t}+\beta_{2,t}x_{2,t}+\cdots+\beta_{m,t}x_{m,t}+\epsilon_t=\boldsymbol{\beta}_t'\mathbf{x}_t+\epsilon_t
\label{eq:model}
\end{equation}
be a model for the return on allowance prices $r_{EUA,t}$, where $\boldsymbol{\beta}_t=\left(\beta_{0,t},\beta_{1,t},\dots,\beta_{m,t}\right)'$ is a vector of coefficients and $\mathbf{x}_t=\left(1,x_{1,t},\dots,x_{m,t}\right)'$ is a set of potential (stationary) price drivers. The error term $\epsilon_t$ may be correlated and may exhibit changing variance over time. This simple framework allows for constant coefficients by imposing the restrictions $\beta_{j,t}=\beta_j$, for $j=0,\dots,m$ and for all $t=1,\dots,T$. We estimate this linear regression specification by ordinary least squares (OLS). 

Without restrictions on the coefficients, Equation \eqref{eq:model} represents a time-varying coefficient model including a deterministic time trend $\beta_{0,t}$ as well as covariates with time-varying coefficient functions $\beta_{j,t}$ for $j=1,\dots,m$. All coefficients are assumed to be functions of time which need to follow certain smoothness conditions.\footnote{The assumptions require that, for instance, the functions are twice continuously differentiable and bounded. We refer to \citet{Cai} for the complete set of assumptions. In addition, as a necessary step to ensure the consistent estimation of these functions, they have to be defined on the interval $\left(0,1\right)$. Therefore, we need to map all points to this interval such that, formally, we have $\beta_{j,t}=\beta_j\left(t/T\right)$. This is explained, for instance, in \citet{Robinson}.}  

A key advantage of the time-varying coefficient model is that the form of the coefficient functions does not have to be specified in advance. It lets the data determine the form of the relationship. Due to this nonparametric nature the model offers a large degree of flexibility and generality. Estimation of such models is well-established in the econometric literature. We use the local linear kernel estimator of \citet{Cai}. The estimator fits a locally weighted least squares regression to a neighborhood around each time point in the sample. The weighting function is called the kernel function. It has been shown that the estimator has properties in finite sample and it is known to reduce the boundary effects which are a concern in nonparametric estimation. For more information on the estimator and its properties, we refer the interested reader to \citet{FG} for a general overview, and to \citet{Cai} for details on the estimator in the context of model \eqref{eq:model}.

To be able to judge the significance of our results, we construct 95\% confidence intervals around the nonparametric estimates using a bootstrap method. We rely on the autoregressive wild bootstrap which offers robustness to serial correlation as well as heteroskedasticity. Using nonparametric estimation with bootstrapping is a powerful combination which has been studied in the econometric and statistical literature by e.g. \citet{Buhlmann}, \citet{NP} and \citet{FSU}. Details on the estimation technique and the bootstrap algorithm are given in the Technical Appendix. It is particularly suitable for applications in the EU ETS market as in previous studies the residuals of the model experienced heteroskedasticity \citep[e.g.][]{Koch2014a,Lutz2013}. An additional advantage of the method is that it can also be applied when data points are missing such that there is no need to resort to interpolation techniques when some data series are incomplete. Although missing data are not a major concern in our application, it nevertheless occasionally happens that single data points are missing from some series. In this case, we simply delete the corresponding data points from our dataset. All empirical results in this section are obtained using the stationary data as determined in Section \ref{sec:stationarity}.

Both model specifications of \eqref{eq:model} are not designed to explain sudden jumps in allowance prices. To investigate whether this is a serious issue in our application, we remove outliers and subsequently re-estimate the model.\footnote{Results for the original data (before the removal of outliers) are very similar and they can be found in Appendix \ref{sec:outlier}.} For this we apply the impulse indicator saturation (IIS) approach proposed in \citet{Santos2008}. With this method, conditioning on our set of explanatory variables, we find seven outliers in the EUA return series which are removed. More information on the location of the outliers and the IIS application can be found in Appendix \ref{sec:outlier}.

\subsection{Linear regression results}
\label{sec:lin_reg}
To get a first understanding of the data and to be able to compare our results to the previous literature, we obtain a set of results from Model \eqref{eq:model} with constant coefficients.  
Table \ref{tab:lin.reg} displays the results from the corresponding OLS regression. We present the coefficient estimates with corresponding Newey-West standard errors \citep{Newey1987}, which are robust to mild forms of autocorrelation and heteroskedasticity. We consider three specifications, where we include the oil price, the STOXX 50 index and the STOXX 600 index, respectively. In specification (a), the two significant factors are the gas and the oil price. The coal price, as an important driver, does not show a significant effect on the allowance price in this initial regression. The estimates are very similar in specifications (b) and (c) where we include the stock indices.

\begin{table}[h]
\centering
\caption{Linear Regression Results}
\begin{tabular}{lccccccccc}
 \midrule\midrule
 & \multicolumn{3}{c}{(a)} & \multicolumn{3}{c}{(b)} & \multicolumn{3}{c}{(c)} \\\cmidrule(lr){2-10}
 & $\hat{\beta}_j$ & $se_{NW}$ & $p$-value & $\hat{\beta}_j$ & $se_{NW}$ & $p$-value & $\hat{\beta}_j$ & $se_{NW}$ & $p$-value \\\cmidrule(lr){2-4} \cmidrule(lr){5-7} \cmidrule(lr){8-10}
 Coal  & -0.119 & 0.094 & 0.206 & -0.061 & 0.097 & 0.528 & -0.07 & 0.097 & 0.425\\
 Gas  & 0.190 & 0.075 & 0.012 & 0.198 & 0.074 & 0.007 & 0.198 & 0.074 & 0.008 \\
 Oil  & 0.214 & 0.069 & 0.002 & -- & -- & -- & -- & -- & -- \\
 Temp & -0.001 &  0.001 & 0.572 & -0.001 & 0.001 & 0.450 & -0.001 & 0.001 & 0.451\\
 Stoxx 50 & -- & -- & -- & 0.139 & 0.103 & 0.031  & -- & -- & -- \\
 Stoxx 600 & -- & -- & -- & -- & -- & -- & 0.296 & 0.110 & 0.007 \\
 \midrule\midrule
\end{tabular}
\caption*{\textit{Source:} Own calculations using R.\\\textit{Notes:} Results obtained using OLS estimation. The dependent variable is the return on EUAs and the set of (stationary) regressors changes in each specification. The $p$-values are based on Newey-West standard errors.}
\label{tab:lin.reg}
\end{table}

This does not come as a surprise given the results from previous studies, as we do not split the data into sub-periods, nor do we include any dummy variables to take out the effect of major policy announcements. In all specifications, both significant coefficients are positive and thus show the anticipated sign. The coefficient of the coal price also shows the sign predicted by economic theory, while being insignificant.\footnote{In Appendix \ref{sec:add_var} we include each stock index in addition to the oil price and show that they do not have a significant effect.}

\subsection{Time-varying coefficient results}
\label{sec:nonpara}
We now apply the nonparametric approach to our data. The results presented in this section are obtained using the gas, coal and oil price returns as explanatory variables as well as the temperature series after seasonality has been removed. As a robustness analysis, we provide additional results (e.g. from the inclusion the other variables) in Appendix \ref{sec:add_var}.

\begin{figure}[!h]
	\centering
	\subfigure[$\hat{\beta}_0(t)$]
    {
    \includegraphics[width=0.47\linewidth, clip, trim = {0 1cm 0 2cm}]{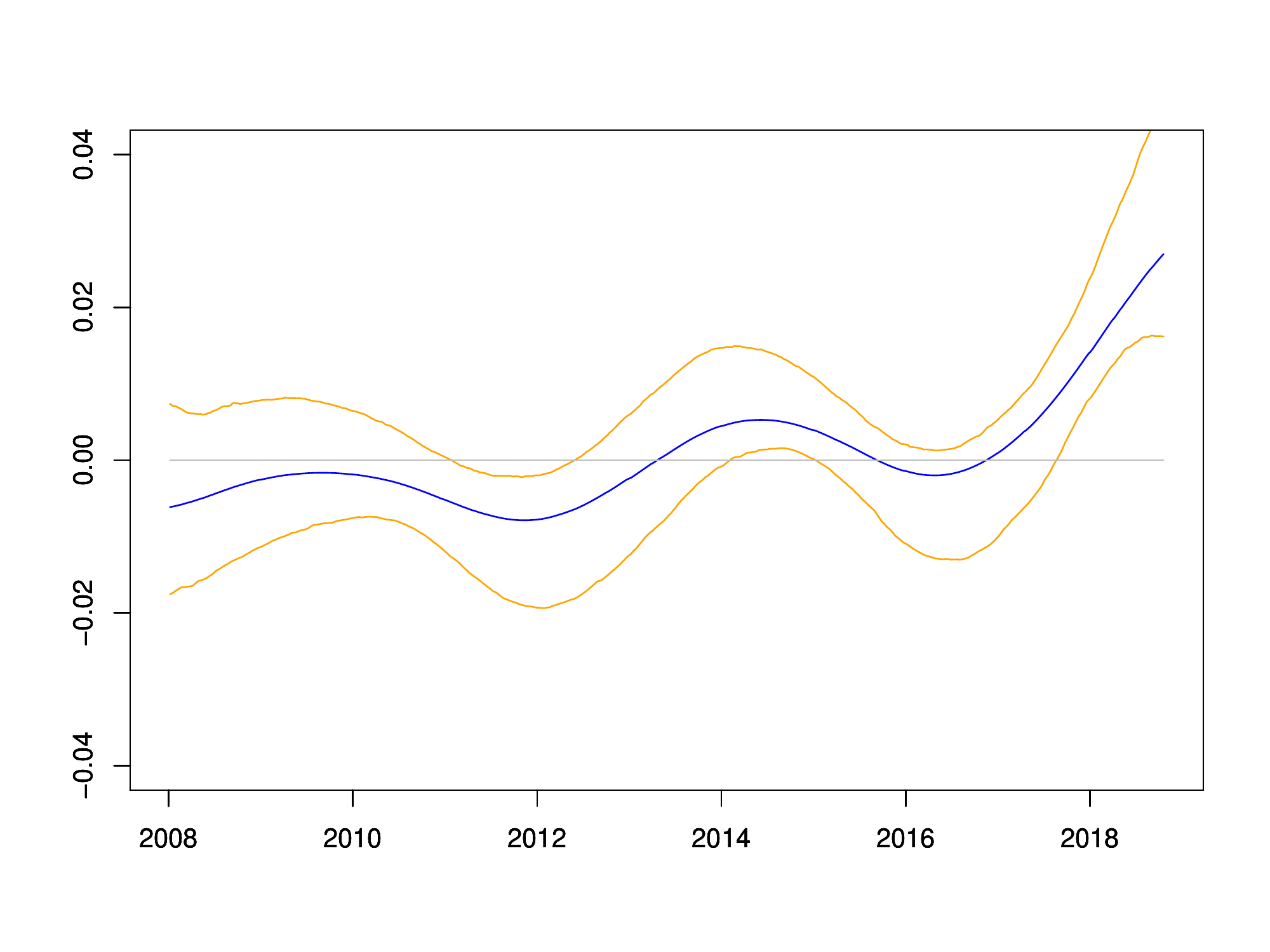}	
    \label{fig:trend}
     }
     \subfigure[$\hat{\beta}_{coal}(t)$]
     {
      \includegraphics[width=0.47\linewidth, clip, trim = {0 1cm 0 2cm}]{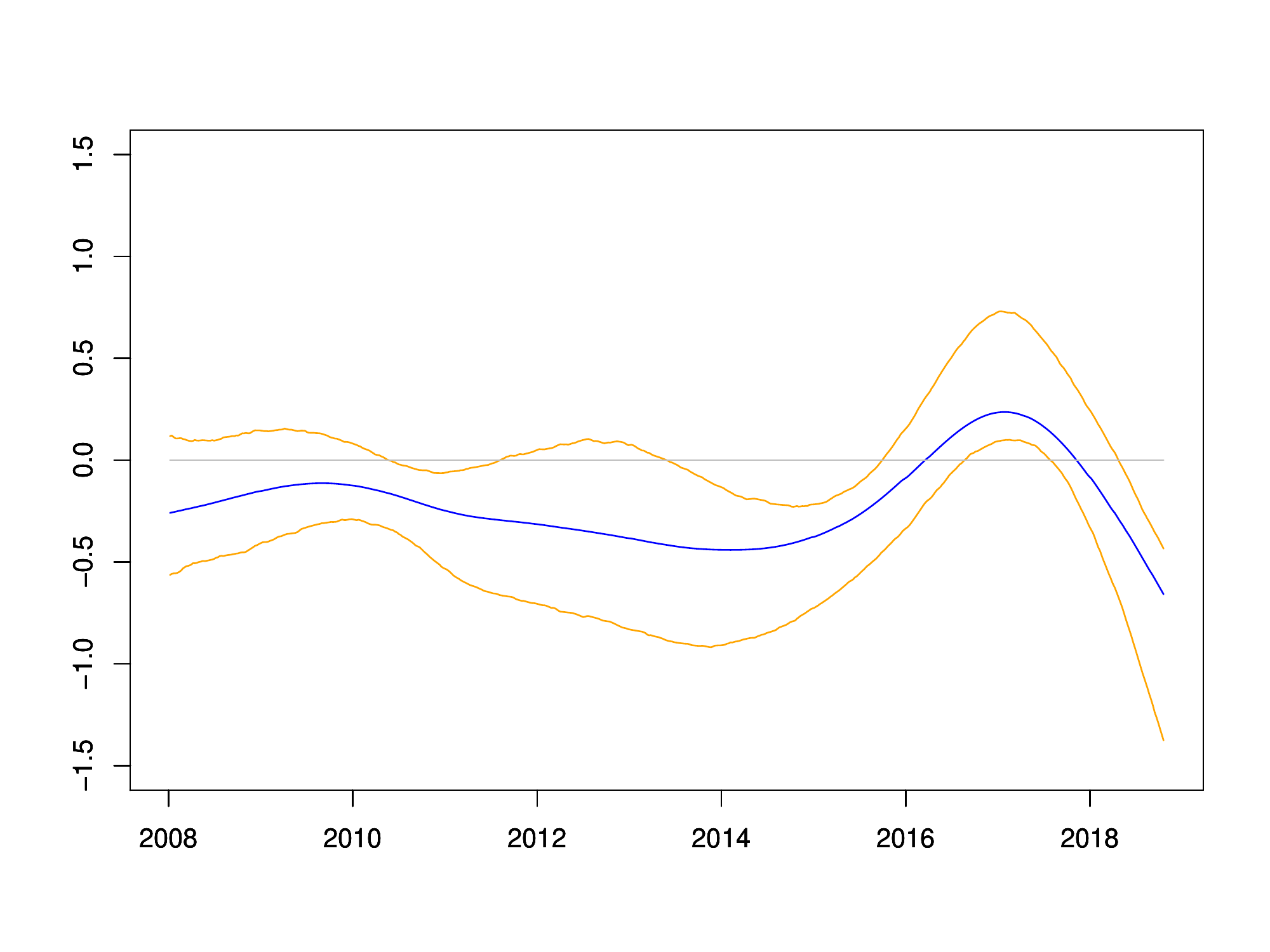}
      \label{fig:coal}
     }\\
		\subfigure[$\hat{\beta}_{gas}(t)$]
    {
    \includegraphics[width=0.47\linewidth, clip, trim = {0 1cm 0 1.5cm}]{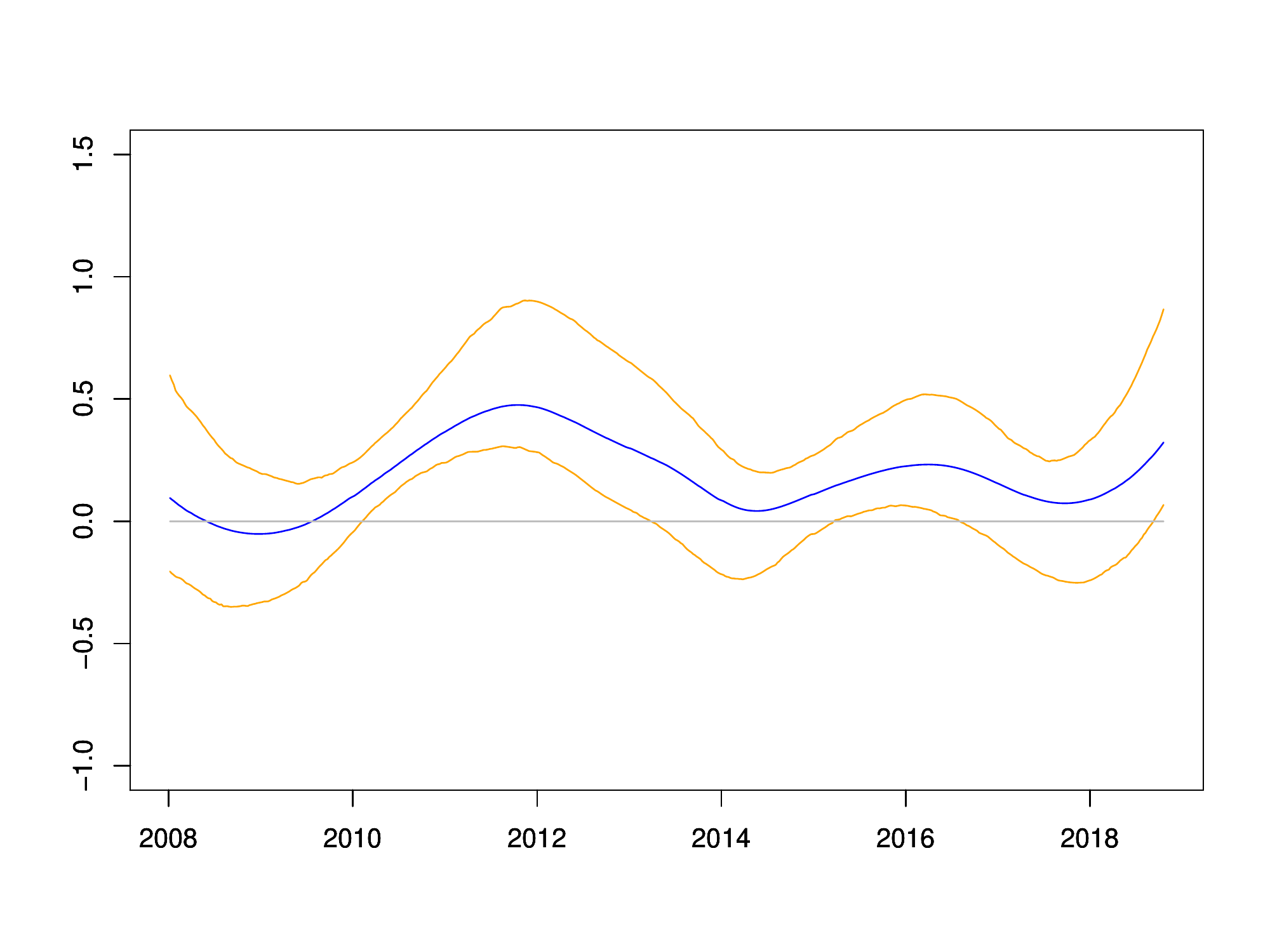}	
    \label{fig:gas}
     }
     \subfigure[$\hat{\beta}_{oil}(t)$]
     {
      \includegraphics[width=0.47\linewidth, clip, trim = {0 1cm 0 1.5cm}]{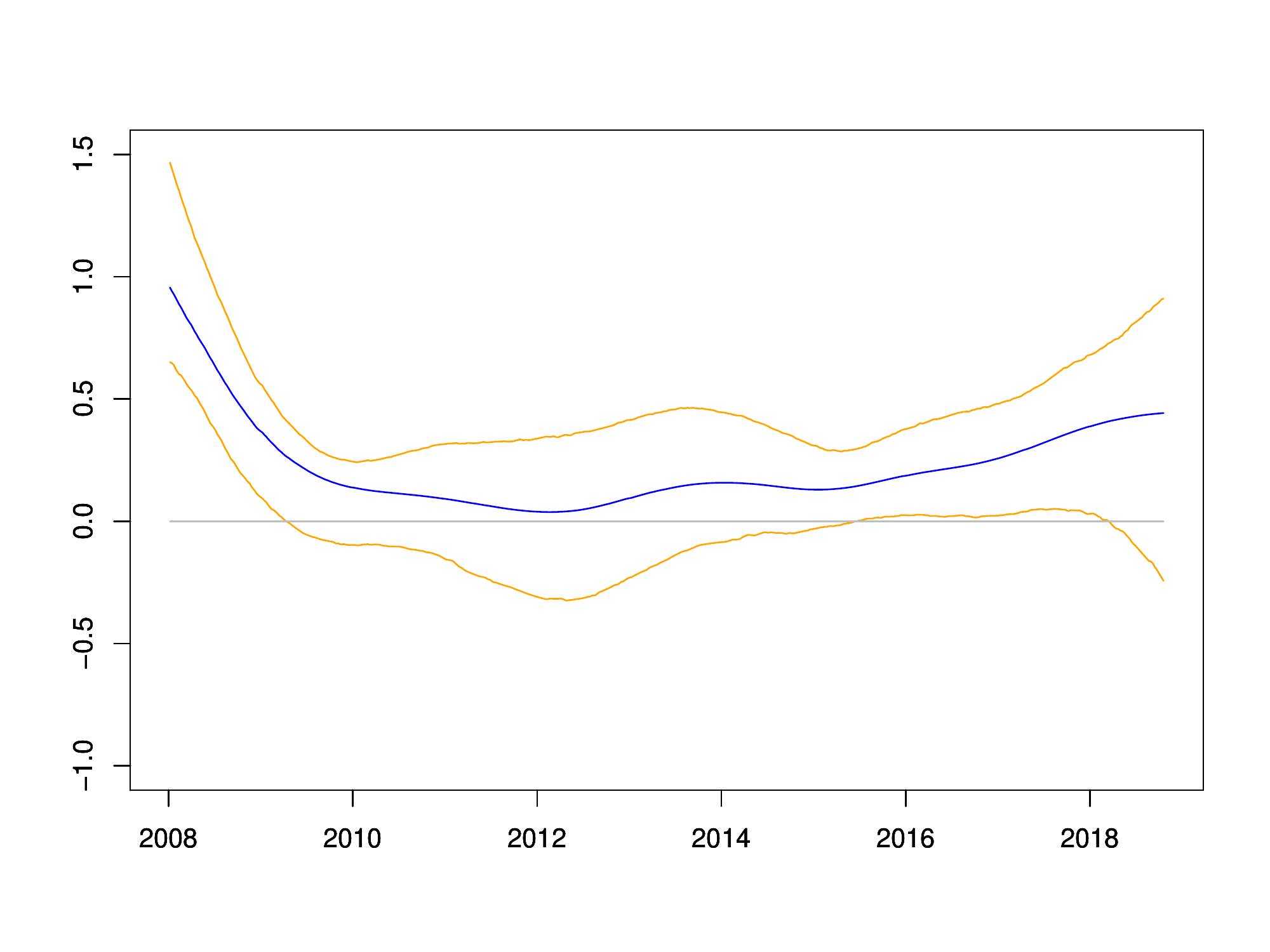}
      \label{fig:oil}
     }
     \subfigure[$\hat{\beta}_{temp}(t)$]
     {
      \includegraphics[width=0.47\linewidth, clip, trim = {0 1cm 0 1.5cm}]{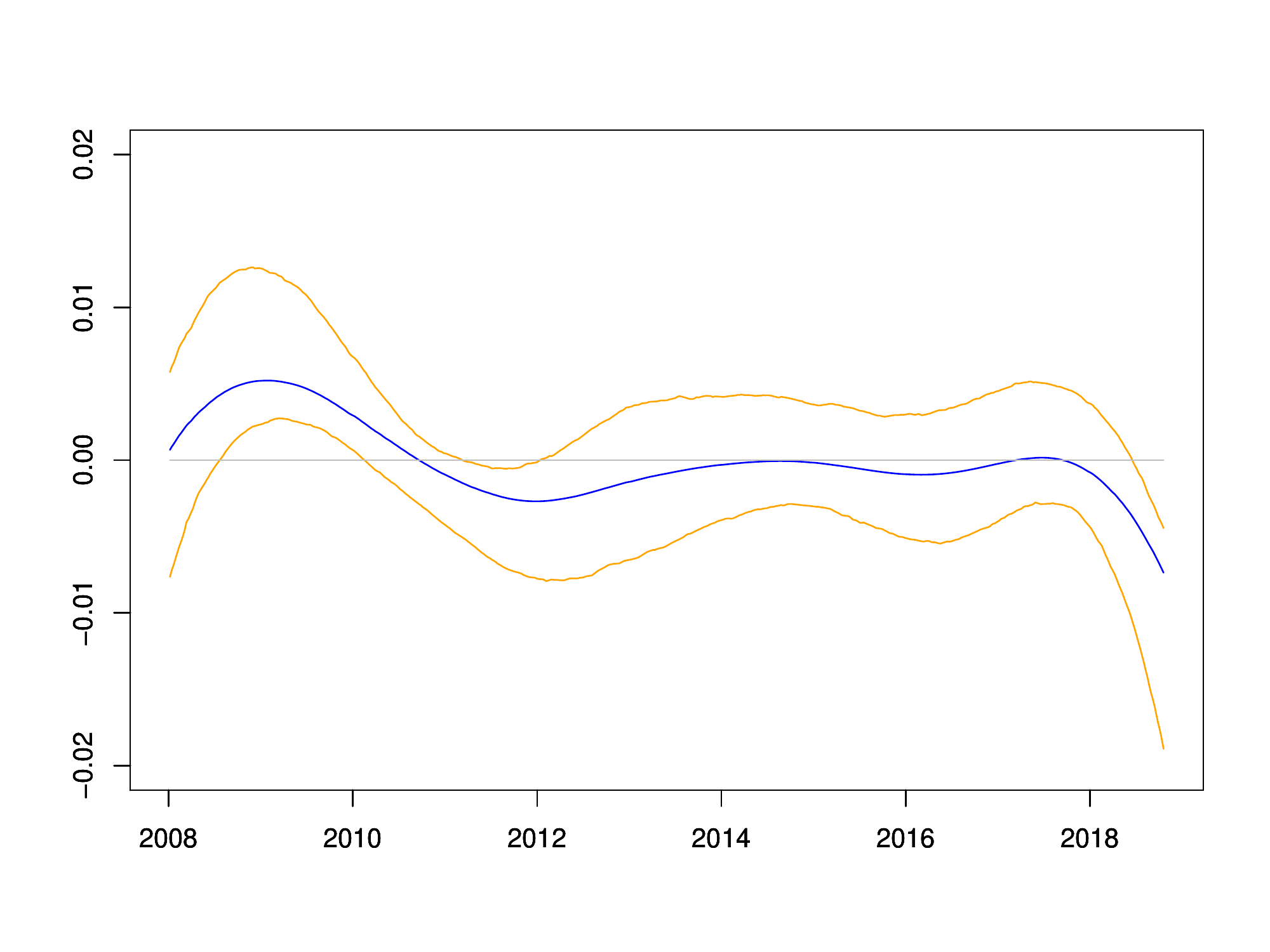}
      \label{fig:temp}
     }
		\caption{Time-varying coefficient estimates\\\textit{Source:} Own calculations.\\\textit{Notes:} Nonparametrically estimated coefficient curves (blue) and 95\% confidence intervals (orange) from a regression of the return on EUA on a trend and the four main price drivers. For the nonparametric estimation, we apply the local linear estimator with the Epanechnikov kernel given by the function $K(x)=\frac{3}{4}(1-x^2)\mathbbm{1}_{\left\{|x|\leq 1\right\}}$. We use a bandwidth parameter of $h=0.09$ which we justify in Appendix \ref{sec:bandwidth_selec}. The bootstrap procedure is applied using 999 replications.}
		\label{fig:nonparametric}
\end{figure}

The estimated trend and coefficient curves (blue) together with their 95\% confidence intervals (orange) can be found in Figure \ref{fig:nonparametric}. As with linear regression, a coefficient is significant if zero (indicated by the gray line) does not fall within the confidence interval. The main difference to parametric regression is that there is not merely one coefficient whose estimate and confidence interval permits a verdict on the question of significance of an explanatory variable. We have a coefficient estimate and a corresponding confidence interval for each point of the sample. Thus, there can be periods of significance and insignificance as well as changes in the sign and magnitude of a coefficient. 

Against this background, we observe in Figure \ref{fig:nonparametric} that the significance of all of the included variables changes over time. For all variables, there are period with a significant effect as well as periods of insignificance. All graphs further have in common that the width of the confidence intervals changes over time and in many cases, become substantially wider at the beginning and the end of the sample. The widening at the beginning and the end is common in the application of the bootstrap method and reflects that the first and last estimates can be inaccurate. As the nonparametric estimator uses a two-sided window for estimation, there are boundary effects which can distort the first and last 20 estimated points on the coefficient curves. This contributes to the widening of the confidence intervals at these points. Widening in the middle of the sample can be an indication of heteroskedasticity or autocorrelation in the data over this time span. In such cases, the bootstrap method accounts for these irregularities by making the confidence intervals wider such that the nominal confidence level (in this case of 95\%) can be maintained. 

In Panel (a), the nonparametric trend fluctuates around zero for most of the considered time span. However, at the end of the sample, it turns significantly positive. This shows that the recent price increase in allowance prices seems to be picked up by the trend component. Panel (b) shows that the coal price has a significant negative effect for two periods: one ranging from 2010 to mid 2011 and one from 2013 to mid 2015. Subsequently, the coefficient becomes positive and significant for a short period in 2016/2017. This is a very interesting finding given that its coefficient was found to be insignificant in the linear regression analysis. This suggests that the time variation might have caused the insignificance in the linear regression results. Moving on to Panel (c), we see the coefficient of the gas price series. It has the expected positive sign and is significant over long periods of the sample. This is in line with previous findings as well as the linear regression results presented in Section \ref{sec:lin_reg}. However, we also find a period of insignificance in the usually stable gas price coefficient which is located in 2014. The coefficient of the oil price, as displayed in Panel (d) is positive and significant over two periods -- until 2009 and from 2015 onward. Finally, from Panel (e) we see that the temperature series shows only two very short periods of significance and the magnitude of the coefficient estimate is, as in the linear regression results, negligible.


Overall, our analysis provides evidence of time variation in the relationship between allowance prices and the considered price drivers. It stresses the need that the relationship has to be modeled with care. Due to the various periods of insignificance, our results offer a potential explanation for insignificant coefficients found with linear regression techniques used in some of the previous work. In addition, although the method offers great flexibility in modeling the relationship, the recent upward trend stays in the trend component. This implies that fundamentals have not gained relevance as drivers during the price run up. We now further examine the recent price increase using a combination of statistical tests and models potentially explosive behaviour.
\vspace{-4mm}
\section{Explosive behaviour in allowance prices}
\label{sec:trend}
In the previous section we saw that the recent upward movement was not picked up by any of the included price drivers and one might suspect that the recent trend could represent a period of explosiveness. It is therefore our goal in this section to further investigate this recent price development. Our method involves two steps. First, we apply the approach by \citet{PSY} which provides a refined way to test for periods of explosive autoregressive behaviour and, in an additional step, to locate them.\footnote{While \textit{explosive autoregressive behaviour} is the most accurate way of describing the phenomenon we test for, we use this terminology interchangeably with \textit{explosive behaviour}, \textit{explosiveness} and sometimes \textit{bubble} in the remainder of this paper. We elaborate further why we have to be careful with the term \textit{bubble} in Sections 5.2 and 6.} The test is based on a right-sided unit root test performed on a forward and backward expanding window. We indeed find a significant explosive period in the allowance price series which overlaps with the recent price increase. In a second step, we further analyse the explosive period we found in the first step. With the help of a novel class of statistical models we predict ex-ante probabilities of collapse of the explosive period in allowance prices which we compare to the out-of-sample price behaviour.

\subsection{Testing for explosive episodes}
\label{sec:bubble_results}
To investigate whether the upward trend in allowance prices constitutes an episode of explosive behaviour, we apply the testing procedure of \citet{PSY}. Similarly to ordinary unit root tests, it tests the null hypothesis of a unit root in the data. The construction of the test is based on the ADF test as performed in Section \ref{sec:stationarity}, but it considers a different alternative hypothesis. Instead of stationarity, the \citet{PSY} test considers mildly explosive behaviour under the alternative. An additional difference is that the test is applied to sub-samples of the data. For rejection of a unit root in favor of explosive episodes it is sufficient if the data show such behaviour in a sub-sample. As the name explosive episode or bubble suggests, this is not a long-term phenomenon. The test applies an ADF test to various subsamples. Subsequently, the test statistic is constructed as the supremum over all ADF test statistics. It is therefore called Generalised Supremum ADF (GSADF) test. In case of rejection, there is evidence of explosive behaviour. For such cases, \citet{PSY} additionally propose a method to locate the beginning and end point of the episode. Details on the testing procedure and the date stamping are available in our Technical Appendix.

A bubble is present where prices diverge from fundamentals. In this context, it means that there is evidence for an explosive period when the price series shows such behaviour but the fundamental drivers do not. Therefore, we also apply the tests to the main price drivers. We test coal, gas and oil prices as well as the two stock indices.\footnote{We do not consider temperature data here given the fact that it is a climatological time series which due to its natural properties will only experience gradual structural change rather than explosiveness (this is also confirmed by the plot of the data in Figure \ref{fig:timeseries}(f)).} The test was proposed under the assumption of uncorrelated and homoskedastic error terms. It has been shown in e.g. \citet{Harvey2016} and \citet{Pedersen2017} that a violation of this assumption leads to overrejection which can be resolved by a bootstrapping approach. In our application, we rely on both, the simulated critical values and the bootstrap approach of \citet{Pedersen2017}.\footnote{We thank the authors for kindly providing us with code for the bootstrap test.} 

\begin{table}[t]
\centering
\caption{GSADF test results}
\begin{tabular}{lccccc}
\midrule\midrule
 \multicolumn{2}{c}{Test statistics} & & \multicolumn{3}{c}{Critical values (90\%, 95\%, 99\%)} \\ \cmidrule(lr){1-2} \cmidrule(lr){4-6}
 Variable & GSADF & & \multicolumn{1}{c}{simulated} & & \multicolumn{1}{c}{bootstrap} \\ \cmidrule(lr){1-1} \cmidrule(lr){2-2} \cmidrule(lr){4-4} \cmidrule(lr){5-6} 
 EUA  & 3.998 & &\multirow{6}{*}{(1.983, 2.175, 2.608)} &  & (2.270, 2.555, 3.201)  \\
 Coal  & 1.676 & &   & & (2.487, 2.795, 3.446)  \\
 Gas  & 1.299 & &   & & (2.383, 2.645, 3.372)  \\
 Oil  & 2.722 & &   & & (2.200, 2.505, 3.104) \\
 Stoxx 50 & 0.782 &  &    & & (2.310, 2.668, 3.099) \\
 Stoxx 600 & 0.953 & & &  &  (2.302, 2.302, 3.170) \\
 \midrule\midrule
\end{tabular}
\caption*{\textit{Source:} Own calculations using Matlab and the R pacakge MultipleBubbles.\\\textit{Notes:} The GSADF test statistics with simulated critical values (2000 repetitions) and bootstrapped critical values (4999 repetitions). The test is applied using a minimum window size of 40 observations, which is chosen according to the rule $r_0 = 0.01+1.8\sqrt{T}$ suggested by \citet{PSY}. For the selection of the number of lags we use the Bayesian Information Criterion.}
\label{tab:GSADF}
\end{table}

Table \ref{tab:GSADF} presents the results; the left part of the table shows the test statistics for each considered time series and the right part gives the two different sets of critical values -- the critical values computed by Monte Carlo simulation as described in \citet{PSY} and the bootstrapped critical values. For both tests, we would like to use significance levels of 10\%, 5\% and 1\%. As the test is right-sided, this can be achieved by the 90\%, 95\% and 99\% critical values, respectively. Note that, as the bootstrap results will be conditional on the original sample, we have a slightly different set of critical values for each series.

\begin{figure}[h!]
	\centering
	\subfigure[EUA prices]
    {
    \includegraphics[width=0.6\linewidth, clip, trim = {0 1cm 0 2cm}]{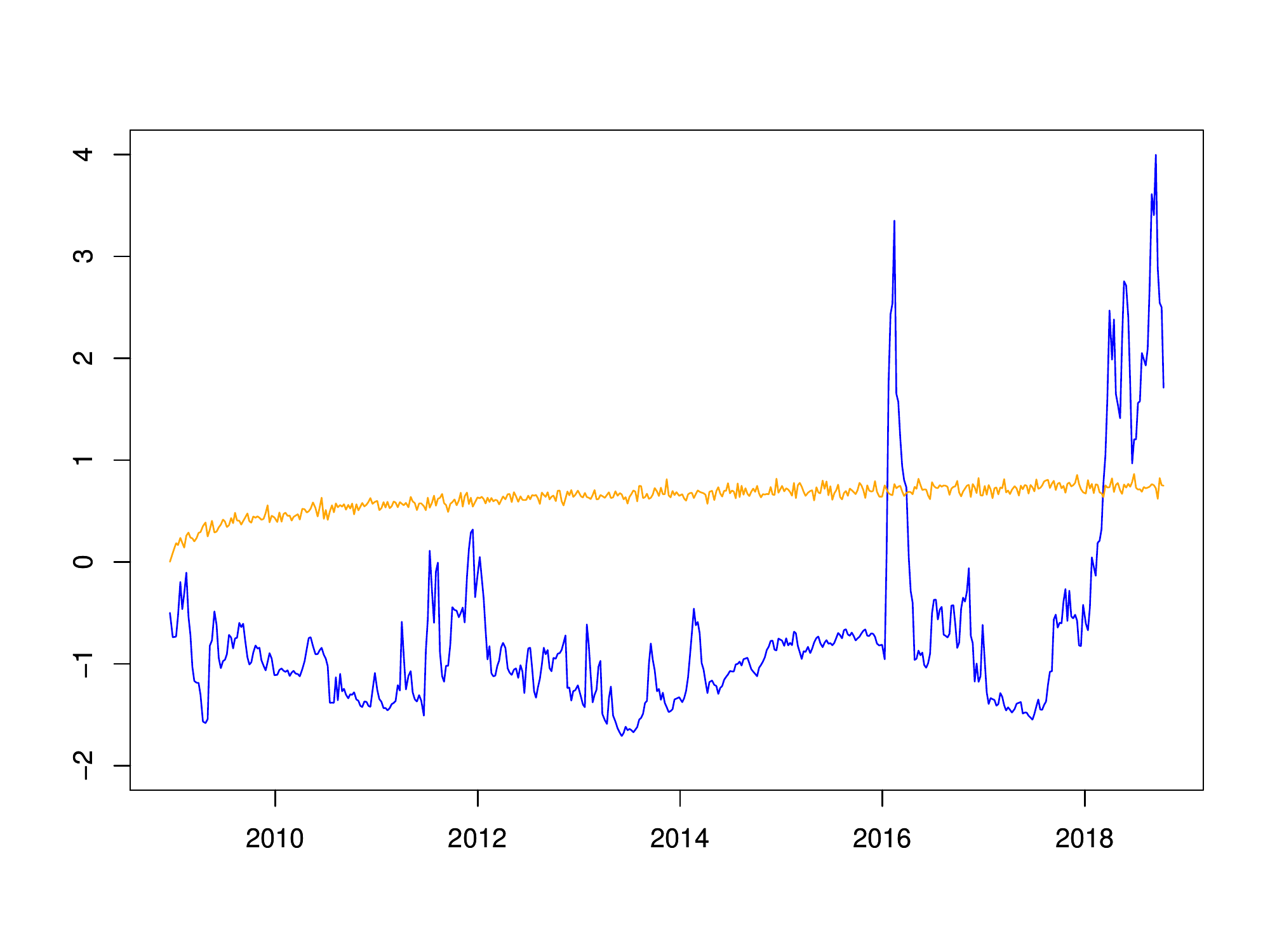}	
    \label{fig:BSADF_eua}
     }\\
     \subfigure[Oil prices]
    {
    \includegraphics[width=0.6\linewidth, clip, trim = {0 1cm 0 1.5cm}]{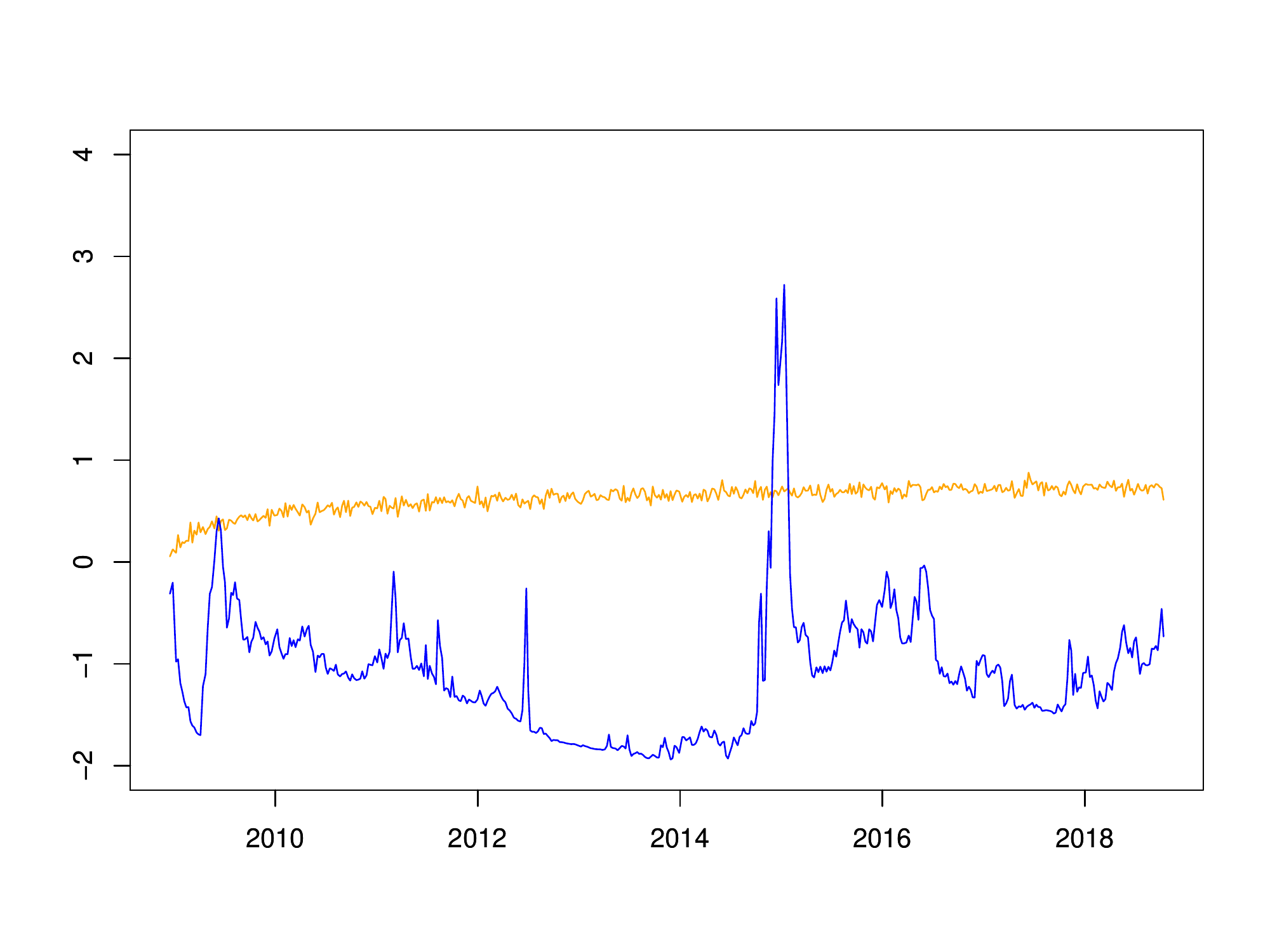}	
    \label{fig:BSADF_oil}
     }
		\caption{Date stamping of explosive episodes\\\textit{Source:} Own calculations.\\\textit{Notes:} Results from the BSADF test as proposed by \citet{PSY}, applied to the EUA prices in Panel (a) and the oil prices in Panel (b). Each panel shows the bootstrapped critical value series (orange) and the test statistics (blue).}
		\label{fig:BSADF}
\end{figure}

Both tests come to the same conclusion. They present evidence of explosive behaviour not only in the allowance price series but also in the oil price series. For the coal and gas prices and the stock indices, the test is not rejected and hence, these factors are excluded as possible drivers of the movement. In contrast, the oil price cannot be excluded. Therefore, for the two cases of rejection we move to the date stamping procedure, which is determined by the BSADF test sequence which is short for Backward Supremum ADF test. We again obtain critical values from simulation and bootstrapping using the same specifications as for the GSADF test. The results using bootstrapped critical values are presented in Figure \ref{fig:BSADF}. It simultaneously plots the series of critical values (orange) and the test statistics (blue). Panel (a) gives results for the allowance price series and Panel (b) for the oil prices. Using simulated critical values leads to very similar conclusions. The results can be found in Appendix B.

There is clear evidence for a long period of explosive behaviour in allowance prices, starting at the beginning of 2018. The series of test statistics starts exceeding the critical value series in early March of 2018 and does not cross it again until the end of the sample. Before 2018, we can see several such episodes in Panel (a) with the most pronounced one located in 2016. Starting in mid January 2016, the BSADF statistic lies above the critical value sequence for 11 weeks until the beginning of April 2016. This exceeds the established minimum duration of 7 weeks. Observing the allowance price development at the time, we see that this period is caused by a price drop in spring 2016. \footnote{ Some analysts have portrayed the price drop as a potential consequence of the COP15 outcome. But to the best of our knowledge none of the scientific event studies on EUA prices has covered this period.} The spikes in the test statistic before 2016 do not last longer than 3 weeks and therefore, cannot be taken as evidence for explosive behaviour. 

In 2018, the length of the explosive episode clearly exceeds the minimum duration. Comparing this to the oil price results in Panel (b), we do not find overlapping explosive behaviour in this series, although the test detects a potential explosive period and we observe an upward trend in oil prices at the end of the sample. The period in oil prices which caused the GSADF test to reject is located in 2014. Starting mid 2014 oil prices fell for a period of more than 6 months indicating that it was a period of collapse and not exuberance that is picked up by the test.

Specifically, the period of collapse is first detected by the test in mid October 2014 where the test statistic lies above the critical values for three weeks, then drops for one week before it exceeds it again for 11 consecutive weeks. This period lasts from November 2014 until the beginning of February 2015. This is in line with the findings in \citet{Sharma2018} who investigate several different oil price series. The authors find a significant period of collapse with similar timing in all investigated oil prices. In addition, the absence of evidence for periods of exuberance or collapse in the gas price series is also confirmed by \citet{Sharma2018}.\footnote{For completeness, we also apply the GSADF test to the fuel switching price. Unsurprisingly, we do not find evidence for explosive behaviour in this series.} 

Combining these results provides statistical evidence that the allowance price experienced an explosive period which is not accompanied by similar behaviour in any other series we considered. Although the gas and the oil price experienced a simultaneous upward trend at the end of our sample period, the formal test allows us to exclude both factors as potential drivers of the rapid EUA price increase. This is to our knowledge a first empirical result pointing in the direction of a period of exuberance in EUA prices which is not driven by abatement-related fundamentals. While this finding strongly suggests the existence of a price bubble, it should be noted that there could be other explanations which are in line with these results. For instance, non-explosive changes in fundamentals could lead to new equilibrium price levels and the path to the new levels can potentially appear explosive. This has been pointed out by \citet{Harvey2016}. As shown in Section \ref{sec:fundamentals}, changes pointing towards an increasing role of fundamentals as a result of the reform could not yet be found in the data. 

\subsection{Predicting probabilities of collapse}
\label{sec:crash_odds}
The statistical evidence points to an explosive episode whose drivers are not related to fundamentals. Non-fundamental bubbles in prices are often seen as a result of widely shared exuberant beliefs among market participants  regarding price outlooks, which trigger inflationary trends later sustained by a self-confirming mechanism \citep{diba1988explosive,diba1988theory}.
During such explosive events, it might be rational even for traders informed of the discrepancy between the inflated and fundamental prices to participate in the bubble in an attempt to time the market: making a profit on the growth of the bubble and exiting before the exuberance phases out and downward pressure on prices induces a crash down to the fundamental value \citep{abreu2003bubbles}.

Recently, rational expectation specifications of financial bubbles have been shown to exhibit as particular solutions so-called noncausal, or non-fundamental, autoregressive time series models \citep{gourieroux2016stationary}, that is, autoregressive processes with roots located inside the unit circle. 
These statistical models, compatible with rational expectations economic models, have the ability to parsimoniously mimic explosive episodes in financial time series data and are increasingly applied in the economic and statistical literature to analyse time series with forward-looking components. 

\cite{fries2018conditional,fries2018path} derives analytical expressions for the ex ante crash odds of bubbles generated by noncausal models driven by extreme, power-law distributed shocks, which we exploit here to quantify the (un)sustainability of the recent explosive episode in the EUA price. 
We focus on the arguably simplest noncausal model -- the noncausal AR(1) -- which mimics bubbles with an exponentially-shaped inflation phase culminating in a peak before collapsing to pre-bubble levels. There are two main reasons for this choice: (i) the obtained exponential fit of the recent upward trend in the EUA prices appears excellent, (ii) bubbles generated by the noncausal AR(1) feature a memory-less, or non-aging, property 
which implies that the crash date cannot be known with certainty by market participants and indicates compatibility of the model with a no-arbitrage setting. Further details on the approach are provided in the Technical Appendix.\\
\indent Modeling the explosive episode in the EUA price as an ongoing realisation of a noncausal AR(1) bubble, the sustainability of the upward trend can be completely characterised by the given of two parameters: the growth rate of the exponential inflation phase, and a parameter describing the likely height of the incoming peak -- the tail exponent of the shocks' power-law distribution. 
A higher growth rate and a likely lower peak (i.e. higher tail exponent) entail a less sustainable explosive episode.
The growth rate can be conveniently obtained by fitting an exponential trend on the recent data -- which boils down to a linear regression of the log-prices on time.  To this end, we include all data points post-March 2018 into the exercise, as this date marks our earliest statistical evidence of explosive episode based on \cite{PSY} test. For sensitivity analysis, we also consider the estimate of the growth rate using the data points since the beginning of the upward trend, mid 2017. 
The second parameter, the power-law tail exponent, is more elusive and cannot be easily retrieved from the data. We prefer to remain as agnostic as possible regarding the value of this second parameter, and only presume it lies within the widest range of reasonable and admissible values under \cite{fries2018conditional}
framework.\footnote{\label{footnote:tailexponent_range} This range encompasses values of the power-law tail exponent, denoted $\alpha$, from $\alpha=0.5$ to $\alpha$ arbitrarily close to 2: that is, from very heavy-tailed shocks as extreme as a L\'evy distribution which have even infinite expectation, to much milder shocks close to have finite variance.
Most studies in the financial literature report values from significantly below one up to four for financial series (\cite{ibragimov2016heavy}
and the references therein), with reported values above two furthermore not necessarily evidence against the infinite variance hypothesis \citep{mcculloch1997measuring}, 
which is used in Fries (2018a,b).} 
\begin{figure}
	    \centering
	    \includegraphics[scale=0.125]{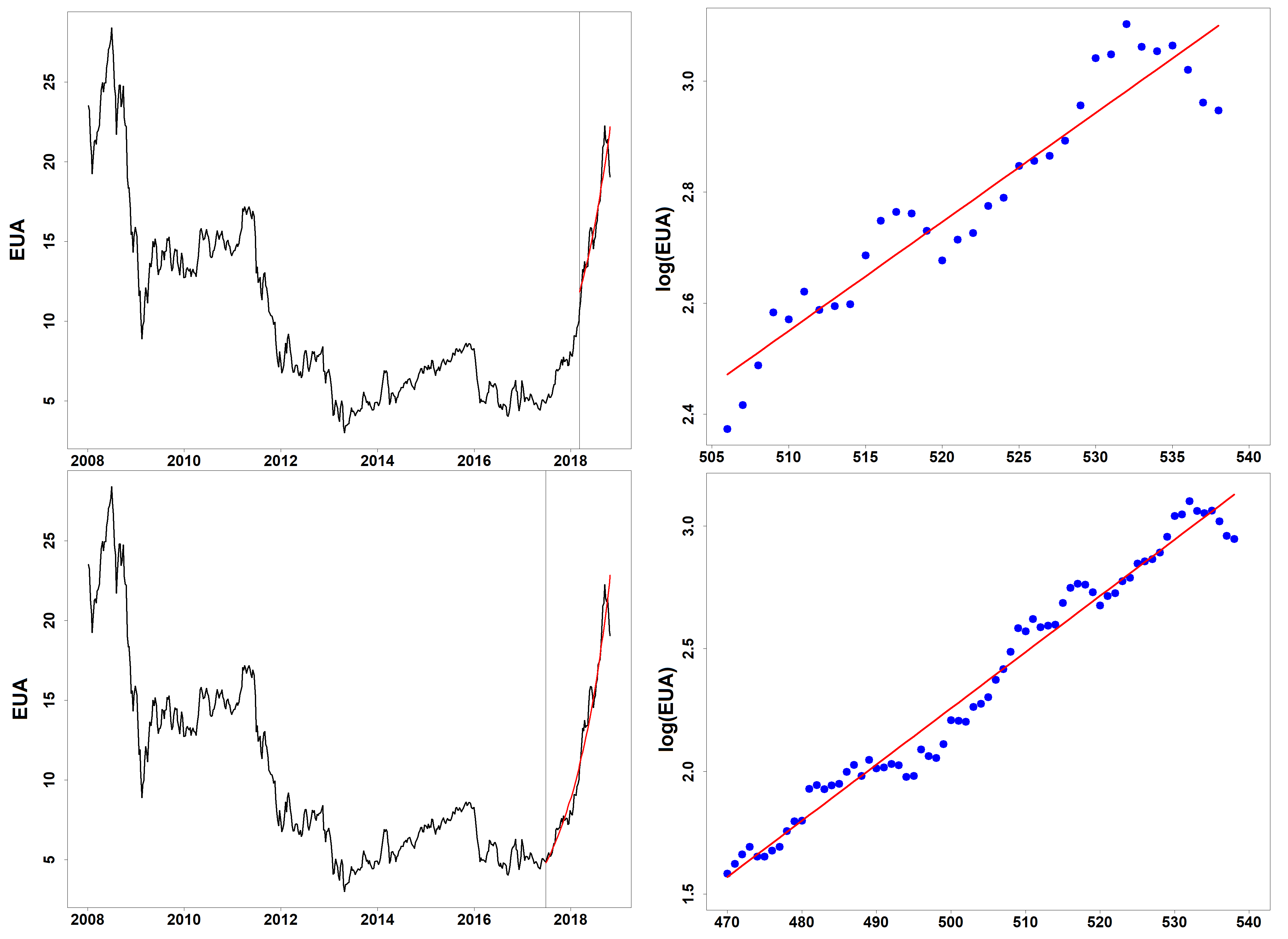}
	    \caption{Exponential trend fitted on the EUA prices (left panels) based on the linear regressions of log-EUA prices on time and intercept $\log(\text{EUA}_t)=at+b$ (right panels). Upper panels: fit obtained using the data points post-March 2018 ; Lower panels: fit obtained using the data points post-June 2017.}
	    \label{fig:log_EUA_fits}
	\end{figure}
	
		\begin{table}\footnotesize
		\centering
		\begin{tabular}{cc|cc|cc}
		    & & \multicolumn{2}{c}{\makecell{Based on EUA data\\ from 03/2018 to 10/2018}} & \multicolumn{2}{c}{\makecell{Based on EUA data\\ from 06/2017 to 10/2018}}\\
		    \hline\hline
		    \makecell{Exponential trend\\ monthly growth rate $\hat{a}$} & & \multicolumn{2}{c|}{$8.1\%$} & \multicolumn{2}{c}{$9.5\%$}\\
		    & & & & & \\
		    \makecell{Corresponding noncausal\\ AR coefficient $\hat{\rho}$} & & \multicolumn{2}{c|}{0.92} & \multicolumn{2}{c}{0.91} \\
		    & & & & & \\
		    \makecell{Plausible range for\\ tail exponent $\alpha$} & & \multicolumn{2}{c|}{[0.5 -- 2]} & \multicolumn{2}{c}{[0.5 -- 2]}\\
		    & & & & & \\
		    & & Lower est. & Upper est. & Lower est. & Upper est. \\
		    & & & & & \\
		     & $m = 1$ \hspace{0.65cm} & 4.0\% & 15\% & 4.6\% & 17\%\\
		     Odds of crash & $3$ & 11\% & 39\% & 13\% & 44\%\\
		     within $m$ months & $6$ & 22\% & 62\% & 25\% & 68\% \\
		     & $12$ & 39\% & 86\% & 44\% & 90\%\\
		     & & & & & \\
		     \makecell{Expected explosive\\ episode duration} & (in months) & 7 & 25 & 6 & 22\\
			\hline\hline
		\end{tabular}
		\caption{\footnotesize Estimated monthly growth rates $\hat{a}$ of exponential trends fitted on the EUA log-prices $\ln(\text{EUA}_t) = a t + b$, based on data since March 2018 (left) and since mid-2017 (right) ; Corresponding noncausal AR(1) coefficients $\hat{\rho}=\exp(-\hat{a})$ ; Uniform prior on power-law tail exponent $\alpha\in[0.5,2]$ ; Corresponding lower and upper estimates for crash odds within $m=1,3,6,12$ months computed as $1-(1/2)^{m/\hat{h}_{0.5}}$, where $\hat{h}_{0.5}=-\frac{\ln 2}{\alpha \ln(\hat{\rho})}$ is the half-life of an $\alpha$-stable noncausal AR(1) bubble with parameters $\hat{\rho}$ and $\alpha$ ; Expected duration of the explosive episode computed as $\frac{1}{1-\hat{\rho}^\alpha}$.
		}
		\label{tab:EUA_crash_odds}
	\end{table}
\indent Table \ref{tab:EUA_crash_odds} gathers the results and Figure \ref{fig:log_EUA_fits} presents the exponential fits of the recent EUA data. 
The obtained growth rates using either the data post-March 2018 or post-June 2017 both point to a similar pace of around 8-9\% price growth per month. Moreover, the fits on the log-prices appear at least adequate on the data post-March 2018 and even excellent on the data post-June 2017.
This comforts the assumption of an exponentially-shaped bubble \textit{\`a la} noncausal AR(1), which appears very much compatible with the price trajectory.
Because we assume a range of possible values for the tail exponent parameter rather than a specific value (see Footnote \ref{footnote:tailexponent_range}), we obtain lower and upper estimates of the odds of a crash occurring at future horizons. We report these lower and upper estimates for 1, 3, 6 and 12 months horizons beyond the end of our sample.
The crash odds that we find are relatively important: already at 1 month horizon, the likelihood of a crash ranges from 4\% (lower estimate) up to 17\% (upper estimate).
According to our lower estimates, the likelihood of a crash occurring within 12 months are high but nevertheless below 50\%, whereas our upper estimates appear to place the 50\% chance tipping point much earlier, between 3 and 6 months horizon.
Another equivalent measure of sustainability of an explosive episode (in the framework of the assumed model) is that of its expected duration. According to our same estimates, the expected duration of the EUA explosive episode should range from 6 (lower est.) to 25 months (upper est.).

A posteriori, in 2020, it is apparent on updated data that the explosive trend wore out very quickly after the end of our sample. 
Even though an abrupt vertical crash down to pre-bubble level did not occur, the ex-post explosive trend (un)sustainability appears highly compatible with our estimates.
The fact that the exponential trend leveled off and yet the prices did not collapse could be an indication that the drivers of explosive growth vanished. It is, however, unclear whether this implies a long-term stabilization of prices or a postponement of the collapse. 

\vspace{-4mm}
\section{Conclusion and policy implications}
\label{sec:conclusion}
This paper studies the recent upward price movement in allowance prices in the EU ETS, which is widely attributed to the ETS reform enacted in 2018. The analysis is divided in two parts. In the first part, we investigate the relationship between allowance prices and their fundamental price drivers using a time-varying coefficient model. In particular in the presence of changing market conditions, such as during a period of major reforms, a flexible approach seems inevitable. Results confirm that the recent upward trend cannot be explained by movements in the considered price drivers. Additionally, our results corroborate the main findings of previous studies that coal and gas prices are the most important explanatory variables. The particular feature of time-variation in our approach allows us to identify periods of significance and insignificance. We find periods of both for all variables, which partly explains why the explanatory power of the estimated models in the literature is low. Even the gas price, which is usually the strongest price driver, displays a period of insignificance around 2014. This underlines that -- even using more flexible methods -- abatement-related fundamentals have little explanatory power and it remains difficult to find a strong model for allowance prices. 

In the second part, we look for potential explosive behaviour in allowance prices. Using the testing procedure of \citet{PSY}, we find evidence for a period of exuberance coinciding with the recent upward trend. This could be an indication that the market is in the inflationary phase of a bubble. The fundamental price drivers do not show the same behaviour: we merely detect an explosive period in the oil price series which does not coincide with the EUA price increase. The other considered price drivers do not show any evidence of exuberance. This leads us to conclude that, similar to the price drop around 2011, this development cannot be explained by the abatement-related fundamentals. It suggests that anticipation of the new reform has triggered market participants into speculation, resulting in the steep upward trend.  

Subsequent modeling of the explosive episode allows us to predict probabilities of collapse for different time horizons, starting in October 2018 where we see a first slow down of the rapid price increase. The upper estimates for one year ahead are as high as 86\%. However, the current development of prices has not shown a collapse. At the beginning of 2020, prices are still at a level of 25\euro{}/tCO2. Combining this observation with our predicted crash odds motivates two potential explanations: (i) market participants have stabilized their beliefs about the upcoming cancellation of allowances and they regard the current price level as reasonable; (ii) a price drop to pre-bubble level is overdue. 

These findings speak to two major concerns that have been raised about the reform. First, there is high uncertainty whether the new price level will sustain. The mechanism to cancel allowances is extremely complex \citep{Perino2018}, and the volume of allowances that will be canceled is difficult to predict. If the market has indeed overreacted to the reform and beliefs about its impact have stabilised at a too high level, then it is currently experiencing a bubble bound to collapse. Our work does not imply this will undoubtedly be the case; rather it challenges the high confidence with which policy makers rule out such a turn of events.


Second and related, there are concerns that once policy makers start to make discretionary intervention in the market as they did with this reform, regulatory uncertainty will increase. This is because discretionary changes in policy induce changes in the economic system, which in turn necessitate re-estimation and future changes in policy \citep{Kydland1977}. The resulting regulatory uncertainty distorts prices upward or downward depending on whether firms anticipate injection or withdrawal of allowances trough future regulatory intervention \citep{Hasegawa2014, Salant2016}. This leads to inefficient decision making by firms. Our findings suggest that the discretionary reform of the EU ETS has changed the economic system by sparking speculation about the price impact of the reform - and possibly also formed beliefs among market participants that policy makers will not shy away from intervening again. If that were to happen, the EC's intention to stabilise the market would had the adverse effect of instigating a period of regulatory instability instead, which ultimately also defies making credible long-term commitments \citep{Edenhofer2014}. 

Finally, it has been argued that both in terms of price predictability and intrinsic capacity to make long-term commitments, a carbon tax is superior to an ETS \citep{Hassler2016}. If anything, our paper supports this case in terms of providing evidence for the problems of the latter. If indeed a carbon tax would facilitate long-term commitments, EU policy makers should thus replace the ETS with a tax. For the time being, implementing a price collar could at least solve one of the problems, i.e. make prices more predictable \citep{Fuss2018}. For all we know from ETSs with such a mechanism in place, e.g. California's programme, this has worked out as intended.

\clearpage
\FloatBarrier
\onehalfspacing
\bibliography{BibFileEUETS.bib}
\begin{appendices}
\section{Technical Appendix}
\subsection{Time-varying coefficient estimation and confidence intervals}
Given an observed set of data on the response series $\left\{y_t\right\}$ and the regressors $\left\{\mathbf{x}_t\right\}$, the coefficient curves $\beta_j(\cdot)$ in \eqref{eq:model} can be estimated via local linear nonparametric kernel estimation as in \citet{Cai}. Underlying this method is a first-order Taylor approximation of each $\beta_j(\cdot)$, for $j=0,...,d$, at a fixed time point $\tau\in\left(0,1\right)$. For $t/n$ in a neighborhood of $\tau$ it holds that
\begin{equation}
\beta_j(t/n)\approx\beta_j(\tau)+\beta_j^{(1)}(\tau)(t/n-\tau).
\label{eq:taylor}
\end{equation}
with $\beta_j^{(1)}(\cdot)$ denoting the first derivative of the coefficient function $\beta_j(\cdot)$. If we replace $\beta_{j,t}=\beta_j(t/n)$ by approximation \eqref{eq:taylor} for every $j=0,...,d$, equation \eqref{eq:model} can now be rewritten as:
\begin{equation}
y_t\approx\boldsymbol{\beta}(\tau)'\mathbf{x}_t+\boldsymbol{\beta}^{(1)}(\tau)'\mathbf{x}_t(t/n-\tau)+z_t,
\label{eq:model2}
\end{equation}
where $\boldsymbol{\beta}^{(1)}(\tau)=\left(\beta_0^{(1)}(\tau),\cdots,\beta_d^{(1)}(\tau)\right)'$ denote the stacked first derivatives of trend functions evaluated at $\tau$. The local linear estimator of this model is found by minimizing the following weighted sum of squares with respect to $\boldsymbol{\theta}$, where $\tilde{\mathbf{x}}_t(\tau)=(\mathbf{x}_t,\mathbf{x}_t(t/n-\tau))'$:
\begin{equation}
\widehat{\boldsymbol{\theta}}(\tau)=\operatorname{argmin}_{\boldsymbol{\theta}}\sum_{s=1}^n \left\{y_s-\tilde{\mathbf{x}}_s(\tau)'\boldsymbol{\theta}\right\}^2K\left(\frac{s/n-\tau}{h}\right),
\end{equation}
where $K(\cdot)$ is a kernel function and $h>0$ is a bandwidth. As $n\rightarrow \infty$, the bandwidth is assumed to satisfy $h\rightarrow 0$ while $nh\rightarrow \infty$. The solution to this minimisation problem gives the estimator of $(d+1)$ coefficient functions $\boldsymbol{\beta}(\cdot)$ as well as their $(d+1)$ first derivatives $\boldsymbol{\beta}^{(1)}(\cdot)$. Let $\boldsymbol{\theta}(\tau)=(\boldsymbol{\beta}(\tau),\boldsymbol{\beta}^{(1)}(\tau))'$ denote the vector of stacked coefficient functions and first derivatives. Then, the estimator $\widehat{\boldsymbol{\theta}}(\tau)=(\widehat{\boldsymbol{\beta}}(\tau),\widehat{\boldsymbol{\beta}}^{(1)}(\tau))'$ can be expressed as,
\begin{equation}
\widehat{\boldsymbol{\theta}}(\tau)=\left( \begin{array}{cc}
\mathbf{S}_{n,0}(\tau) & \mathbf{S'}_{n,1}(\tau) \\
\mathbf{S}_{n,1}(\tau) & \mathbf{S}_{n,2}(\tau)  \\ \end{array} \right)^{-1}\left( \begin{array}{c}
\mathbf{T}_{n,0}(\tau) \\
\mathbf{T}_{n,1}(\tau) \\ \end{array} \right)\equiv\mathbf{S}_n^{-1}(\tau)\mathbf{T}_n(\tau),
\label{eq:cai_estimator}
\end{equation}
for $\tau\in\left(0,1\right)$, where for $k=0,1,2$,
\begin{align*}
\mathbf{S}_{n,k}(\tau)&=\frac{1}{nh}\sum_{t=1}^n\mathbf{x}_t\mathbf{x}_t'(t/n-\tau)^kK\left(\frac{t/n-\tau}{h}\right),\\
\mathbf{T}_{n,k}(\tau)&=\frac{1}{nh}\sum_{t=1}^n\mathbf{x}_t(t/n-\tau)^kK\left(\frac{t/n-\tau}{h}\right)y_t.
\end{align*}
This estimator can be seen as a weighted least squares estimator of a model of the form \eqref{eq:model2}. The fitted values are thus obtained by
\begin{equation}
\hat{y}_t=\mathbf{x}_t'\widehat{\boldsymbol{\beta}}(t/n),
\end{equation}
which shows that we are only interested in the estimates of the coefficient curves, the estimated first derivatives only serve as a by-product of the specific choice of estimator.

For the construction of confidence intervals, we rely on the autoregressive wild bootstrap as proposed in \citet{FSU} for a nonparametric trend model. We extend the method to model \eqref{eq:model} by using the following bootstrap algorithm:
\begin{algorithm}[Autoregressive Wild Bootstrap]
$\phantom{A.1}$
\begin{enumerate}
\item Estimate model \eqref{eq:model} and form a residual series. This means, calculate \begin{equation*}
\hat{z}_t = y_t-\mathbf{x}_t'\widetilde{\boldsymbol{\beta}}(t/n),\;\;\;\;\;t=1,...,n,
\end{equation*}
where the estimate $\widetilde{\boldsymbol{\beta}}(t/n)$ is obtained by bandwidth $\tilde{h}>h$.
\item For $0 < \gamma < 1$, generate $\nu_1^\ast,\ldots,\nu_n^\ast$ as i.i.d.~$\mathcal{N}(0,1-\gamma^2)$ and let $\xi_t^* = \gamma \xi_{t-1}^* + \nu_t^*$ for $t=2,\ldots,n$. Take $\xi_1^* \sim \mathcal{N}(0,1)$ to ensure stationarity of $\{\xi_t^*\}$.
\item Calculate the bootstrap errors $z_t^*$ as $z_t^* = \xi_t^* \hat{z}_t$ and generate the bootstrap observations by $y^{\ast}_t = \mathbf{x}_t'\widetilde{\boldsymbol{\beta}}(t/n) + z^{\ast}_t$ for $t=1,\ldots,n$, where $\widetilde{\boldsymbol{\beta}}(t/n)$ is the same estimate as in the first step.
\item Obtain the bootstrap estimator $\hat{\boldsymbol{\beta}}^*(\cdot)$ as defined in \eqref{eq:cai_estimator} using the bootstrap series $\{y_t^*\}$, with the same bandwidth $h$ as used for the original estimate $\hat{\boldsymbol{\beta}}(\cdot)$.
\item Repeat Steps 2 to 4 $B$ times, and let 
\begin{equation*} 
\hat{q}_{j,\alpha} (\tau) = \inf\left\{u \in \mathbb{R}:\mathbb{P}^* \left[\hat{\beta}_j^*(\tau) - \tilde{\beta}_j(\tau)\leq u\right] \geq \alpha\right\}
\end{equation*}
denote, for all $j=0,\dots,d$, the $\alpha$-quantile of the $B$ centered bootstrap statistics $\hat{\beta}_j^*(\tau) - \tilde{\beta}_j(\tau)$. These bootstrap quantiles are then used to construct confidence bands as described below.
\end{enumerate}
\end{algorithm} 
Note that in Step 1 of the above algorithm, a different bandwidth is used to perform the nonparametric estimation. We follow the practical implementation in \citet{FSU} using $\tilde{h}=0.5h^{5/9}$. Compared to the original bandwidth $h$, this bandwidth is larger and produces an oversmoothed estimate as starting point for the bootstrap procedure. The reason for this is the presence of an asymptotic bias whenever local polynomial estimation is applied. The bias contains the second derivatives of the coefficient functions, which can only be consistently estimated using a larger bandwidth $\tilde{h}$.

To construct confidence intervals for every $\beta_j(\cdot)$ for a confidence level of $1-\alpha$, we use the quantiles as obtained in Step 5 of the bootstrap procedure. The confidence intervals are given by
\begin{equation}
I^{\ast}_{j,\alpha}(\tau)=\left[\hat{\beta}_j(\tau)-\hat{q}_{j,1-\alpha/2},\hat{\beta}_j(\tau)-\hat{q}_{j,\alpha/2}\right].
\label{eq:interval}
\end{equation}

\subsection{Testing procedure of the bubble detection test}
\label{sec:bubble_test}
\citet{PSY} propose a recursive procedure which tests for the presence of explosive episodes and locates the starting (and potentially termination) point of such episodes. The pricing equation underlying the approach can be written as
\begin{equation}
EUA_t=\sum_{i=0}^\infty\left(\frac{1}{1+r_f}\right)^i\mathbb{E}_t(U_{t+i})+B_t,
\label{eq:fundamentals_bubble}
\end{equation}
where $r_f$ is the risk-free interest rate, $U_t$ represents the underlying fundamentals and $\mathbb{E}_t$ is the conditional expectation given the information available at time $t$. $B_t$ is the bubble component which is assumed to satisfy $\mathbb{E}_t(B_{t+1})=(1+r_f)B_t$. When $B_t=0$, there is no bubble present in the price series and the degree of nonstationarity is determined by the fundamentals which are assumed to at most $I(1)$. In the presence of bubbles, when $B_t\neq0$, the price series shows explosive behaviour.

Under the null hypothesis of the test, the bubble component is zero such that the the price follows and $AR(1)$ process with a drift:
\begin{equation*}
    EUA_t=dT^{-\eta}+\theta EUA_{t-1} + \epsilon_t,
\end{equation*}
where $T$ is the sample size, $d$ is a constant and $\eta>\frac{1}{2}$ controls the magnitude of the drift component. Following \citet{PSY}, we consider the case where $d=\eta=\theta=1$ such that the price process will be $I(1)$.

Since the fundamentals are unobservable, it might be challenging to estimate this component of equation \eqref{eq:fundamentals_bubble}. In an ideal application, a widely used (theoretical) model exists with which one can obtain the fundamental value of this asset. In this case, the model can be calibrated and parameters can be estimated based on past data. This is done, for instance, in \citet{Shi2017}. In our application, however, no such model exists. As we have seen in Section \ref{sec:nonpara}, the relationship between allowance prices and the abatement-related fundamental price drivers is not stable over time. Given these results, we do not explicitly model the fundamental component but look for simultaneous explosive periods in the most established price drivers. We do this by applying the testing procedure separately to these series as it is done, for instance, in \citet{Corbet2018}.

As explained in \citet{PSY}, in practice, it can also be difficult to distinguish bubbles from periods of price run ups, caused for instance by temporary changes in the discount rate. The latter can mimic bubble behaviour and will therefore also be detected by the test, although it is not incorporated in the theoretical framework where a constant risk-free rate is used. This is why \citet{PSY} stress the importance of specifying in advance a minimum duration for an episode to qualify as bubble.

In general, the recursive testing procedure detects periods of mildly explosive behaviour and market exuberance in time series, and it is able to identify the location. The test applies a series of right-tailed ADF tests on a (backward and) forward expanding sub-sample. The regression model on which the test is based is closely related to the standard ADF test regression. It is a version of the same regression, but on a particular window:
\begin{equation}
\Delta EUA_t=\alpha_{r_1,r_2} + \beta_{r_1,r_2} EUA_{t-1}+\sum_{i=1}^k\psi_{r_1,r_2}^i\Delta EUA_{t-i}+\epsilon_t,
\end{equation}
where $r_1$ denotes the start of the window and $r_2$ the end, both expressed as a fraction of the sample size $T$. The ADF test is concerned with the null hypothesis $H_0:\;\beta_{r_1,r_2}=0$ and the statistic will be denoted by $ADF_{r_1}^{r_2}$. The minimum window size is $r_0$ and the actual window size is $r_w=r_2-r_1$.

The above regression is run multiple times on $\left\lfloor Tr_w\right\rfloor$ observations. The SADF test, which is the first version of the test, was introduced by \citet{PWY}. The regression is estimated on a forward expanding sample, starting at $r_1=0$, whose length increases such that $r_w$ runs from $r_0$ to 1. The test is the supremum over all ADF statistics, hence named the SADF test. Formally, it can be written as
\begin{equation}
SADF(r_0)=\sup_{r_2\in\left[r_0,1\right]}ADF_0^{r_2}.
\end{equation}
More powerful in the case of multiple bubbles is a variant of the test called generalised SADF (GSADF) test in which not only the end point of the window is varied but also the starting point $r_1$. Different windows are considered, for $r_1$ varying from 0 to $(r_2-r_0)$. The test statistic is defined as
\begin{equation}
GSADF(r_0)=\sup_{\substack{r_2\in\left[r_0,1\right]\\r_1\in\left[0,r_2-r_0\right]}} ADF_{r_1}^{r_2}.
\end{equation}
Both tests are right-sided tests. Hence, if the test statistic exceeds the critical value, there is evidence for the existence of an explosive period. It does not provide any indication how many such episodes there are and where they are located. To achieve this, \citet{PSY} develop a date-stamping strategy based on a third version of the sup ADF statistic, the backward SADF statistic (BSADF). This test proceeds in a similar way than the SADF and GSADF tests with the main difference of being obtained for every point in the sample. Fix a point in the sample as the end point of the window, $r_2$, and vary the starting point from 0 to $(r_2-r_0)$, then the BSADF test is obtained as
\begin{equation}
BSADF_{r_2}(r_0)=\sup_{r_1\in\left[0,r_2-r_0\right]} ADF_{r_1}^{r_2}.
\end{equation}
Applying this test to each point in the sample results in a sequence of test statistics. To draw conclusions, we need to compare this sequence to a corresponding sequence of critical values. Before we can identify explosive episodes, it is important to define the minimum duration of a period to qualify as evidence for explosive behaviour. If the test statistic lies above the critical value merely for a few observations, this does not provide sufficient evidence for the existence of a bubble. It is rather a short-lived blip, as \citet{PSY} call it. They suggest to chose a minimum duration which is dependent on the sample, such as $L_T=\log{T}$. We can then identify explosive episodes if we find periods for which the BSADF statistic exceeds the critical values for at least $L_T$ consecutive observations. In our case the minimum duration according to this formula would be 7 weeks. This procedure can also identify ongoing bubbles and serve as an early warning system.

We apply the GSADF test and the date-stamping procedure in the next section using a minimum window size that is set according to the rule suggested by \citet{PSY}, $r_0 = 0.01+1.8\sqrt{T}$, resulting in 40 observations. For the selection of the number of lags we use the Bayesian Information Criterion. We apply two versions of the test. Firstly, the test as originally proposed by \citet{PSY}. For this, we obtain critical values from 2000 replications of their Monte Carlo simulation exercise to mimic the finite sample distribution given our sample size. The results are obtained using the \textsf{R} package MultipleBubbles which is accompanying the \citet{PSY} paper. Secondly, we use a sieve bootstrap approach as proposed in \citet{Pedersen2017}. This paper shows that the GSADF test is critically oversized in the presence of serial correlation which implies that there is a serious risk of finding false evidence for explosive behaviour. The bootstrap corrects the size but it comes at the cost of a lower power. This is why we use both tests as complementary and show that they come to the same conclusion making our results more robust.

\subsection{Modelling bubbles with noncausal $\alpha$-stable processes}
\label{sec:noncausal}

\begin{figure}
    \centering
    \includegraphics[scale=0.24]{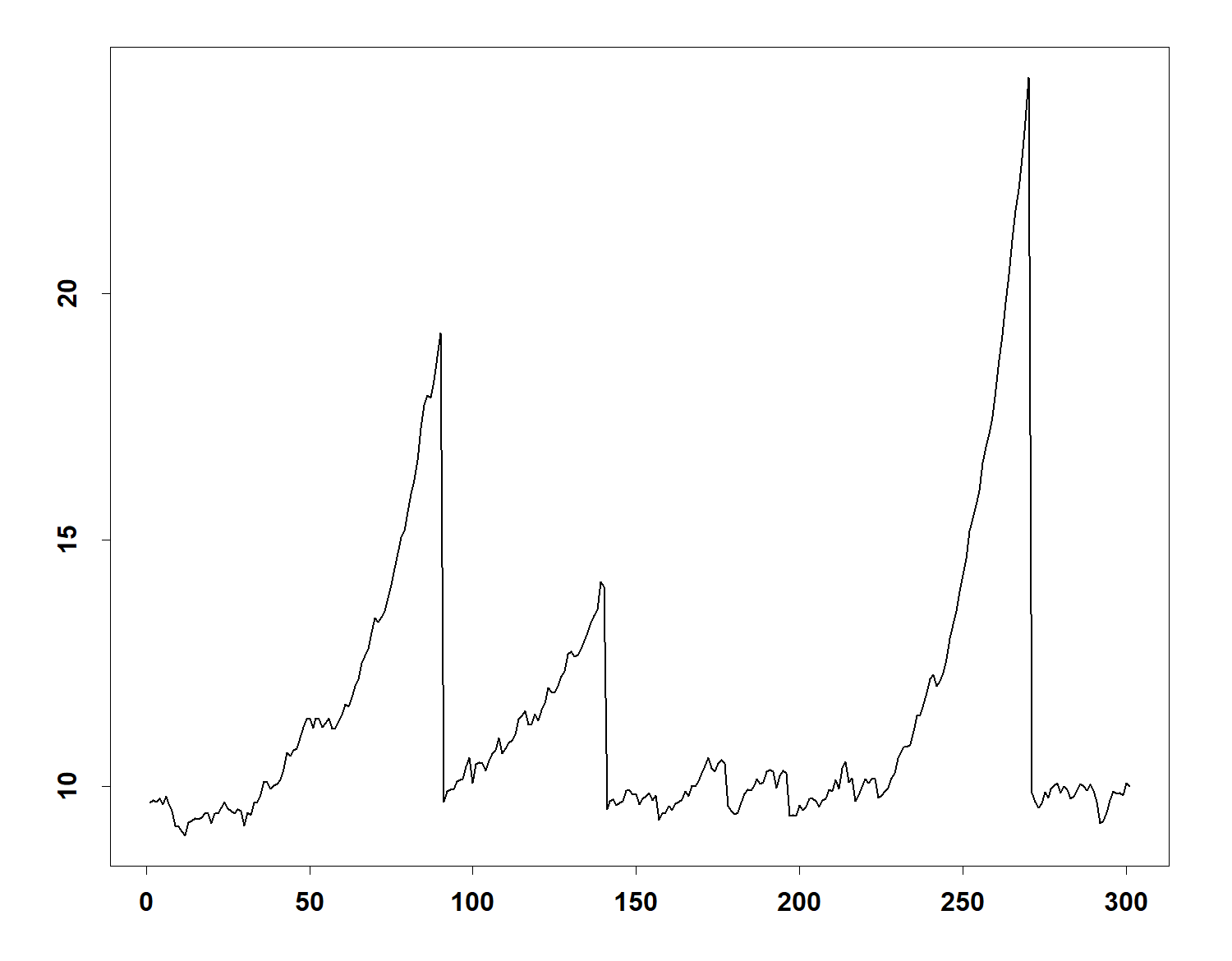}
    \caption{Sample path of an $\alpha$-stable noncausal AR(1) process with parameters $\rho=0.95$, $\alpha=1.7$, $\beta=1$, $\sigma=0.1$, $\mu=0.5$.}
    \label{tab:ar1_sample_path}
\end{figure}

The $\alpha$-stable noncausal AR(1) was introduced and partially studied by \cite{gourieroux2017local} as a candidate model to analyse and forecast bubbles in financial time series, and was studied in details by \cite{fries2018conditional}.
This bubble-generating process features several surprising properties sharply contrasting with classical linear time series models; we start by recalling some of these properties before presenting the results regarding the prediction of bubble crash odds.\\
A noncausal AR(1) process is defined as the strictly stationary solution of the stochastic recurrence equation:
\begin{align*}
X_t = \rho X_{t+1} + \varepsilon_t, 
\end{align*}
where $0<|\rho|<1$, and $(\varepsilon_t)$ is an i.i.d. error sequence.
The appellation \textit{noncausal}, standard in the literature (as well as \textit{anticipative}), refers to the fact that the stationary solution $(X_t)$ of the above equation admits a moving average representation in terms of <<future>> error terms as $X_t = \sum_{k=0}^{+\infty} \rho^k \varepsilon_{t+k}$.
The above process is said to be an $\alpha$-stable noncausal AR(1) when $\varepsilon_t\stackrel{i.i.d.}{\sim}\mathcal{S}(\alpha,\beta,\sigma,\mu)$, where $\mathcal{S}(\alpha,\beta,\sigma,\mu)$ denotes the univariate $\alpha$-stable distribution with tail index $\alpha\in(0,2)$, asymmetry $\beta\in[-1,1]$, scale $\sigma>0$ and location $\mu\in\mathbb{R}$, i.e., the distribution with characteristic function:
\begin{align*}
\mathbb{E}[e^{iu\varepsilon_0}] = \exp\Big\{-\sigma^{\alpha}|u|^{\alpha}\Big(1-i\beta \,\text{sign}(u)w(\alpha,u)\Big)+iu\mu\Big\},
\end{align*}
with $w(\alpha,s) = \text{tg}\left(\frac{\pi \alpha}{2}\right)$, if $\alpha \ne 1$, and $w(1,s) = - \frac{2}{\pi}\ln|s|$ otherwise, for $s\in\mathbb{R}$. 
The case $\alpha=2$ (excluded in the noncausal bubble modelling framework) corresponds to the Gaussian distribution, which is a particular stable distribution -- and the only instance which does not feature heavy-tails. 
For $\alpha<2$, the case of interest for bubble modelling, $\alpha$-stable distributions have power-law decaying tails:
\begin{align*}
\mathbb{P}(|\varepsilon_0| > x) \underset{x\rightarrow\infty}{=} O(|x|^{-\alpha}).
\end{align*}
For more details on univariate and multivariate $\alpha$-stable distributions we refer to the comprehensive monograph by \cite{samorodnitsky29stable}. Figure \ref{tab:ar1_sample_path} illustrates a typical sample path of a noncausal AR(1) process featuring multiple bubbles.

Although seemingly defined in reverse time, $(X_t)$ is actually a Markov process in the classical sense: $X_{t+h}|X_t,X_{t-1},\ldots \stackrel{d}{=} X_{t+h}|X_t$ \citep{cambanis1995prediction}. 
As a linear combination of $\alpha$-stable errors, $X_t$ is itself marginally $\alpha$-stable; it is thus heavy-tailed and admits very little marginal moments: $\mathbb{E}[|X_t|^s]<+\infty$ only for $s<\alpha$.
Despite featuring infinite marginal variance, the conditional distribution of future realisation of the process, say, $X_{t+h}$ for some horizon $h\ge1$, given the past history $X_t,X_{t-1},\ldots$ admits finite moments at higher orders, namely:
$$
\mathbb{E}[|X_{t+h}|^\gamma|X_t] < +\infty, \hspace{1cm} \text{ for } \gamma < 2\alpha+1.
$$
Thus, the predictive distribution of the process always admits a well-defined conditional expectation, even though marginally $\mathbb{E}[|X_t|]=+\infty$ for $\alpha\le1$, and can admit up to finite conditional variance, skewness and kurtosis, provided the tail exponent $\alpha$ is close enough to 2. \cite{gourieroux2017local} obtained closed form expressions for the conditional expectation and variance in the special cases of symmetric ($\beta=0$) and Cauchy-distributed ($\alpha=1$, $\beta=0$) errors, highlighting that $(X_t)$ can feature GARCH effects, and even presents a stationary unit root under special parameterisation (Cauchy errors and $\rho>0$) in the sense that $\mathbb{E}[X_{t+1}|X_t]=X_t$.

The stationary solution $(X_t)$ admits an infinite forward-looking moving average representation in terms of the future errors as $X_t=\sum_{k=0}^{\infty} \rho^k \varepsilon_{t+k}$. 
By a change of variable in the moving average representation, one can rewrite $X_t$ as $X_t=\sum_{\tau \in \mathbb{Z}} \rho^{\tau-t}\mathds{1}_{\tau-t}\varepsilon_\tau$, and it can be noticed that the trajectory of the process $(X_t)$ is a linear combination of so-called baseline paths \citep{gourieroux2017local,fries2019mixed}: deterministic functions of time $t\mapsto \rho^{\tau-t}\mathds{1}_{\tau-t\ge0}$, which are shifted in time by $\tau$ and scaled by the random error $\varepsilon_{\tau}$. 
Due to the heavy-tails of the distribution, realisations of the errors $\varepsilon_t$'s often take extreme values.
Intuitively, if $\varepsilon_{\tau_0}$ is extreme for a particular date $\tau_0$, then the trajectory of $(X_t)$ can be locally approximated by an exponential trend culminating in a peak and ending in a crash down to pre-bubble levels: $X_t\approx \rho^{{\tau_0}-t}\mathds{1}_{{\tau_0}\ge t} \varepsilon_{\tau_0}$, for $t$ in the vicinity of $\tau_0$. 

This intuition is formalised and shown to hold in \cite{fries2018conditional,fries2018path}, and provides a convenient analytical framework to formulate predictions of likely future paths for trajectories of noncausal processes. 
In particular, it allows a formal quantification of the crash odds at future horizons of bubbles generated by the noncausal AR(1).
\cite{fries2018conditional} obtained functional forms for the conditional expectation, variance, skewness and kurtosis of $X_{t+h}$ given $X_t=x$ for any admissible parameterisation and for the general class of \textit{two-sided} infinite moving average processes of the form $X_t=\sum_{k\in\mathbb{Z}}a_k\varepsilon_{t+k}$. Although the conditional moments have a complex form in general, 
their expressions drastically simplify when one considers the important case of interest of explosive episodes.
For the noncausal AR(1) with $\rho>0$, it is shown 
that during explosive episodes (i.e., large conditioning values $X_t=x$) the conditional moments of $X_{t+h}$ given $X_t=x$ become equivalent to that of a weighted Bernoulli random variable charging probability $\rho^{\alpha h}$ to the value $\rho^{-h}x$, and probability $1-\rho^{\alpha h}$ to 0.
In other words, conditionally on an explosive episode having reached the level $X_t=x$, either:
\begin{enumerate}
    \item with probability $\rho^{\alpha h}$, the bubble will survive at least until horizon $h$ and follow an exponential trend of growth rate $\rho^{-1}$: from $X_t=x$, the trajectory will grow to $\rho^{-h}x$.
    \item with probability $1-\rho^{\alpha h}$, the bubble will crash at some intermediate future date before horizon $h$: from $X_t=x$, the trajectory will crash to $X_{t+h}$, with $X_{t+h}/X_t \approx 0$.
\end{enumerate}
This  asymptotic behaviour of $X_{t+h}|X_t$, initially deduced from the form of first four conditional moments, has been confirmed to hold in distribution \citep{fries2018path}.
Surprisingly, it can be noticed that for the $\alpha$-stable noncausal AR(1), the survival probability of bubbles at any horizon $h$, $\rho^{\alpha h}$, does not depend on the conditioning level $X_t=x$. Since the noncausal AR(1) is a Markov process, this implies that the survival probabilities of bubbles generated by the model do not depend at all on their past history: the bubbles feature a memory-less, or non-aging, property characterised by an exponential survival probability function.
They can thus be fully characterised by their so-called half-life, the duration $h_{0.5}$ such that their survival probability at horizon $h_{0.5}$ is 50\%. For the noncausal AR(1) with autoregressive parameter $\rho$ and tail index $\alpha$, the half-life reads:
\begin{align*}
h_{0.5} = -\dfrac{\ln 2}{\alpha \ln \rho}.
\end{align*}
From the latter quantity, the crash odds (CO) at any horizon $h\ge1$ can be expressed as
\begin{align*}
CO(h) = 1 - (1/2)^{\frac{h}{h_{0.5}}},
\end{align*}
while the expected duration (ED) of an explosive episode reads:
\begin{align*}
ED(h) = \dfrac{1}{1-\rho^{\alpha}}.
\end{align*}

\section{Robustness Analysis}
In addition to the final results shown in the paper, we conducted a formal outlier detection as well as an extensive bandwidth selection procedure, both serve as a robustness check for our results. In addition, we present results using additional/alternative data series.

\subsection{Outlier detection}
\label{sec:outlier}
As mentioned in Section \ref{sec:nonpara} we look for outliers in the dependent variable as our methods are not designed to explain sudden jumps in allowance price returns. The results presented in the main text are the results after outliers have been removed. We now explain how we detect them and how the results look without the removal.

We apply the impulse indicator saturation (IIS) approach proposed in \citet{Santos2008}. This approach includes a dummy variable at every possible time point and performs expanding and contracting multiple block searches to determine which dummy variables should be retained. It is applied here to the EUA return series with a nominal significance level of $\alpha=0.005$. 

\begin{figure}[h!]
\centering
\includegraphics[width=0.8\textwidth,trim=0cm 0cm 0 0cm, clip]{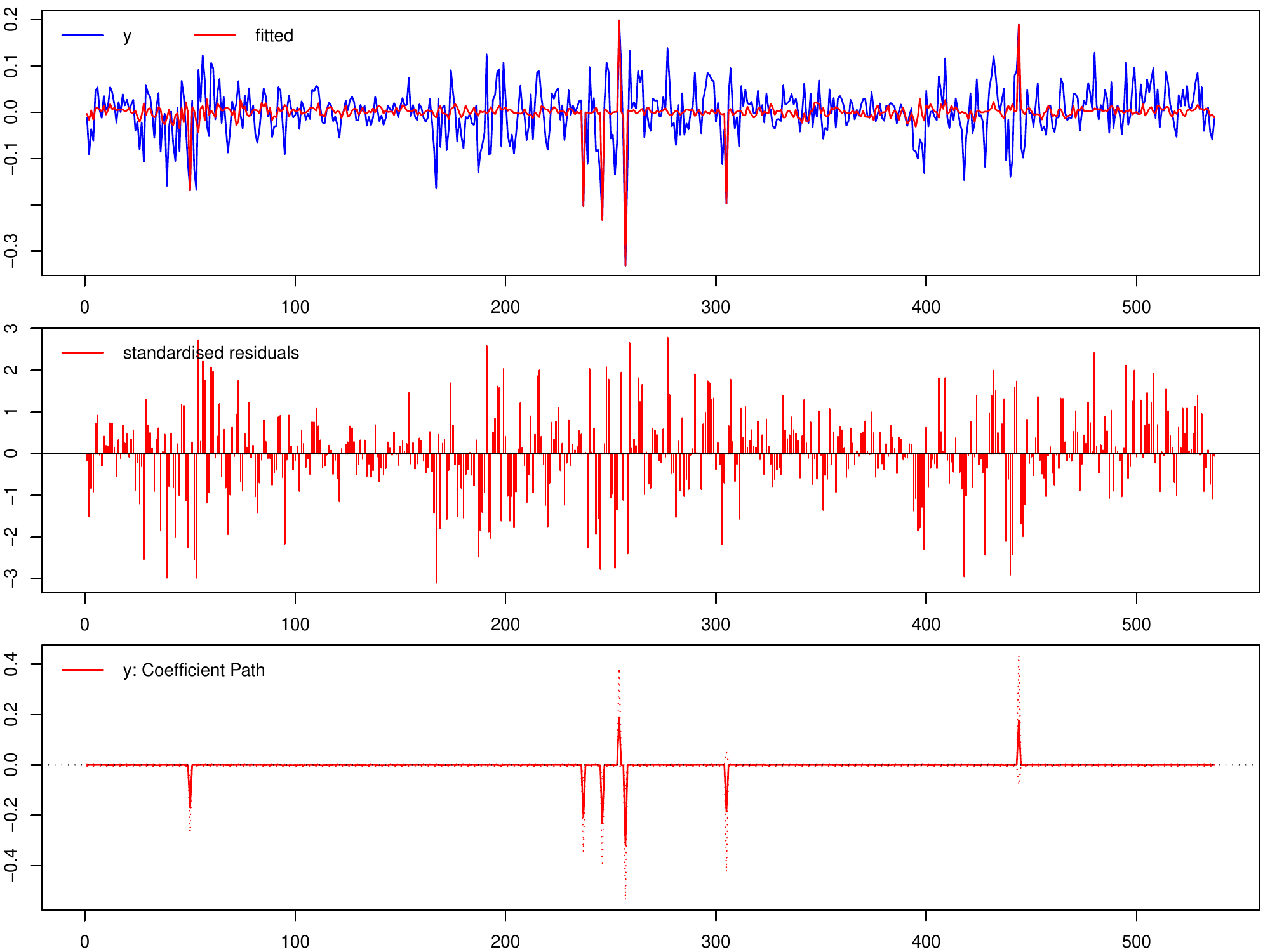}
\caption{Summary of impulse indicator saturation results. Figure produced with R package \textit{gets}. Top: EUA returns (blue) and fit of dummy regression (red); middle: standardised residuals after regression on dummies; bottom: retained dummy variables}
\label{fig:IIS}
\end{figure}

\begin{figure}[!h]
	\centering
	\subfigure[$\hat{\beta}_0(t)$]
    {
    \includegraphics[width=0.47\linewidth, clip, trim = {0 1cm 0 2cm}]{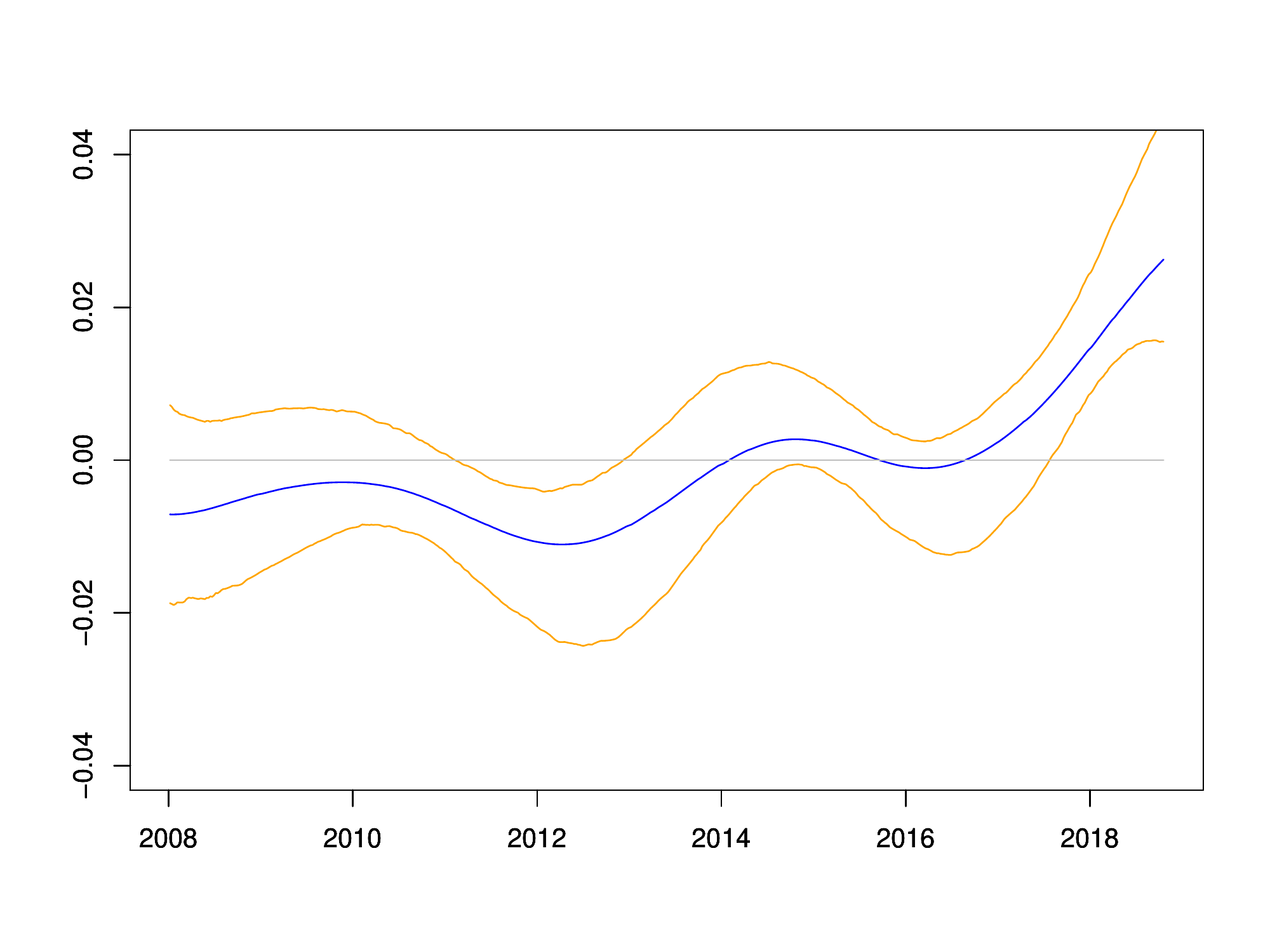}	
    \label{fig:trend2}
     }
     \subfigure[$\hat{\beta}_{coal}(t)$]
     {
      \includegraphics[width=0.47\linewidth, clip, trim = {0 1cm 0 2cm}]{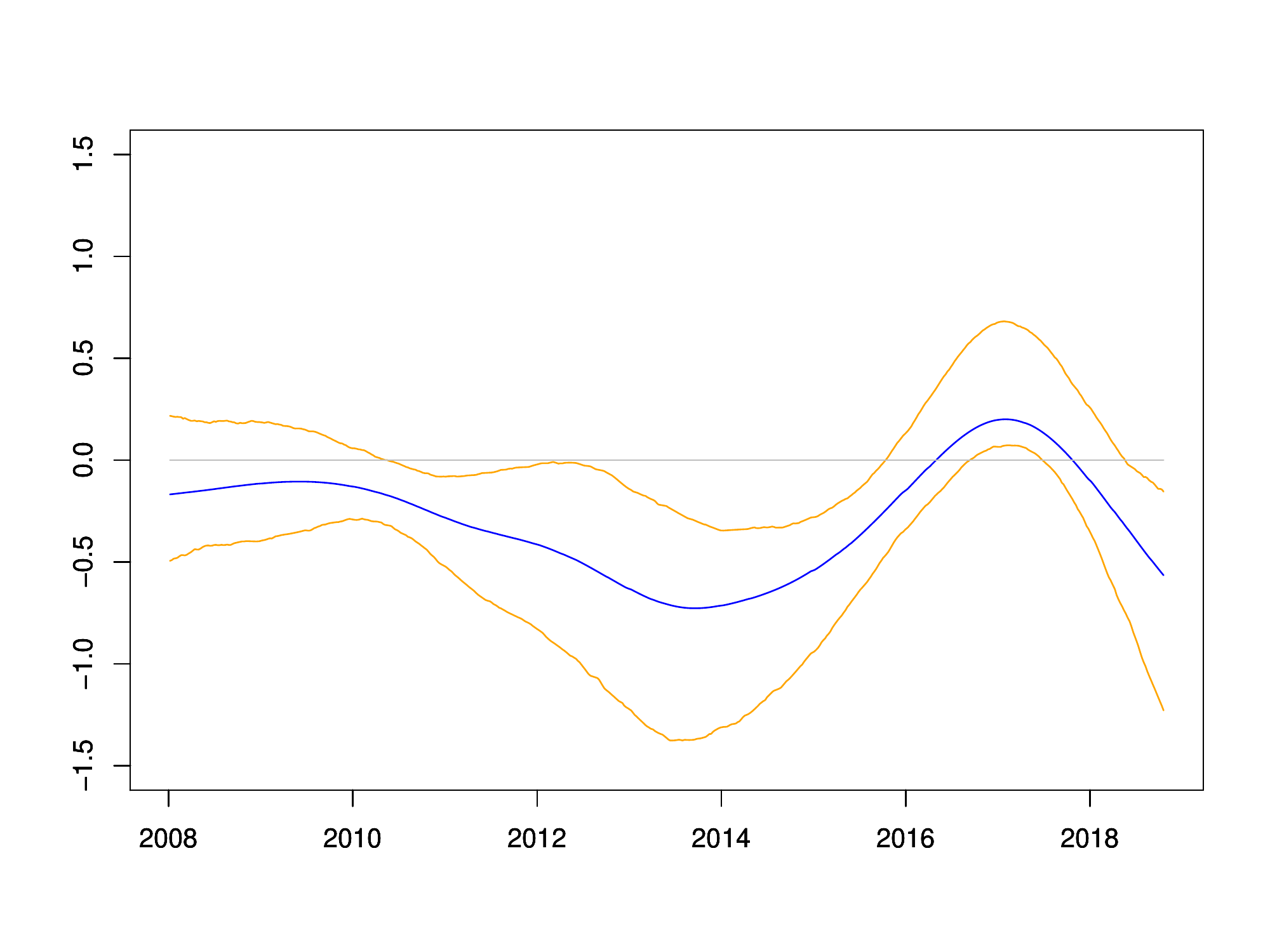}
      \label{fig:coal2}
     }\\
		\subfigure[$\hat{\beta}_{gas}(t)$]
    {
    \includegraphics[width=0.47\linewidth, clip, trim = {0 1cm 0 1.5cm}]{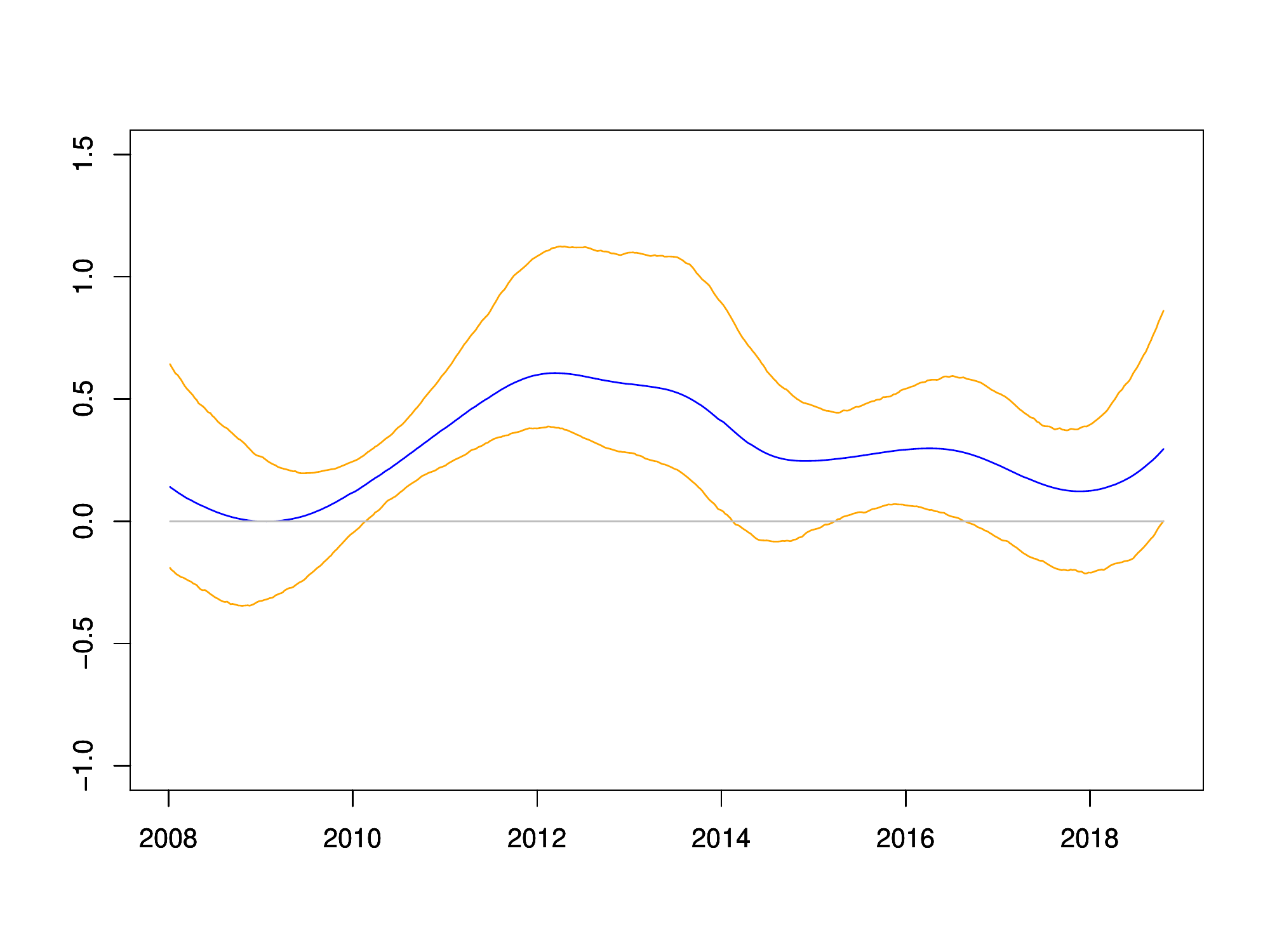}	
    \label{fig:gas2}
     }
     \subfigure[$\hat{\beta}_{oil}(t)$]
     {
      \includegraphics[width=0.47\linewidth, clip, trim = {0 1cm 0 1.5cm}]{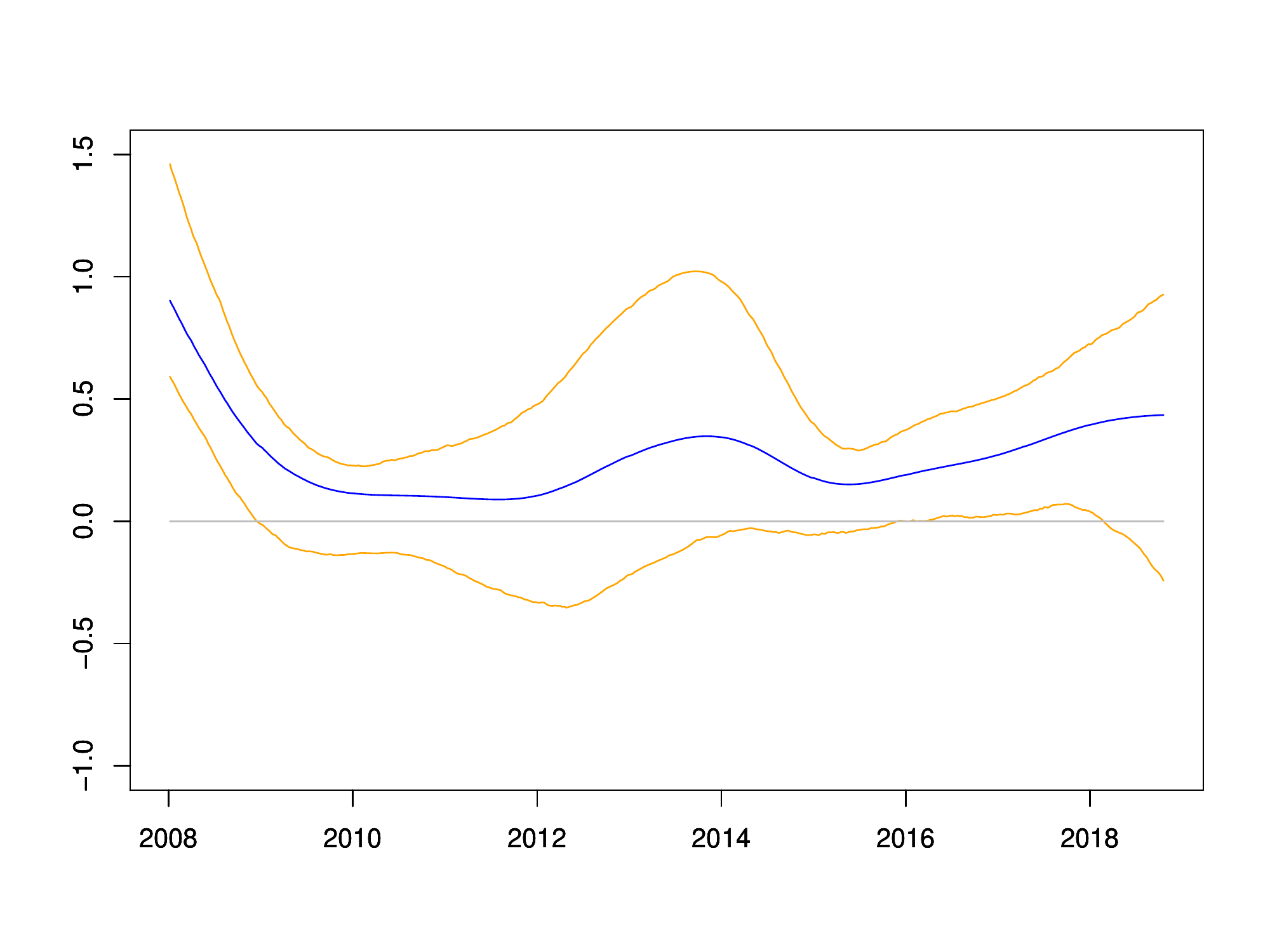}
      \label{fig:oil2}
     }
     \subfigure[$\hat{\beta}_{temp}(t)$]
     {
      \includegraphics[width=0.47\linewidth, clip, trim = {0 1cm 0 1.5cm}]{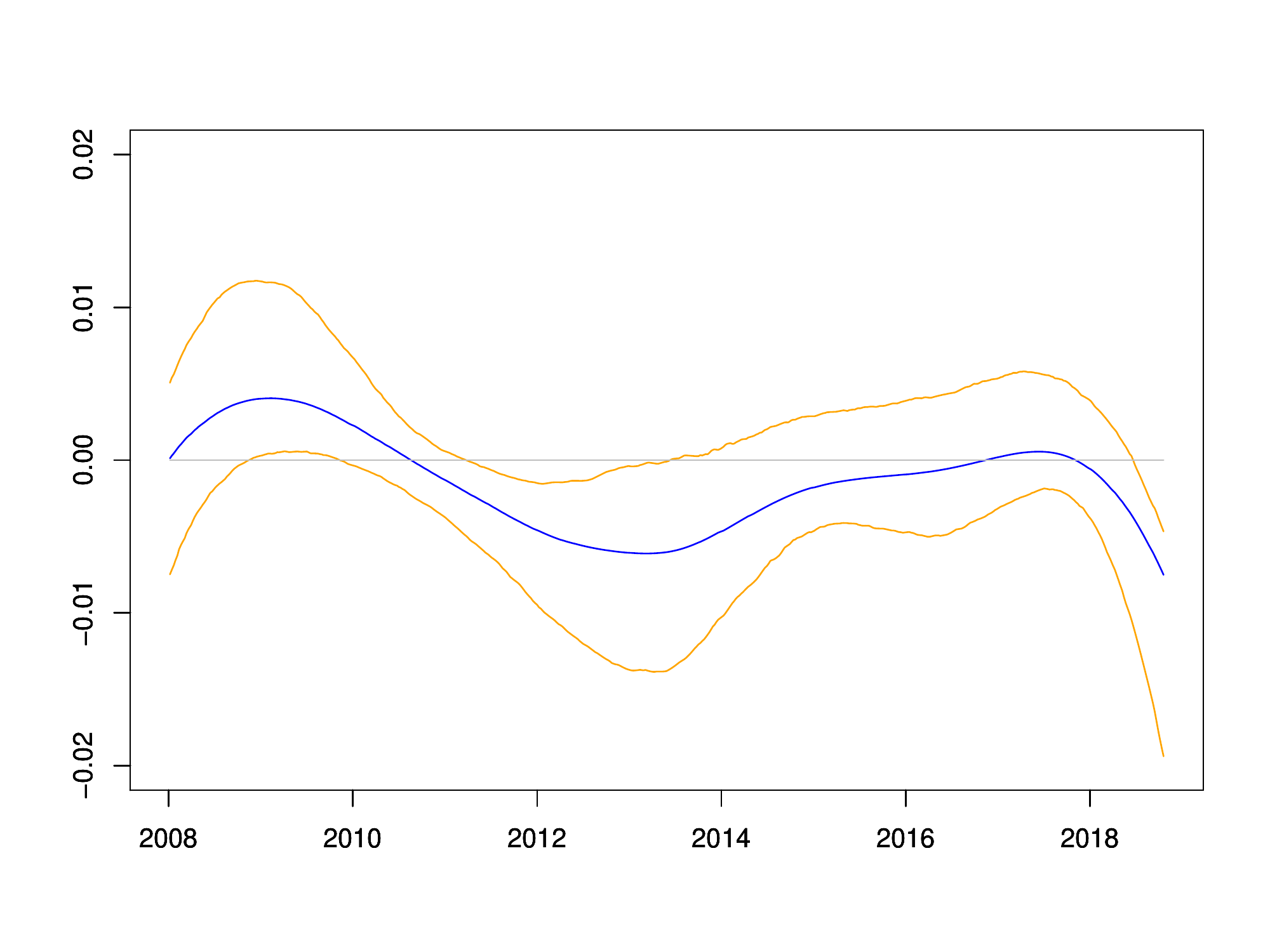}
      \label{fig:temp2}
     }
		\caption{Nonparametrically estimated coefficient curves and 95\% confidence intervals before the removal of outliers}
		\label{fig:nonparametric2}
\end{figure}

The IIS method applied to our dependent variable retains 10 dummy variables corresponding to 10 outliers located at observations 39, 50, 53, 167, 237, 246, 254, 257, 305 and 444. Corresponding to time points in October 2008, January and February 2009, June 2011, November 2012, January 2013, March and April 2013,  March 2014 and December 2016. If we control for the impact of our most important explanatory variables -- coal, gas, oil prices and temperature data -- three of these outliers can be explained (at observation 39, 53 and 167). Figure \ref{fig:IIS} depicts the results of this IIS application. With this approach we retain 7 outliers which we delete from the EUA price series and subsequently repeat our analysis. The content of Figure \ref{fig:nonparametric} was produced with the outlier-corrected data. In Figure \ref{fig:nonparametric2} we repeat this figure but the raw data. Comparison of the two figures shows that there are no major differences. A noteworthy difference is that the confidence intervals get more narrow over some periods. In particular, this effect is visible for the gas, coal and oil price coefficient in the period from end 2012 to the beginning of 2014, where for the original data the confidence intervals experienced a widening which disappears with the removal of outliers. This does not change the results in a substantial way nor does it affect the conclusions drawn regarding the significance of coefficients.

\subsection{Bandwidth selection}
\label{sec:bandwidth_selec}
A crucial aspect of any local fitting method is the choice of the bandwidth parameter $h$. It simultaneously controls the amount of smoothing and the complexity of the estimation. A small value of the bandwidth stands, on one hand, for a small approximation error and a resulting small modeling bias. On the other hand, it means that only a few data points are included in the local neighborhood and therefore, the variance of the estimator is large. In addition, the estimated curve is less smooth and the model is more complex than with a larger bandwidth.  

In contrast, a large value of the bandwidth can create a large modeling bias, but the model is less complex and the amount of smoothing will be large. For the extreme values $h\rightarrow 0$ and $h\rightarrow \infty$, the estimate coincides, respectively, with interpolation of the data points or a parametric linear regression estimate. Therefore, bandwidth selection controls the complexity of the model and in addition, there is a bias-variance tradeoff. 

Comparing parametric and nonparametric fitting, the bandwidth parameter can be seen as an additional dimension. Parametric fitting is like nonparametric fitting, where the choice of the bandwidth parameter is constant and different families of models are considered. Taking a bandwidth of $h\rightarrow \infty$ is unquestioned in all situations of parametric modeling, while with nonparametrics, the bandwidth is seen as an additional parameter, which is carefully selected so that the estimation outcome fits the data well. 

A theoretically optimal bandwidth can be obtained but it is infeasible for practical use as the expression depends on several unobservable quantities, e.g. the second derivative of $\boldsymbol{\beta}(\cdot)$. We refer the interested reader to \citet{FG} for more details. We focus on rules how to select bandwidths in practice.

Many data-driven methods for bandwidth selection are based on  the principle of cross validation (CV). The basic idea is to find the value of the bandwidth that provides the best fit in terms of minimizing the sum of squared residuals without overfitting. Simply finding $h$ that minimises the sum of squared residuals creates a problem of overfitting, since for $h\rightarrow 0$ a perfect fit is obtained in the limit. The result of this obviously problematic procedure would the smallest considered bandwidth in all cases. The leave-one-out estimator provides a way to circumvent overfitting. The first step is to construct the leave-one-out estimator by leaving out the observation $t$ that receives the highest weight in the local estimation. The second step in the least-squares CV approach is to look at the weighted average of the leave-one-out squared residuals and minimise them with respect to $h$. 

\begin{table}[bh!]
\centering
\caption{Bandwidth selection}
\begin{tabular}{lcc}
\midrule\midrule
 & Bandwidth & Reference\\ \cmidrule(lr){2-2} \cmidrule(lr){3-3}
 Cross Validation  & 0.0874 &  e.g. \citet{FG} \\
 Generalised CV & 0.0800 &  \citet{ZWu} \\
 AIC  & 0.0866 & \citet{Cai} \\
 LLO CV  & 0.0894 & \citet{ChuMarron}  \\ \cmidrule(lr){2-3}
 Average & 0.0859 & \\ 
 \midrule\midrule
\end{tabular}
\caption*{\textit{Source:} Own calculations using R.\\\textit{Notes:} Optimal bandwidth chosen by different methods. The leave-($2l+1$)-out CV approach is applied with $l=4$.}
\label{tab:bandwidths}
\end{table}

Cross validation, however, was originally designed for independent data and can therefore potentially lead to problems in time series applications. \citet{ChuMarron} show that cross validation systematically chooses bandwidths that are too small (too large) in the presence of positive (negative) autocorrelation. They propose a modification of the criterion called modified cross validation and show that it works well in time series applications. It follows the same general principle as CV, but it is based on a different estimator. \citet{ChuMarron} use a leave-($2l+1$)-out version of the leave-one-out estimator.

\begin{figure}[h]
	\centering
	\subfigure[CV]
    {
    \includegraphics[width=0.47\linewidth, clip, trim = {0 1cm 0 2cm}]{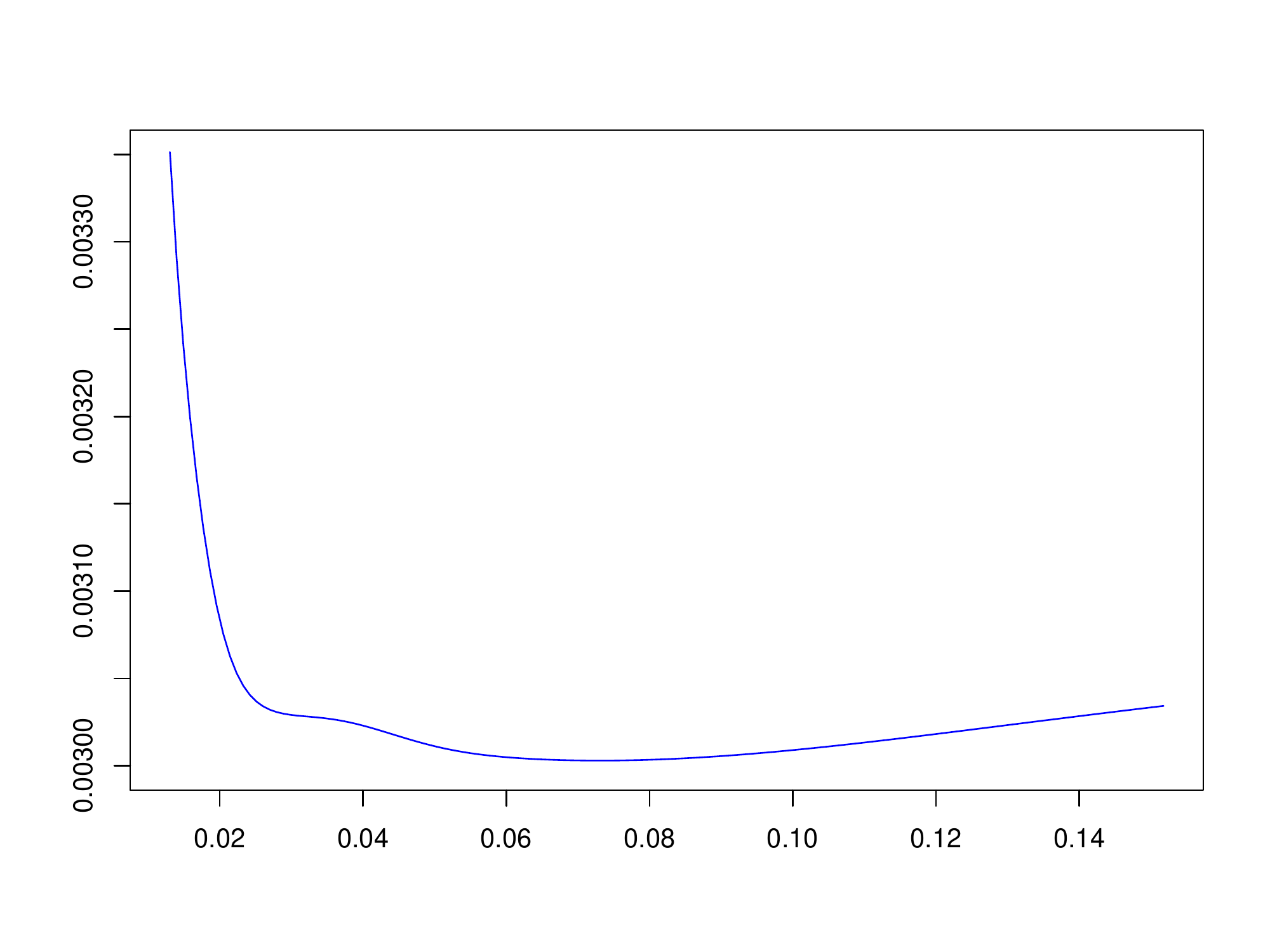}	
    \label{fig:CV}
     }
     \subfigure[GCV]
     {
      \includegraphics[width=0.47\linewidth, clip, trim = {0 1cm 0 2cm}]{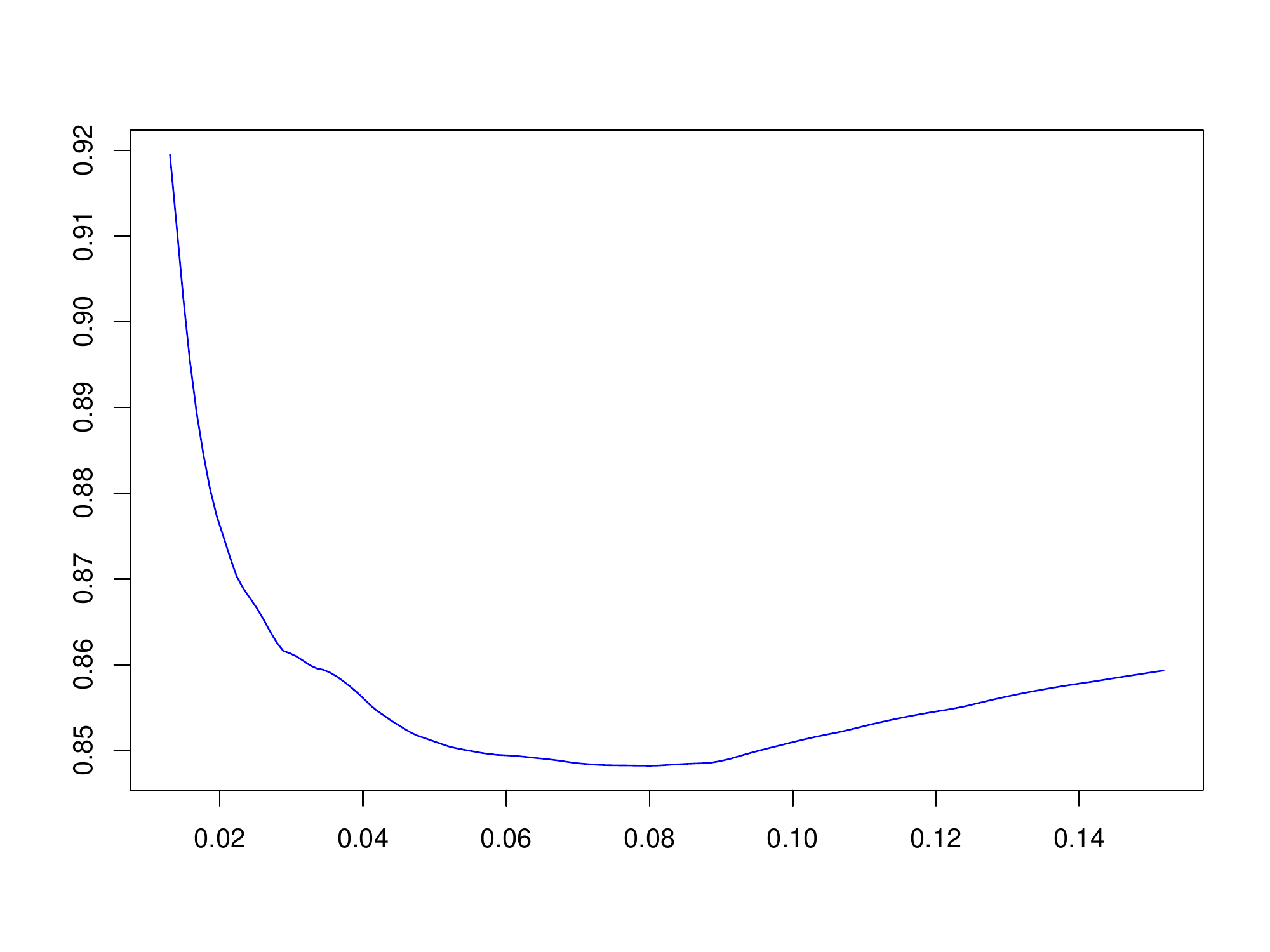}
      \label{fig:GCV}
     }\\
		\subfigure[LLO CV]
    {
    \includegraphics[width=0.47\linewidth, clip, trim = {0 1cm 0 1.5cm}]{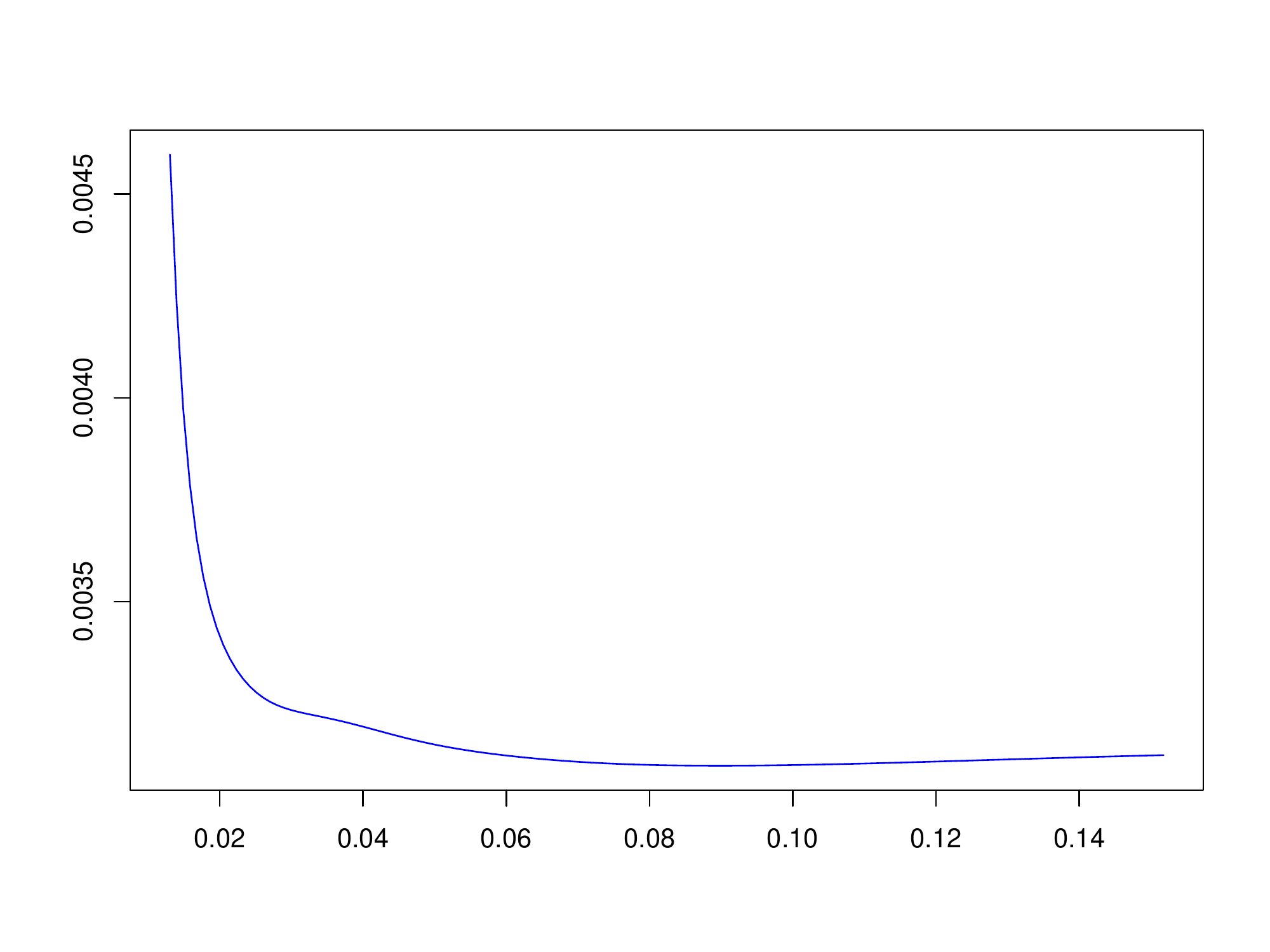}	
    \label{fig:L4O}
     }
     \subfigure[AIC]
     {
      \includegraphics[width=0.47\linewidth, clip, trim = {0 1cm 0 1.5cm}]{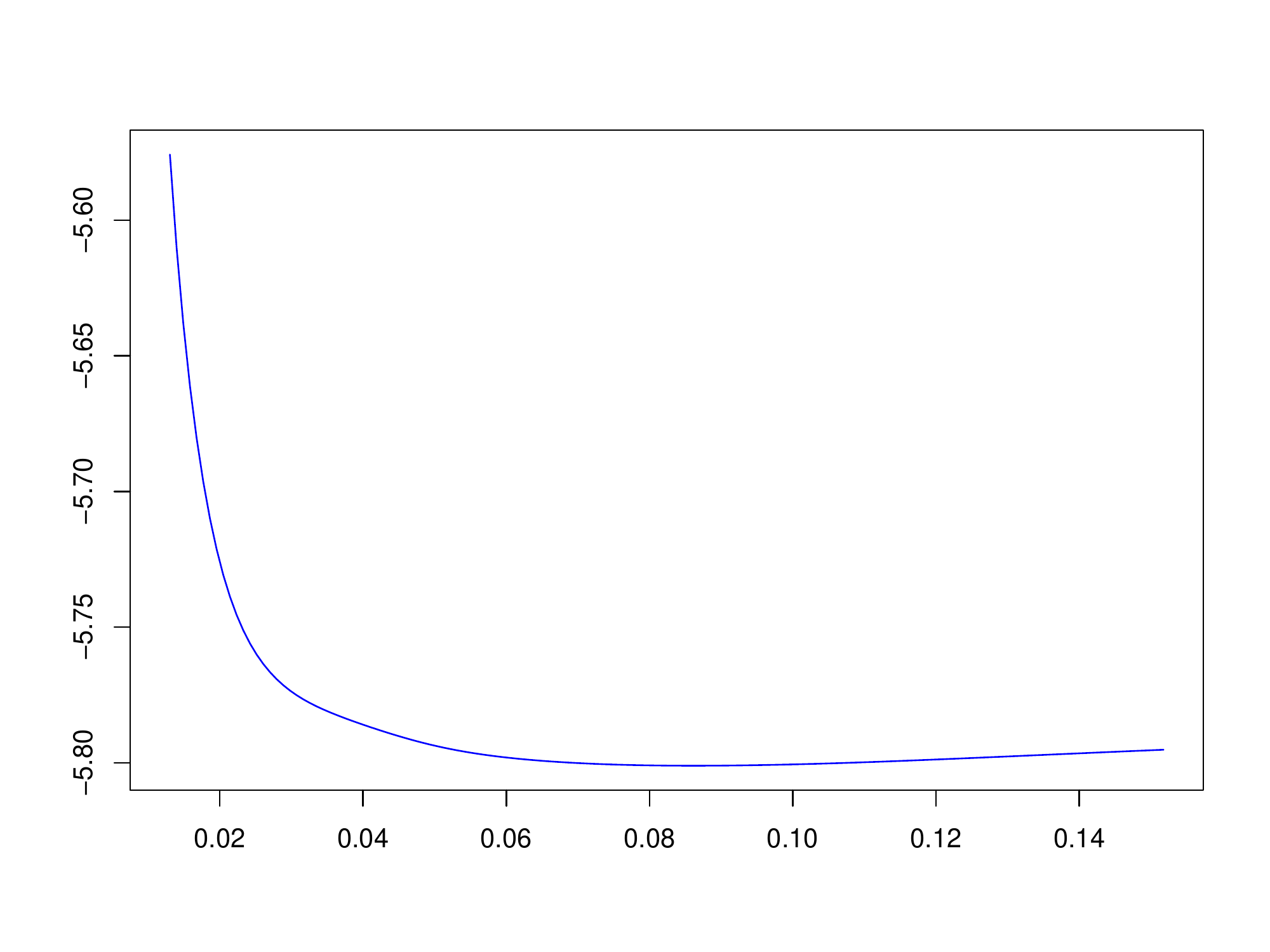}
      \label{fig:AIC}
     }
		\caption{The four different data-driven bandwidth selection criteria: (a) Cross validation, (b) Generalised cross validation, (c) Leave-$(2l+1)$-out CV, (d) Akaike-based bandwidth selection criterion}
		\label{fig:bandwidth_selection}
\end{figure}

In addition to cross validation and modified cross validation, we follow the suggestions in \citet{Cai} as well as \citet{ZWu} and make use of two additional bandwidth selection methods. \citet{Cai} considers an approach based on the Akaike information criterion (AIC) while \citet{ZWu} use generalised cross validation (GCV) originally proposed by \citet{Craven1978}.

\begin{figure}[h]
	\centering
	\subfigure[$h=0.07$]
    {
    \includegraphics[width=0.47\linewidth, clip, trim = {0 1cm 0 2cm}]{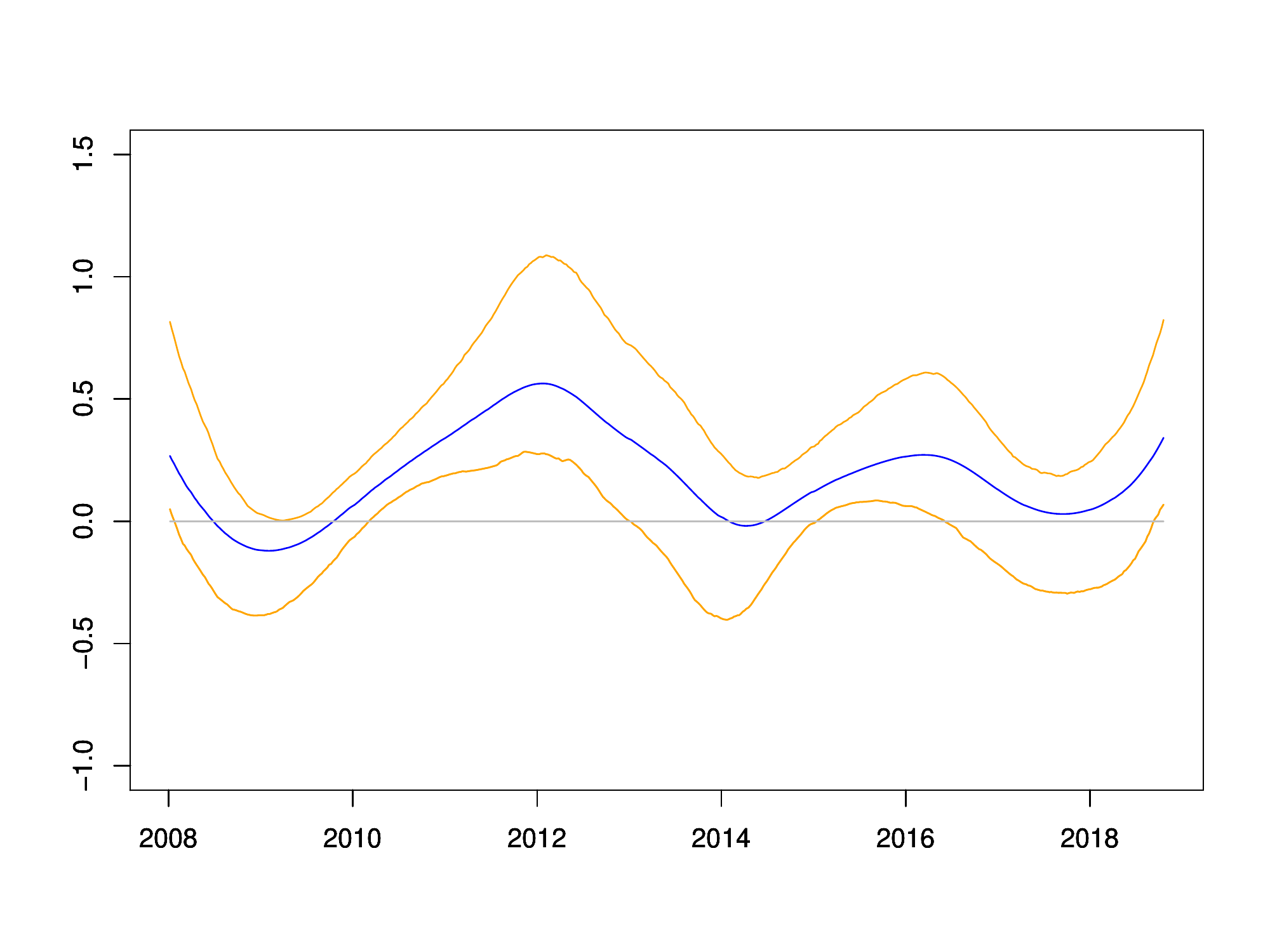}	
    \label{fig:007}
     }
     \subfigure[$h=0.09$]
     {
      \includegraphics[width=0.47\linewidth, clip, trim = {0 1cm 0 2cm}]{Gas_iis.pdf}
      \label{fig:009}
     }\\
		\subfigure[$h=0.11$]
    {
    \includegraphics[width=0.47\linewidth, clip, trim = {0 1cm 0 1.5cm}]{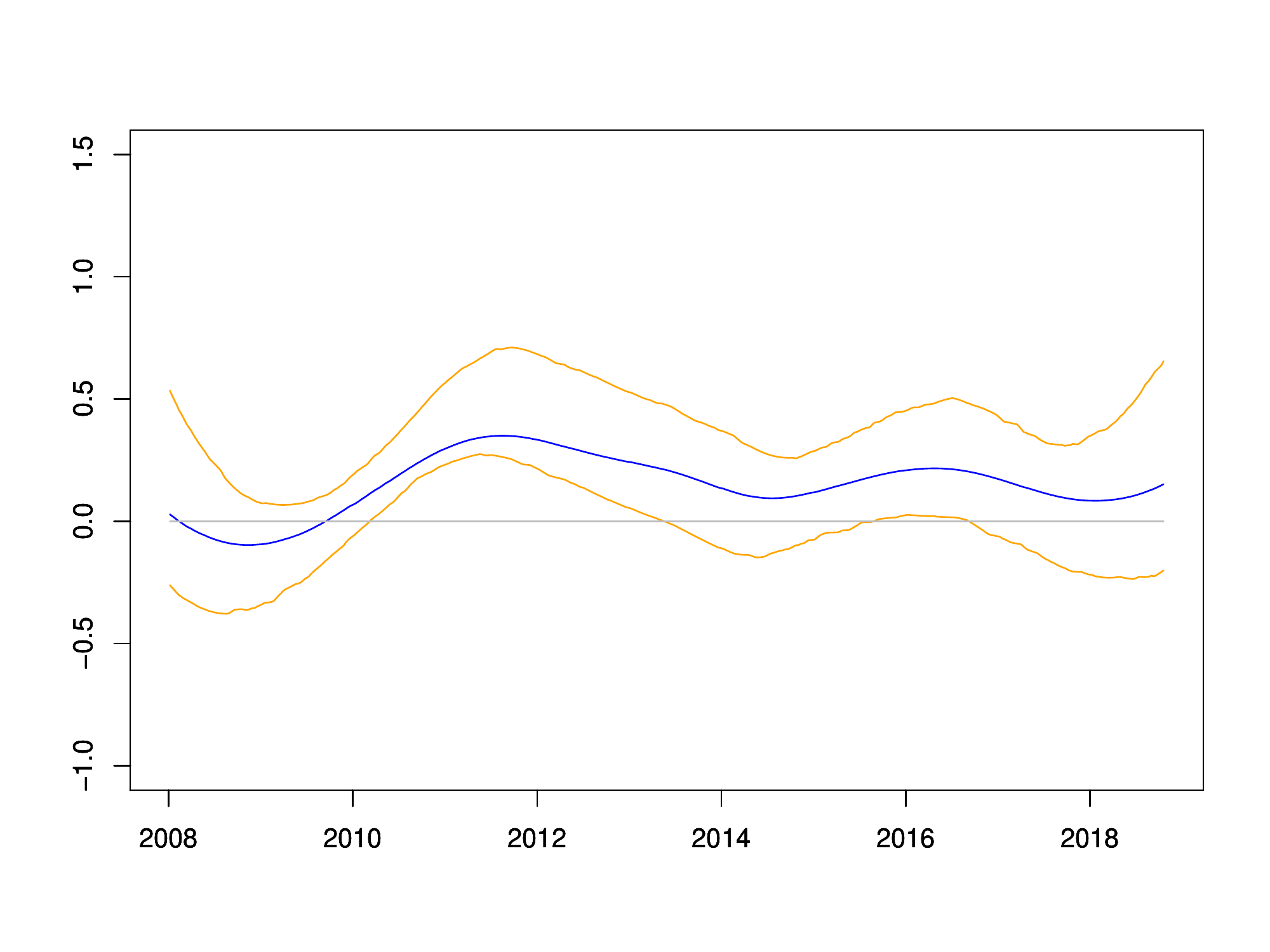}	
    \label{fig:011}
     }
	\caption{Three different estimates of the gas price coefficient, using $h=0.07,0.09,0.11$}
	\label{fig:gas_h}
\end{figure}

The chosen bandwidths are presented in Table \ref{tab:bandwidths} while the criteria are plotted in Figure \ref{fig:bandwidth_selection}. All bandwidths are in a similar range, with the GCV criterion selecting the smallest and the leave-($2l+1$)-out CV the largest bandwidth. In the main text we decide to use a bandwidth of 0.09 which corresponds to the average. We also run the analysis with a smaller and a larger bandwidth, $h=0.07$ and $h=0.11$. We observe that the results do not change much. For illustration, we plot the gas price coefficient obtained with the three different bandwidths in Figure \ref{fig:gas_h}. The results for the other estimates can be obtained from the authors upon request. 

\subsection{Additional explanatory variables}
\label{sec:add_var}
This section is two-fold. First, we replace the oil price as indicator of economic activity by the two stock indices in our data set -- Euro STOXX Europe 50 and STOXX Europe 600. Second, we include several other variables (among them the fuel switching price) in the time-varying coefficient model which we did not include in the final results presented in the main paper. The effect of the fuel switching price is already captured in our model and the remaining additional variables did not show a period of significance. Excluding them from the regression does not alter the shape of the remaining coefficient curves.  

\begin{figure}[!h]
	\centering
     \subfigure[$\hat{\beta}_{Stoxx50}(t)$]
     {
      \includegraphics[width=0.47\linewidth, clip, trim = {0 1cm 0 1.5cm}]{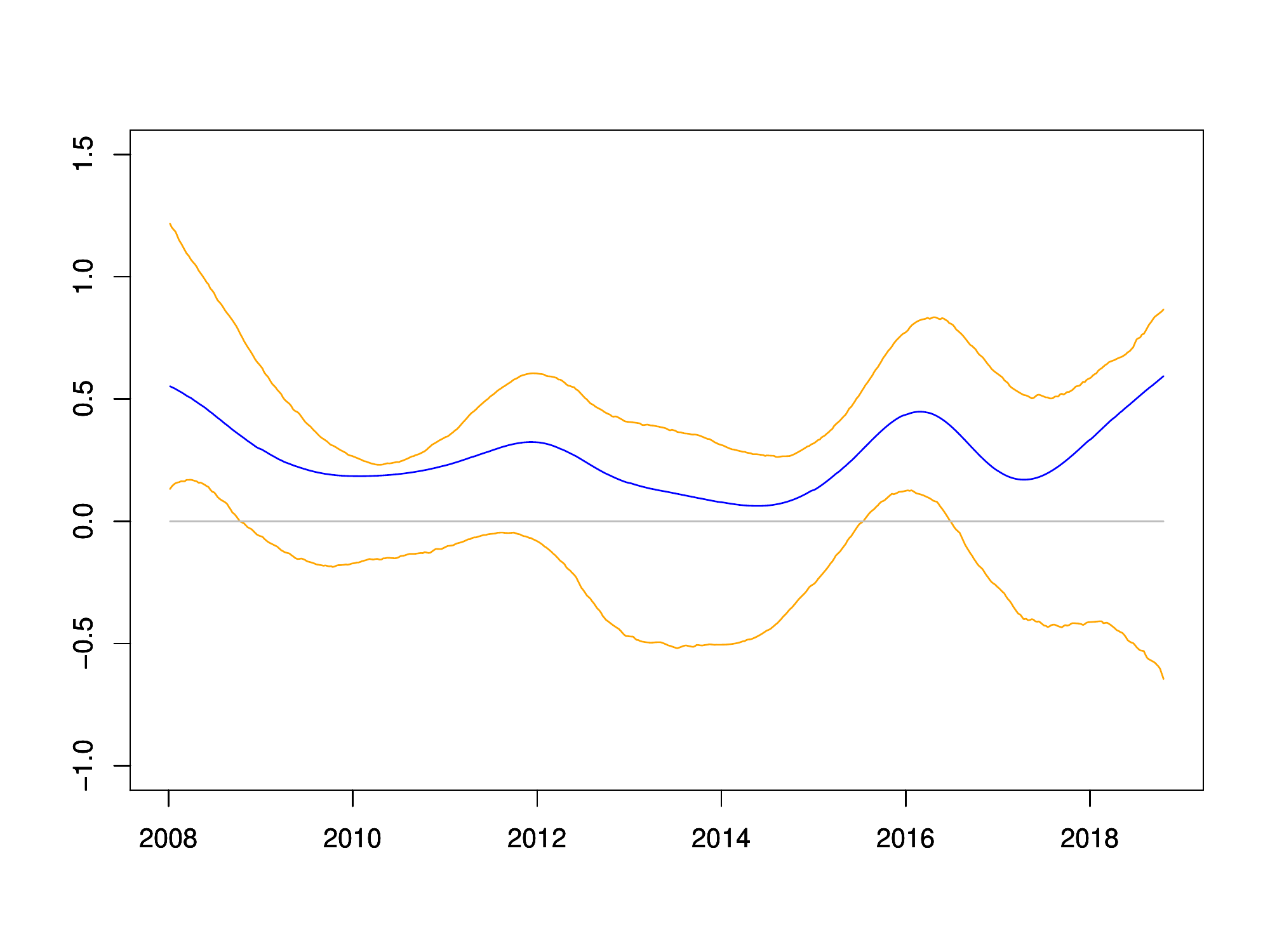}
      \label{fig:stoxx50}
     }
     \subfigure[$\hat{\beta}_{Stoxx600}(t)$]
     {
      \includegraphics[width=0.47\linewidth, clip, trim = {0 1cm 0 1.5cm}]{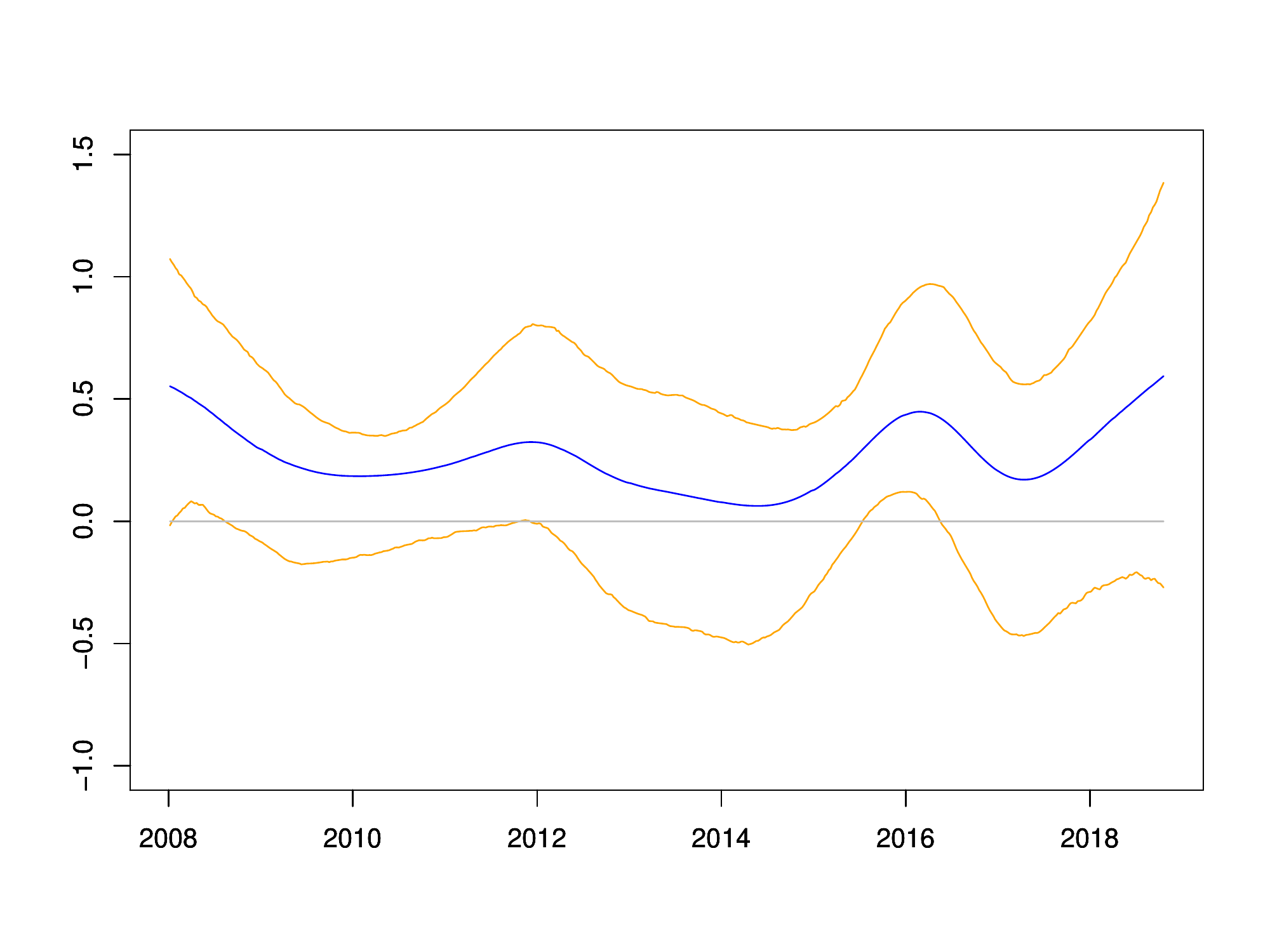}
      \label{fig:stoxx600}
     }
		\caption{Nonparametrically estimated coefficient curves and 95\% confidence intervals for the two stock indices}
		\label{fig:nonparametric_stoxx}
\end{figure}

Starting with the first point, we plot in Figure \ref{fig:nonparametric_stoxx} the estimated parameter curves for the two stock indices. They were entered separately into the nonparametric regression, replacing the oil price. We give here only the coefficient of the replaced data series and not the entire set of regressors. Switching from the oil price to either one of the stock indices does not change the parameter estimates for the remaining regressors. Their coefficient curves are almost identical to Figure \ref{fig:nonparametric}. For the sake of brevity we do not plot them again. They can be obtained from the authors upon request.

Figure \ref{fig:nonparametric_stoxx} reveals that the two stock indices produce very similar coefficient estimates. Given the shape of the two series (cf. Figure \ref{fig:timeseries}) this does not come as a surprise, as both indices show a comparable development over our sample period. Comparing the shape of the two curves in Figure \ref{fig:nonparametric_stoxx} with the oil price coefficient plotted in Figure \ref{fig:oil} shows that there are also some similarities. All three parameter estimates are positive and show some periods of significance which are longest for the oil price coefficient. While the oil price coefficient becomes significant at the end of 2014 and stays significant until the end of the sample (disregarding the boundary effect), the stock indices have a significant coefficient only from the end of 2014 to the beginning of 2015 and the STOXX 600 index in 2012 for a very short period. The period of significance is therefore overall shorter than for the oil price.   
Adding the two indices in addition to the oil price to the linear regression yields the results presented in Table \ref{tab:lin.reg2}. Neither of the indices has a significant effect on allowance prices when the oil price is already accounted for. 
\begin{table}[h!]
\centering
\caption{Linear Regression Results}
\begin{tabular}{lccccccccc}
 \midrule\midrule
 & \multicolumn{3}{c}{(d)} & \multicolumn{3}{c}{(e)} & \multicolumn{3}{c}{(f)} \\\cmidrule(lr){2-10}
 & $\hat{\beta}_j$ & $se_{NW}$ & $p$-value & $\hat{\beta}_j$ & $se_{NW}$ & $p$-value & $\hat{\beta}_j$ & $se_{NW}$ & $p$-value \\\cmidrule(lr){2-4} \cmidrule(lr){5-7} \cmidrule(lr){8-10}
 Coal  & -0.143 & 0.094 & 0.127 & -0.124 & 0.109 & 0.255 & -0.145 & 0.097 & 0.136\\
 Gas  & 0.176 & 0.078 & 0.024 & 0.213 & 0.074 & 0.004 & 0.177 & 0.077 & 0.022 \\
 Oil  & 0.206 & 0.068 & 0.003 & 0.175 & 0.088 & 0.048 & 0.178 & 0.086 & 0.040 \\
 Stoxx 50 & -- & -- & -- & 0.07 & 0.158 & 0.671  & -- & -- & -- \\
 Stoxx 600 & -- & -- & -- & -- & -- & -- & 0.112 & 0.165 & 0.498 \\
 \midrule\midrule
\end{tabular}
\caption*{\textit{Source:} Own calculations using R.\\\textit{Notes:} Extension of Table 2. Results obtained using OLS estimation. The dependent variable is the return on EUAs and the set of (stationary) regressors changes in each specification. The $p$-values are based on Newey-West standard errors.}
\label{tab:lin.reg2}
\end{table}

\begin{figure}[!h]
	\centering
     \subfigure[$\hat{\beta}_{Hydro}(t)$]
     {
      \includegraphics[width=0.47\linewidth, clip, trim = {0 1cm 0 1.5cm}]{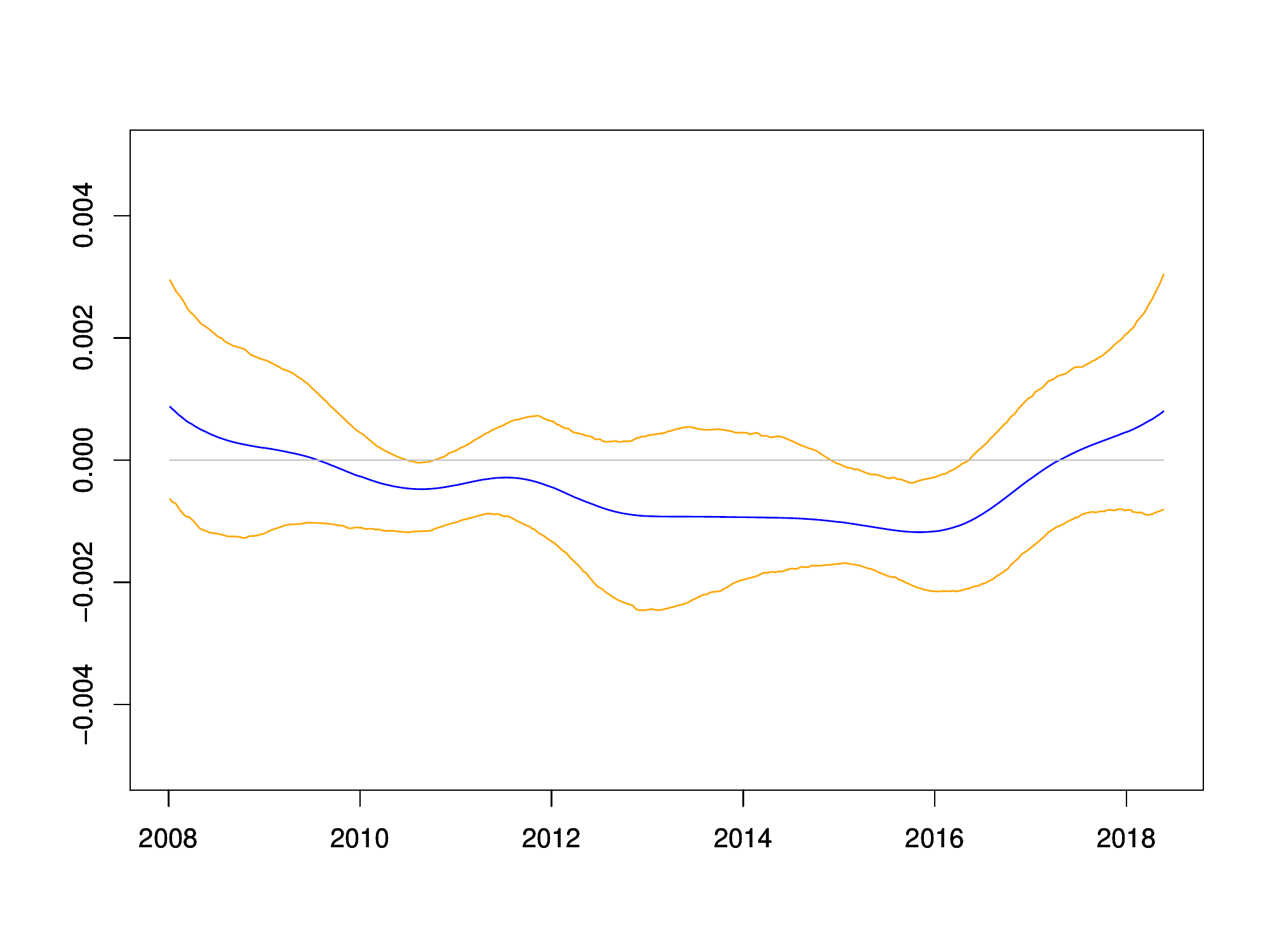}
      \label{fig:hydro}
     }
     \subfigure[$\hat{\beta}_{Wind}(t)$]
     {
      \includegraphics[width=0.47\linewidth, clip, trim = {0 1cm 0 1.5cm}]{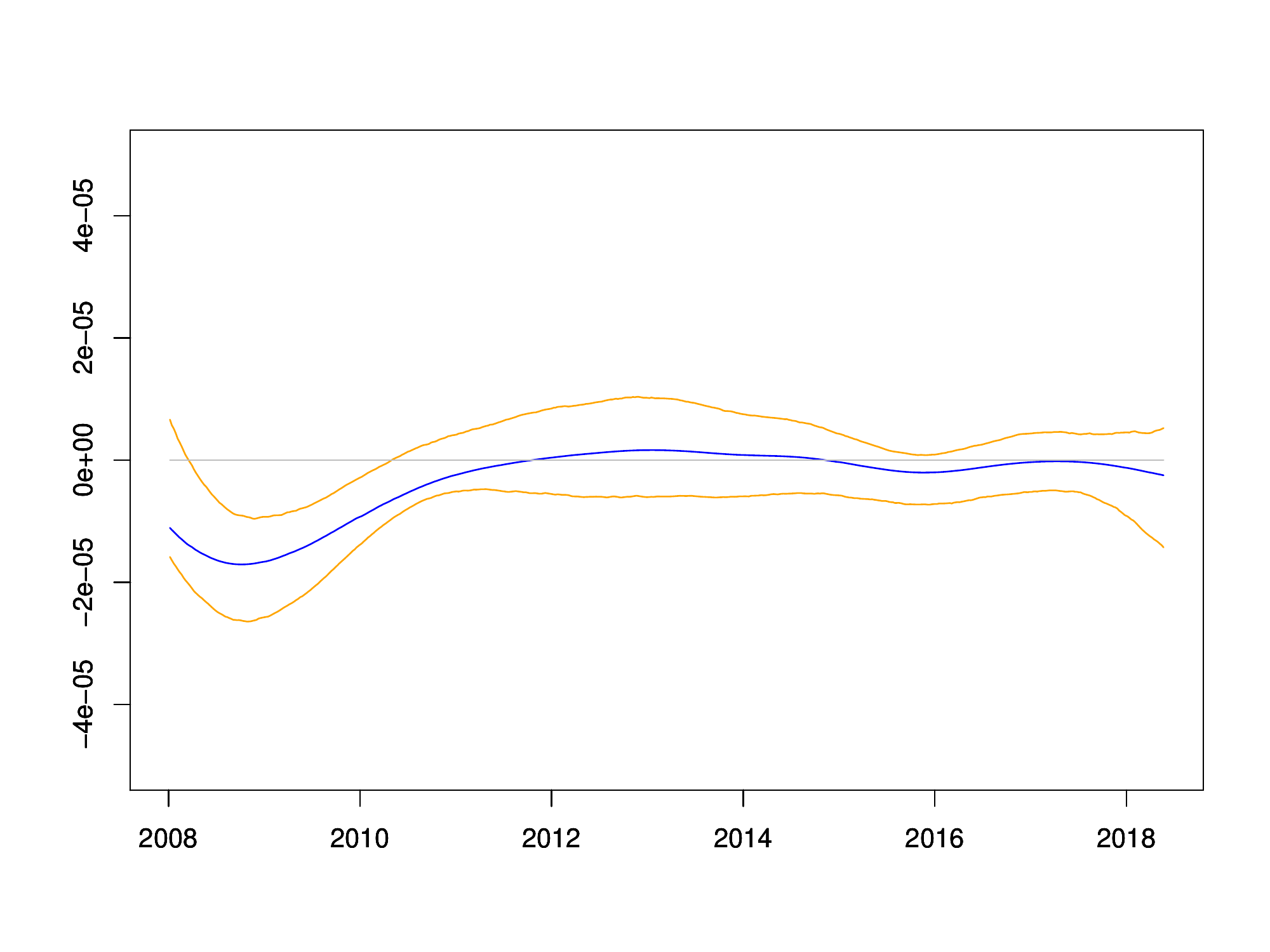}
      \label{fig:wind}
     }
		\caption{Nonparametrically estimated coefficient curves and 95\% confidence intervals for wind and hydro power generation}
		\label{fig:wind_hydro}
\end{figure}

In the second and final part of this section, we add two additional explanatory variables to our nonparametric regression and we replace coal and gas prices by the fuel switching price. We obtain data on energy supply from hydro power in Norway from the Norwegian Water Resources and Energy Directorate as well as data on electricity generation from wind for Germany obtained from the database of the European Network of Transmission System Operators for Electricity (ENTSO-E). Both variables should have a negative effect on allowance prices, as generation from renewable energy sources reduces emissions and therefore the demand for allowances. The hydro power data are weekly data which contain a strong seasonal component which is removed with the help of Fourier terms. This approach is also applied to the temperature data and it is explained in Section \ref{sec:temp_data} below. 
The wind generation data is only available until the end of May 2018 which reduces our sample size for this regression exercise compared to the main analysis. The sample comprises now 517 instead of 538 weekly observations. Added to the nonparametric regression, both new regressors have a coefficient estimate which is extremely low in magnitude for the whole sample. Both estimated coefficient curves are plotted in Figure \ref{fig:wind_hydro}. From panel (a) we see that hydro power is significant over a very brief period in 2016. Panel (b) shows a period of significance for wind at the beginning of the sample until 2010. Although this period is quite long, given the small magnitude of the coefficient (in the order of $10^{-5}$), we consider this effect as negligible. 

\begin{figure}[!h]
	\centering
     \subfigure[Switching price]
     {
      \includegraphics[width=0.47\linewidth, clip, trim = {0 1cm 0 1.5cm}]{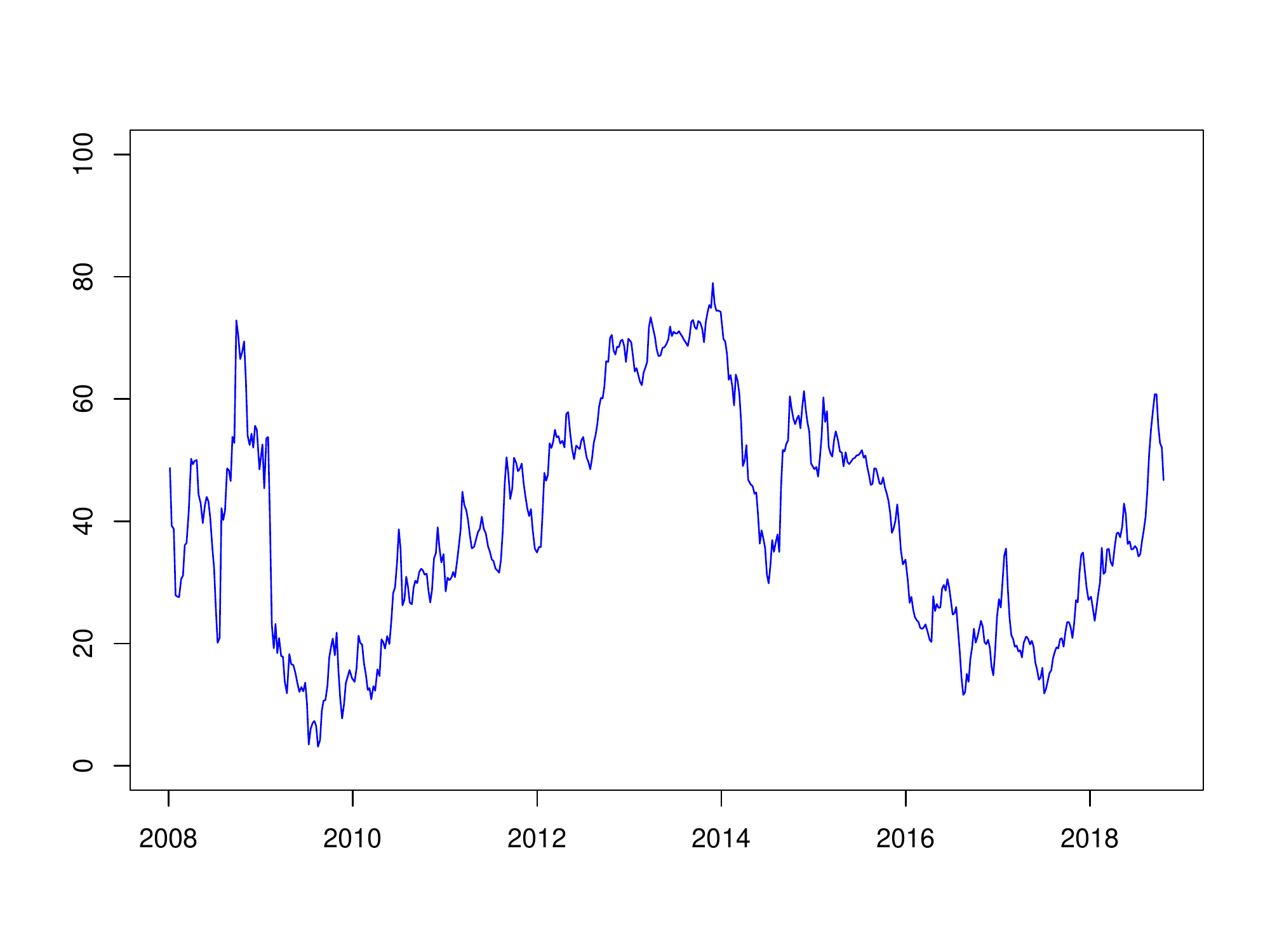}
      \label{fig:switch_price}
     }
     \subfigure[$\hat{\beta}_{Switch}(t)$]
     {
      \includegraphics[width=0.47\linewidth, clip, trim = {0 1cm 0 1.5cm}]{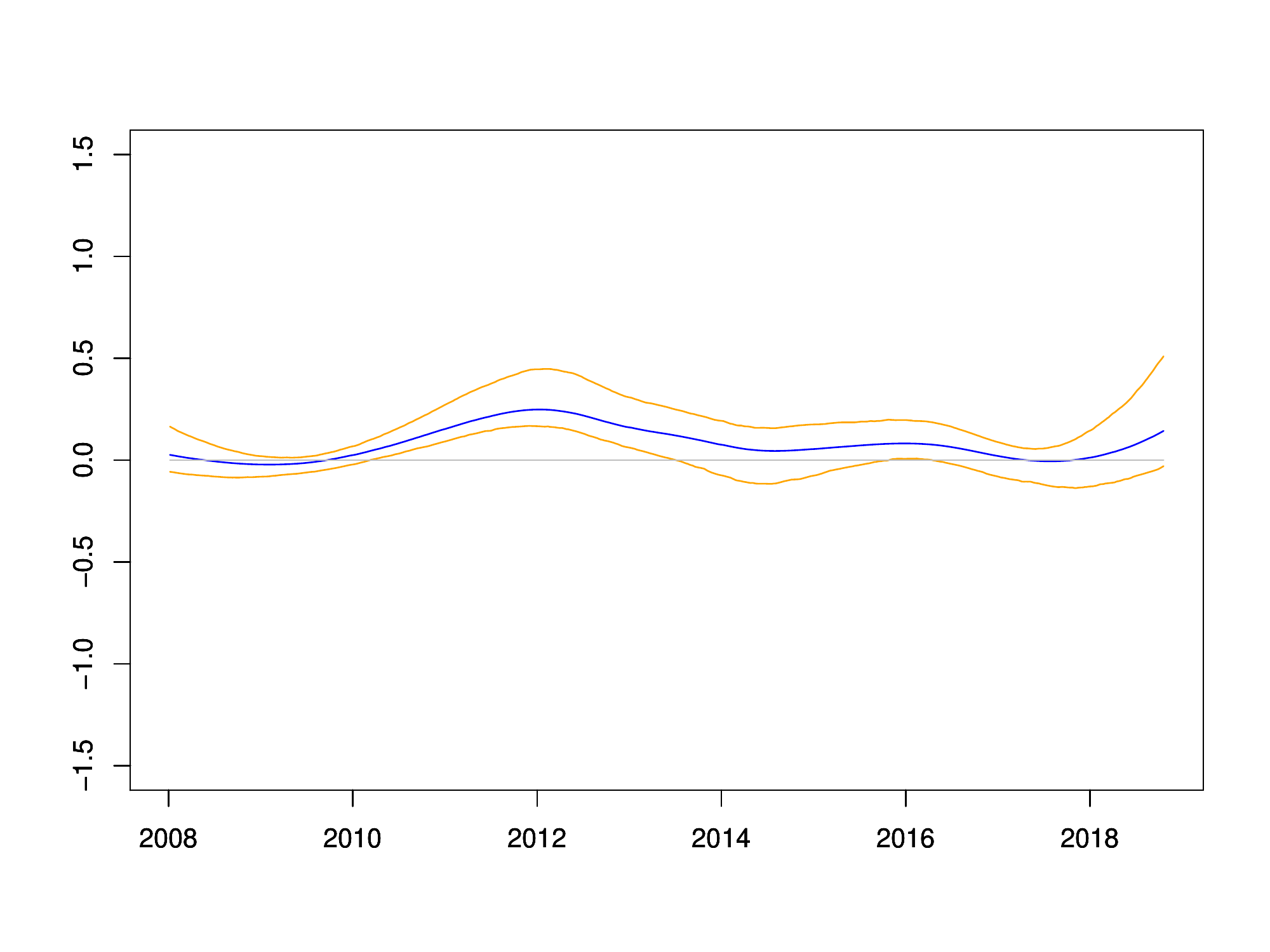}
      \label{fig:beta_switch}
     }
		\caption{Fuel switching price (Panel (a)) and nonparametrically estimated coefficient curves and 95\% confidence intervals for the fuel switching price (Panel (b))}
		\label{fig:switch}
\end{figure}

We calculate the fuel switching price for the switch from coal to gas for electricity generation. The switching price can be obtained from the coal and gas price series together with some additional values: the efficiency of coal and gas plants as well as the respective GHG emission factors. The emission factors are obtained from \citet{Juhrich2016}. The switching price is defined as follows
\begin{equation}
    switch_t=\frac{\eta_{coal}p_{gas}-\eta_{gas}p_{coal}}{\eta_{gas}f_{coal}-\eta_{coal}f_{gas}},
\label{eq:switch}
\end{equation}
where $\eta_{i}$, $f_{i}$ and $p_{i}$ are the plant efficiency, emission factor and fuel prices for $i=coal,gas$, respectively. For details on this we refer to the review by \citet{Delarue2008}. Figure \ref{fig:switch} plots the resulting switching price as well as the estimated coefficient curve when replacing the gas and coal price series by the switching price. 
The switching price displayed in Panel (a) is obtained using our coal and gas price data as well as $\eta_{gas}=0.47$, $\eta_{coal}=0.36$, $f_{gas}=0.202$ and $f_{coal}=0.338$. Note that the gas and the coal price must both be denoted in EUR/MWh. This is why we need to divide our coal price by the conversion factor 8.14.

The general development of the switching price is similar to the gas price as plotted in Figure \ref{fig:gas_prices}. There is a visible upward trend at the end of the sample, which could again be seen as a potential cause of the explosive period in allowance prices. In Section \ref{sec:trend}, we already excluded the coal and the gas price as drivers of this behaviour. Therefore, we do not expect the results for the switching price to be different. For completeness, we obtain the GSADF statistic, which lies with 0.9647 below the critical values indicating no evidence for explosive periods in the switching price. 

From \eqref{eq:switch} we can see that the switching price is a linear combination of the gas and the coal price. Hence, we include it in place of the two price series in our regression. The result is presented in Panel (b) of Figure \ref{fig:switch}. The shape of the estimated coefficient curve closely resembles the coefficient of the gas price which is not surprising given the stronger effect of the gas price on allowance prices compared to the coal price. The estimated coefficient is, however, smaller and the second period of significance vanishes. Including the gas and the coal price separately and not as a linear combination is less restrictive and we therefore do not consider the switching price in our main model.

\subsection{Temperature data}
\label{sec:temp_data}
The temperature data were obtained from the European Climate Assessment \& Dataset (ECA\&D) which provides surface air temperature for 199 measurement stations in Europe. It is provided by The Royal Netherlands Meteorological Institute (KNMI). We refer to \citet{KleinTank} for more details on the temperature series and measurement stations.

\begin{figure}[]
	\centering
     \subfigure[Daily mean temperature]
     {
      \includegraphics[width=0.47\linewidth, clip, trim = {0 1cm 0 1.5cm}]{Temp_data.pdf}
      \label{fig:temp_raw}
     }
      \subfigure[Fourier fit]
     {
      \includegraphics[width=0.47\linewidth, clip, trim = {0 1cm 0 1.5cm}]{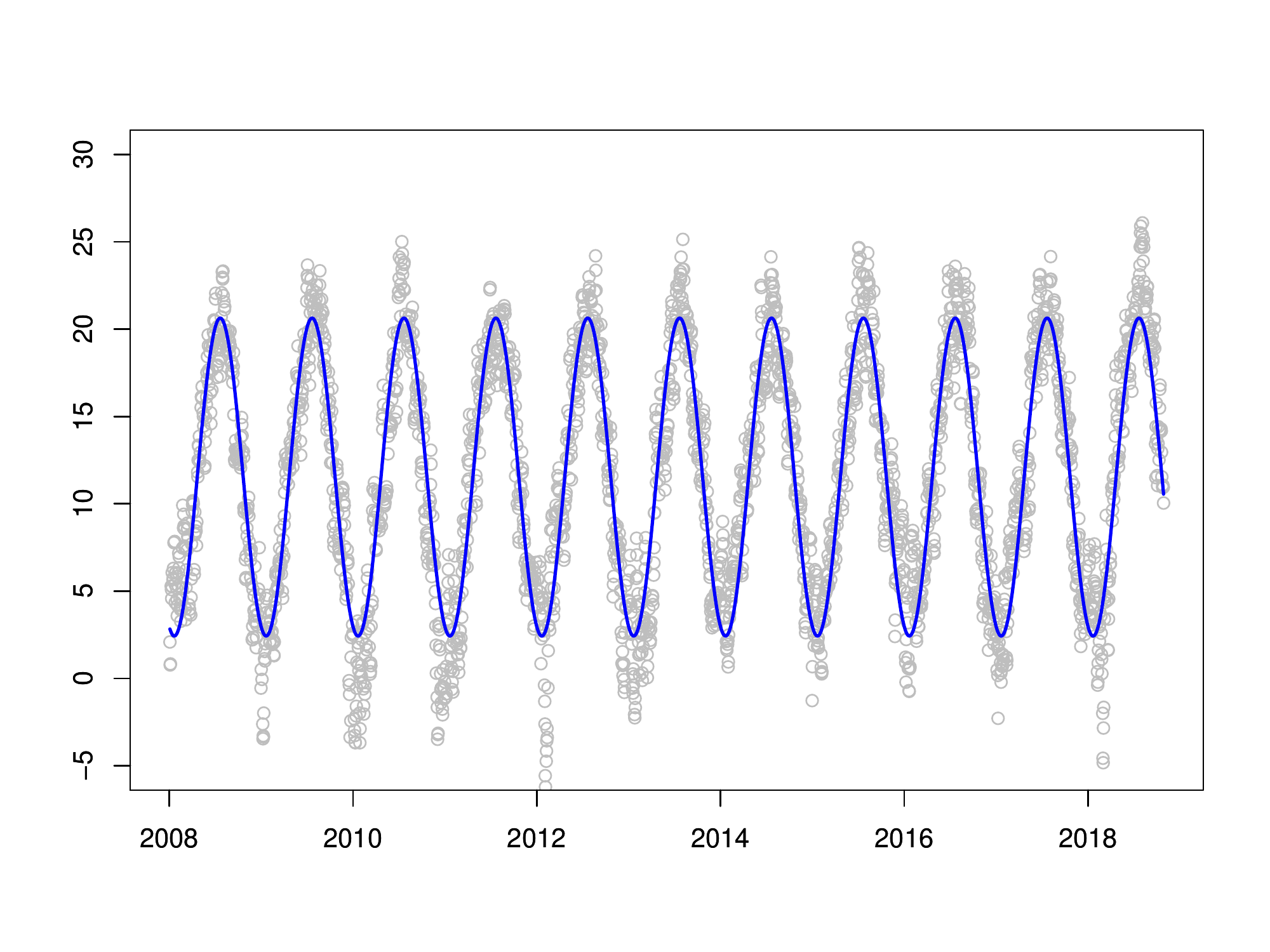}
      \label{fig:temp_fourier}
     }
     \subfigure[Seasonally adjusted]
     {
      \includegraphics[width=0.47\linewidth, clip, trim = {0 1cm 0 1.5cm}]{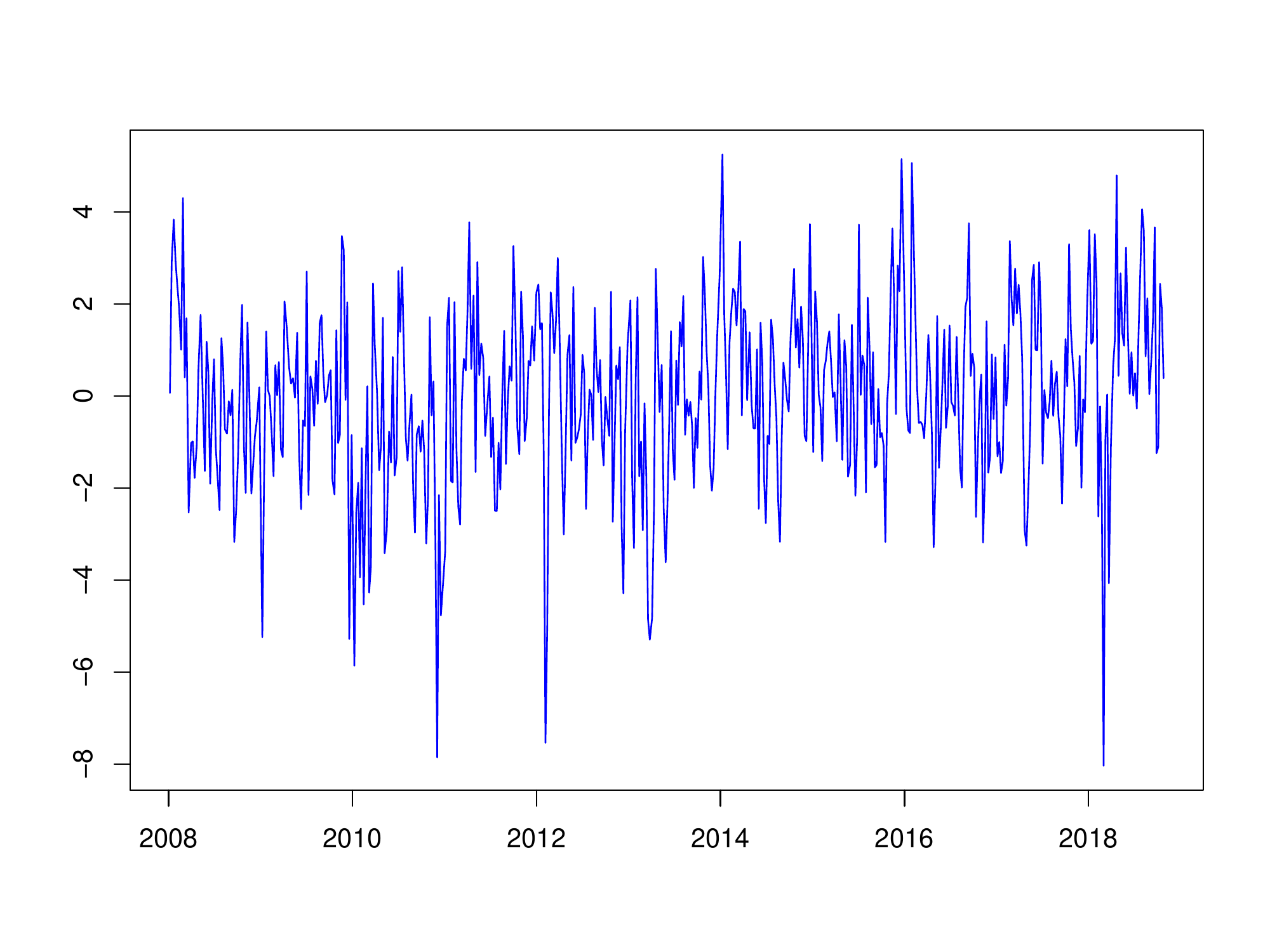}
      \label{fig:temp_seas_adj}
     }
		\caption{Temperature data before and after the removal of the seasonal component.}
		\label{fig:temperature}
\end{figure}

We obtain daily mean temperature series for seven cities, located in seven different countries. They are spread out over Europe: Berlin, Budapest, De Bilt, Dublin, Lyon, Madrid and Stockholm. All series do not contain any missing observations and all were updated until the end of 2018. We take the average over the cities as our temperature series. It is displayed in Figure \ref{fig:temp_raw}. We aggregate the data to weekly means in order to match our sample frequency. In addition, we remove seasonality by fitting a Fourier regression and subsequently, working with the residual series from this regression. This removes the seasonal component with the help of sine and cosine functions. More specifically, we fit the following regression
\begin{equation}
    Temp_t=\alpha_1\cos(2\pi t)+\alpha_2\sin(2\pi t)+\epsilon_t
\end{equation}
and continue to work with the residuals from this regression which are plotted in Figure \ref{fig:temp_seas_adj}. The remaining part (b) of Figure \ref{fig:temperature} plots the daily mean temperature (gray circles) together with the fitted Fourier terms (blue). We can see that the seasonal component is well captured by this method. 

\subsection{Additional BSADF results}
\begin{figure}[t]
	\centering
	\subfigure[EUA prices]
    {
    \includegraphics[width=0.6\linewidth, clip, trim = {0 1cm 0 2cm}]{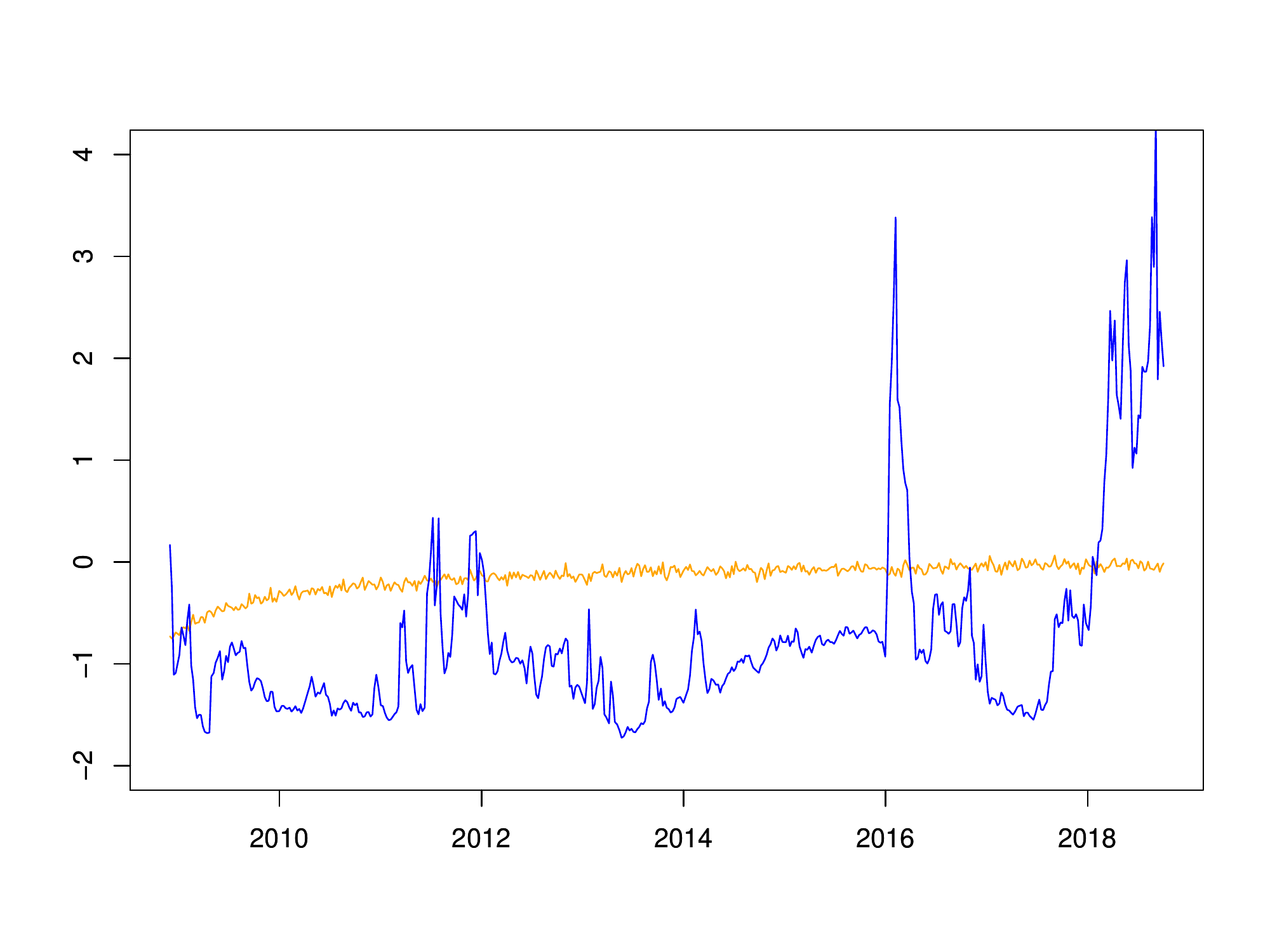}	
    \label{fig:BSADF_eua2}
     }\\
     \subfigure[Oil prices]
    {
    \includegraphics[width=0.6\linewidth, clip, trim = {0 1cm 0 1.5cm}]{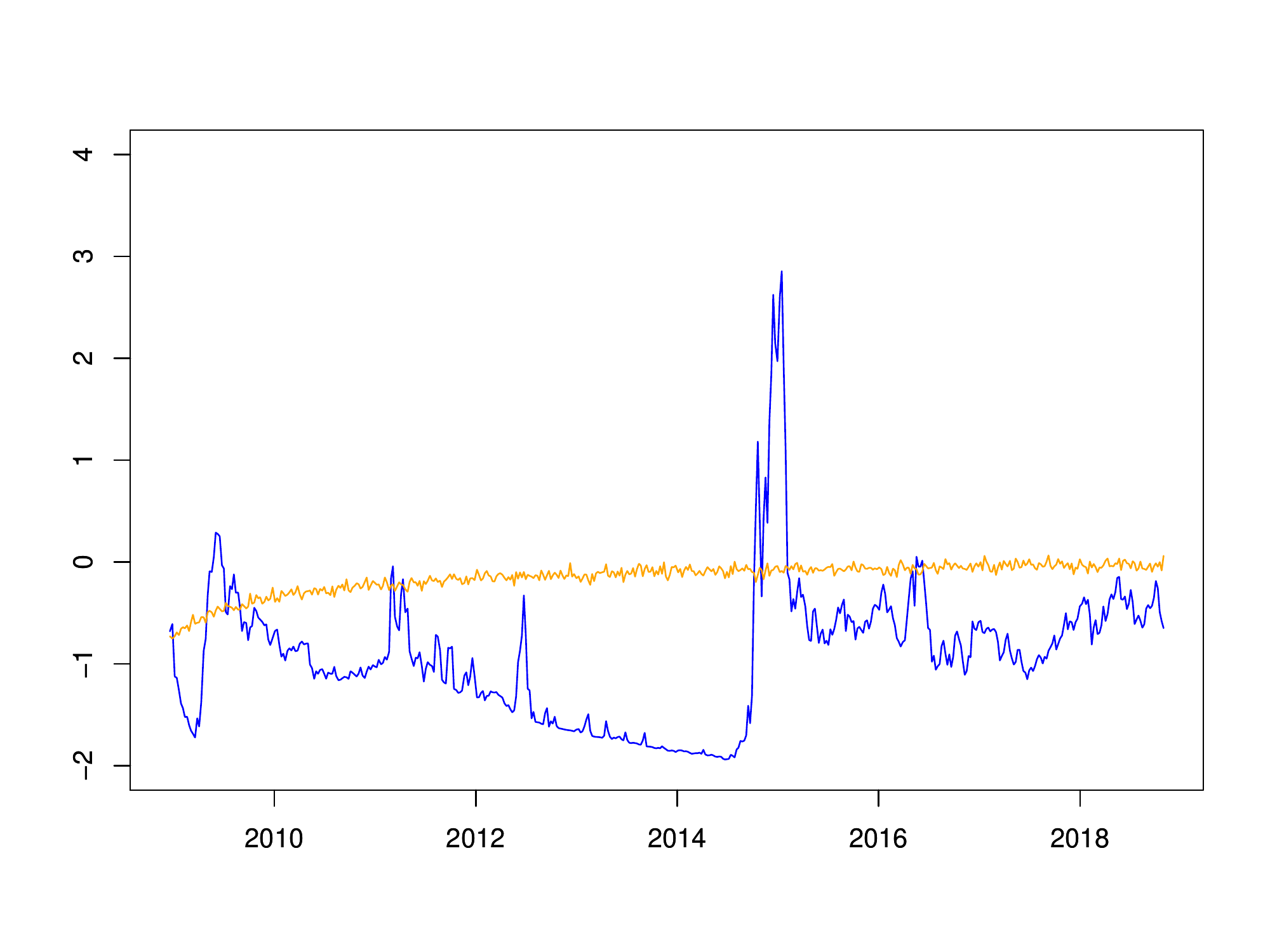}	
    \label{fig:BSADF_oil2}
     }
		\caption{Results from the BSADF test applied to the EUA prices (a) and the oil prices (b). Each panel shows the critical value series (orange) and the test statistics (blue)}
		\label{fig:BSADF2}
\end{figure}

Figure \ref{fig:BSADF2} presents the results from the date-stamping of explosive periods using the BSADF test. Compared to Figure \ref{fig:BSADF} which is the bootstrap version, Figure \ref{fig:BSADF2} plots the series of BSADF test statistics (blue) and the critical value series (orange) obtained by Monte Carlo simulation. Panel (a) presents results from an application on the allowance price series and panel (b) on the oil price series. For both series, the conclusions are the same as presented in the main paper. The only difference lies in the starting point of the explosive period in allowance prices; while it was March 2018 in Figure \ref{fig:BSADF} it is February in Figure \ref{fig:BSADF2}. This minor difference is not surprising given the fact that the bootstrapped critical values are larger due to the size correction.

\end{appendices}

\end{document}